\documentclass[twocolumn,superscriptaddress,amsmath,amssymb,aps,prd]{revtex4-2}
\usepackage{amsthm,amsfonts,mathrsfs,mathtools}
\usepackage{graphicx,subfloat,subfig,color}
\usepackage[justification=raggedright]{caption}
\usepackage{dcolumn}
\usepackage{bm}
\definecolor{PRDblue}{RGB}{46,48,146}
\definecolor{PRDred}{RGB}{220,20,60}
\usepackage[colorlinks,citecolor=PRDred,linkcolor=PRDblue,urlcolor=PRDblue,linktocpage,breaklinks=true]{hyperref}

\ExplSyntaxOn
\cs_new:Npn \pd {\partial}
\cs_new:Npn \ed {\mathrm d}
\cs_new:Npn \im#1 {\mathrm{im}\,{#1}}
\cs_new:Npn \Ric {\mathcal R\mathrm{ic}}
\cs_new:Npn \tr#1{
	\str_if_eq:nnTF{#1}{R}
	{\mathrm{tr}\c_math_subscript_token g\mathcal R\mathrm{ic}}
	{\str_if_eq:nnTF{#1}{K}
		{\mathrm{tr}\c_math_subscript_token h\mathcal K}
		{\mathrm{tr}{#1}}
	}
}
\cs_new:Npn \K {\mathscr K}
\cs_new:Npn \m {1,\cdots,m}
\cs_new:Npn \up#1#2{
\bool_if:nTF{\str_if_eq_p:nn{#1}{V} || \str_if_eq_p:nn{#1}{W}||\str_if_eq_p:nn{#1}{M}||\str_if_eq_p:nn{#1}{P}|| \str_if_eq_p:nn{#1}{O}}
{
{\mathscr #1}^{[#2]}
}
{
\bool_if:nTF{\str_if_eq_p:nn{#1}{K}||\str_if_eq_p:nn{#1}{R}||\str_if_eq_p:nn{#1}{G}||\str_if_eq_p:nn{#1}{T}}
{
{\mathcal #1}^{[#2]}
}
{
{#1}^{[#2]}
}
}
}
\cs_new:Npn \GN {G\c_math_subscript_token {\text N}}
\cs_new:Npn \diag {\mathrm{diag}}
\cs_new:Npn \Span {\mathrm{Span}}
\cs_new:Npn \SF {\mathsf{SF}}
\cs_new:Npn \II {\mathrm{I\!I}}
\cs_new:Npn \AdS {\text{AdS}}
\cs_new:Npn \ads#1 {\text{AdS}_#1}
\cs_new:Npn \Sch#1 {
\mathsf{Sch}\big[{#1}^{[i]}\big]
}
\ExplSyntaxOff
\newtheorem{theorem}{Proposition}

\begin{document}

%\preprint{arXiv:0000.00000}
\noindent\hfill
\texttt{USTC-ICTS/PCFT-24-09}

\title{Multiway Junction Conditions: Booklets and Webs}

%\affiliation{Kavli Institute for Theoretical Sciences (KITS), University of Chinese Academy of Sciences, Beijing 100190, China}
\author{Jia-Yin Shen}
%\email[]{jiayin.shen01@gmail.com}
%\email[]{jiayin.shen@pku.edu.cn}
\affiliation{The Kavli Institute for Astronomy and Astrophysics, Peking University, Beijing 100871, China}
\author{Cheng Peng}
\email[Contact author:~]{pengcheng@ucas.ac.cn}
\affiliation{Kavli Institute for Theoretical Sciences (KITS), University of Chinese Academy of Sciences, Beijing 100190, China}
\affiliation{Peng Huanwu Center for Fundamental Theory, Hefei, Anhui 230026, China}
\author{Li-Xin Li}
%\email[]{lxl@pku.edu.cn}
\affiliation{The Kavli Institute for Astronomy and Astrophysics, Peking University, Beijing 100871, China}

\begin{abstract}
  Junction conditions play a crucial role in constructing new gravity solutions. In this paper, we derive the junction condition for gluing together an arbitrary number of spacetimes along a common interface. We develop a geometric technique of reverse extension and provide precise definitions of geometric quantities at the interface. This leads to a geometric derivation of the multiway junction condition. As a cross-check, we independently re-derive the junction condition by varying the action of some specific gravitational models including both Einstein gravity and dilaton gravity. We demonstrate that the junction condition is invariant under a change of frames, and the form of the junction condition is the same for both 
  spacelike and timelike interfaces.\\
  \textbf{Related DOI:}\href{https://doi.org/10.1103/PhysRevD.110.066021}{10.1103/PhysRevD.110.066021}
\end{abstract}

\maketitle

\section{Introduction\label{sec1}}
Junction conditions at surfaces of discontinuity are one of the central problems in gravitational theories. The junction condition across a singular surface between two bulks in general relativity was formulated in~\cite{Israel1966} based on previous works~\cite{Lanczos1,Lanczos2,MSM1927}. Subsequently, junction conditions between two bulks have been developed in different gravity  theories~\cite{PhysRevD.106.064007,Brassel2023,Deruelle2008junction,Aviles2020,PhysRevD.67.024030,PhysRevD.103.104069,CruzDombriz2014,Rosa:2021mln,Rosa:2023tph}. Recently, junction conditions also play a crucial role in the breakthrough about the black hole information paradox~\cite{Penington:2019npb, Penington:2019kki, Almheiri:2019qdq,Almheiri:2019hni,Almheiri:2019psy,Bousso:2020kmy,Chen:2020uac,Chen:2020hmv,Hernandez:2020nem,Suzuki:2022xwv,PhysRevLett.129.231601,Geng:2020fxl} via its application in various braneworld models~\cite{PhysRevLett.83.3370,PhysRevLett.83.4690,PhysRevD.62.024012,Seahra2009,Geng2022,Wang2016RSII,Li2016}. 

\begin{figure}[htbp!]
	\centering
	\subfloat[]{\raisebox{6.2\height}{\includegraphics[width=0.43\columnwidth]{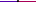}}}
	\label{Bookletgeometry1}
	\hspace{10pt}
	\subfloat[]{\includegraphics[width=0.33\columnwidth]{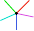}
	}
	\label{Bookletgeometry2}
	\caption{(a)Booklet with two pages where the Israel junction condition applies across the interface. (b)Booklet with six pages where the junction condition we derived applies across the interface. An illustration of the booklet geometries, where the common interface is denoted by black dots and different bulks to be glued together are denoted by the colorful lines; the interface is always of co-dimension $1$ and we suppress all the other dimensions of each bulk so the different bulks appear as curves.}
	\label{Bookletgeometry}
\end{figure}

Existing studies about junction conditions have hitherto focused solely on gluing two bulks, $\up V 0$ and $\up V 1$ along an interface denoted as $\Sigma$. This paper explores the geometry where an arbitrary number of bulks are glued along a common interface. Such multi-boundary configuration receives increasing attention due to its relevance to replica wormholes and the geometric dual of higher point spectral correlation functions that connect multiple boundaries~\cite{Maldacena:2004rf,Penington:2019kki,Almheiri:2019qdq,Saad:2019lba,Witten:2020wvy,Maxfield:2020ale,Mertens:2020hbs}. Concretely, consider gluing together $m+1$ $(d+1)-$dimensional spacetimes $\up V i$, $i=0,1,\cdots,m$, each referred to as a ``page'', along a common codimension-$1$  interface $\Sigma$ to form a ``booklet''. Mathematically, let $\partial \up V i$ be the boundary of $\up V i$ that is to be glued. Let $\psi_i$ be  diffeomorphisms between $\partial \up V i$ and $\partial \up V {i+1}$, $i=0,\ldots, m-1$. An equivalence relation $\sim$ among points on all  $\partial \up V i$ can be induced by composition of $\psi_i$; namely $p\sim q$ if $p\in\partial \up V j$, $q\in\partial \up V i$,  and $p=\psi_{j-1}\circ \psi_{j-2}\circ\cdots \circ \psi_{i}(q)$ assuming $j>i$. This equivalence relation then identifies all $\partial \up V i$ as $\Sigma$, and defines the glued geometry $\mathscr V=\bigsqcup_{i=0}^m \up V i/\sim$, where $\bigsqcup$ represents disjoint union. In addition, the ability to define  gravity consistently on $\mathscr V$ imposes further constraints on the gluing, which are the junction conditions that is the main focus of the rest of this paper. To host gravity, we assume  each page  admits a metric $\up g i$ that are compatible in the tangent space of $\Sigma$, $\mathscr T(\Sigma)$, and induces on $\Sigma$ a metric	$h\coloneqq\up g 0\big|_{\mathscr T(\Sigma)}=\ldots =\up g m\big|_{\mathscr T(\Sigma)}$. Let $\up n i$, pointing from $\Sigma$ into the interior of $\up V i$, be the normal vector of the interface $\Sigma$ within $\up V i$. We denote by $\up K i$ the extrinsic curvature of $\Sigma$ within $\up V i$. The ramification at $\Sigma$ in the glued spacetime makes this construction less straightforward than the case of gluing two bulks. This is solved by noticing  in the derivation of the Israel junction condition an implicit ``continuity condition'' of the normal vectors, i.e. $\up n 0+\up n 1=0$, where $\up n 0$ and $\up n 1$ are the normal vectors with appropriately selected directions in the two bulks, and a key step in our construction is to properly extend the above continuity condition to $\up n 0+\up n 1+\cdots +\up n m=0$ between multiple bulks (see Sec.~\ref{subsec5A} and Appendix~\ref{sec3} for rigorous discussion). This paves the way of writing down a junction condition for gluing together any number of bulks. Since the geometry so constructed have the topology of multiple bulks glued at the same codimension-1 interface, we refer them as ``booklet geometry" and each bulk is refered to as a ``page" of the booklet, as illustrated in Fig.~\ref{Bookletgeometry}.

This paper is organized as follows: in Sec.~\ref{sec2}, we rederive the classical Israel junction condition by defining varous physical quantities at $\Sigma$ and establishing a geometric framework. In Sec.~\ref{sec4}, we develop the method of reverse extension, i.e. introducing an auxiliary bulk $\up V{-0}$ and deriving the junction conditions with the help of it. In Sec.~\ref{sec5}, we explore the gluing of $m+1$ bulks, determine the form the continuity conditions, derive the junction condition between multiple pages with both the methods of reverse extension and the variational principles, and discuss generically the geometric structure of the glued ``booklet'' $\mathscr V$ embedded in a $d+m$-dimensional manifold $\mathscr M$. In Sec.~\ref{sec6}, we derive the junction conditions in dilaton gravity where the tension on the interface could either couple to the dilaton or not. Finally, Sec.~\ref{cmts} contains some discussions. We also discuss the web geometries obtained from gluing multiple bulks on multiple interfaces. 

\section{Review of the Israel junction condition between two bulks\label{sec2}}
The junction condition at the interface, which is a singular hypersurface of  co-dimension $1$,  between two bulks reads~\cite{Israel1966} 
\begin{equation}
	\label{jcfor2bulks}
	\big(\up K 0+\up K 1\big)-\big(\up{\tr K}0+\up{\tr K}1\big)h=-8\pi G\,{ \mathsf s\,}\bar {\mathcal S},
\end{equation}
where $\up K i$ is the extrinsic curvature in the $i^{\text{th}}$ bulk and $\bar {\mathcal S}$ is the surface energy-momentum tensor, and the induced metrics are required to be continuous across the interface. For a timelike interface, the sign $\mathsf s = +1$, and for a spacelike one, $\mathsf s = -1$. In this section, we provide a geometric derivation of this condition as a preparation to study the multi-way junction condition in the subsequent sections. Our motivation is the observation that when it comes to deriving this junction condition, the predominant approach in the literature involves the use of generalized functions/distributions. Assuming two spacetime regions equipped with the metrics $g^+_{\mu\nu}$ and $g^-_{\mu\nu}$ respectively, at their interface $\ell=0$, the Heaviside step distribution $\Theta(\ell)$ splices together these two metrics as $g_{\mu\nu}=\Theta(\ell)g^+_{\mu\nu}+\Theta(-\ell)g^-_{\mu\nu}$. Subsequently, $g_{\mu\nu}$ is used to calculate the connection and curvature, and further, the Einstein tensor. Where the definition of $\Theta(\ell)$ is as follows: when $\ell > 0$, $\Theta(\ell) = 1$; when $\ell < 0$, $\Theta(\ell) = 0$; however, $\Theta(0)$ is indeterminate. The relationship between $\Theta(\ell)$ and the Dirac $delta$ function $\delta(\ell)$ is expressed as $\Theta'(\ell)=\delta(\ell)$. It is crucial to emphasize that the product $\delta(\ell)\cdot\Theta(\ell)$ does not constitute a distribution. More comprehensive information can be found in, e.g.~\cite{Poisson2004}.

In this section, we aim to derive the junction condition in a fully geometrized fashion, which offers several distinct advantages. Firstly, it unveils the geometric essence that the distribution method conceals. Because of the lack of continuity of higher-order derivatives at the interface, the distribution method, in fact, sidesteps a clear definition of geometric quantities. In contrast, the geometrized approach allows for more rigorous definitions of curvature and other relevant quantities. Secondly, the geometric method directly illustrates that the junction condition is independent of coordinate systems or reference frames. This feature is highly convenient. Because our goal is to address the gluing of multiple bulks, in this scenario, a globally consistent coordinate open set cannot exist a priori. Last but not least, the geometrized method provides a framework for extending the junction condition. When dealing with multiple bulks, it is not feasible to categorize them based on $\ell>0$ or $\ell<0$. Therefore, using $\Theta(\ell)$ to splice together geometric quantities will fail.

Let us begin by stating our conventions. We represent two $d+1$-dimensional bulk spacetimes with boundaries as $\up V i$, where $i=0,1$. By identifying the two boundaries through a diffeomorphism $\pd \up V 0\sim \pd \up V 1$, we construct a common $d$-dimensional interface, which is labeled as $\Sigma$. Consequently, we have glued $\up V 0$ and $\up V 1$ together to create a unified structure, denoted as $\mathscr V$. Assume that $\up V i$ is equipped with the metric $\up g i$, and the normal vector of $\Sigma$ within $\up V i$ is $\up n i$. We adopt a further convention that the direction of $\up n i$ points from $\Sigma$ to $\up V i$. While this convention may not be traditional, it proves to be convenient for our future extensions. In the first half of this paper, we eschew the conventional tensor index notation and refrain from utilizing component expressions. This choice is driven by the emphasis in these sections on crafting a theoretical framework, devoid of computational requirements. Introducing both bulk indices $[i]$ and component indices of tensors would create ``symbolic monsters,'' unnecessarily complicating the equations. 

The bulk metrics $\up g i$ induce metrics $\up g i\big|_{\mathscr T(\Sigma)}$ on the interface $\Sigma$. Here the symbol $\big|_{\mathscr T(\Sigma)}$ indicates restricting the operation domain of the metric into the tangent space $\mathscr T(\Sigma)$ of the interface. This immediately gives rise to a natural requirement, we must be able to define a consistent metric $h$ inside $\Sigma$, namely, 
\begin{equation}
	h\coloneqq\up g 0\big|_{\mathscr T(\Sigma)}=\up g 1\big|_{\mathscr T(\Sigma)},
\end{equation}
so that we can have a field theory living on the interface. If we further demand that the metrics $\up g i$ of the bulks are continuous at $\Sigma$, namely $\up g 0\big|_\Sigma=\up g 1\big|_\Sigma$, where the symbol $\big|_\Sigma$ denotes the metric taking value at $\Sigma$, and should not be confused with the symbol $\big|_{\mathscr T(\Sigma)}$, the following relation between normal vectors $\up n i$ can be obtained:
\begin{equation}
	\label{dim2continu}
	\up n 0+\up n 1=0\,,
\end{equation}
which means they are either both spacelike or both timelike. We call this equation as the continuity condition for the normal vector. Implicitly, this condition suggests the differentiability of the spacetime $\mathscr V$ at $\Sigma$. Originally, $\up n 0$ and $\up n 1$ were defined within distinct tangent spaces, i.e., $\up n i\in \mathscr T_\Sigma(\up V i)$; however, to perform algebraic operations on $\up n 0$ and $\up n 1$ as shown in Eq.~\eqref{dim2continu} implies that all $\mathscr T_\Sigma(\up V i)$ are identified as the same space $\mathscr T_\Sigma(\mathscr V)$. 

Let $\bar X$ and $\bar Y$ represent vectors or vector fields tangent to $\Sigma$. Then, the second fundamental form $\up\II i$ of $\Sigma$ within $\up V i$ can be expressed as
\begin{equation}
	\label{IIi}
	\up \II i\big(\bar X,\bar Y\big)=\up\nabla i_{\bar X}\bar Y-\bar\nabla_{\bar X}\bar Y=\lim_{s\to 0}\frac{\bar\tau_{s\bar X}\bar Y-\up \tau i_{s\bar X}\bar Y}{s},
\end{equation}
where $\up\nabla i$ and $\bar\nabla$ represent the covariant derivatives induced by the metric $\up g i$ and $h$ respectively, and $\up\tau i_{s\bar X}\bar Y, \bar{\tau}_{s\bar X}\bar Y$ describes the parallel translations of the vector $\bar Y$ along the integral curve of $\bar X$. Mathematically, it is straightforward to prove that $\up \II i\big(\bar X,\bar Y\big)\propto \up n i$. Then the exterior curvature can be defined as
\begin{eqnarray}
	\notag
	\up\II i\big(\bar X,\bar Y\big)&=&-\mathsf s\,\up K i\big(\bar X,\bar Y\big)\up n i,\\
	 \up K i(\bar X,\bar Y)&=&-\up g i\big(\up\II i(\bar X,\bar Y),\up n i\big).
	 \label{defineKi}
\end{eqnarray}
In fact, the reason for introducing the additional sign $\mathsf s$ in Eq.\eqref{defineKi} is to maintain consistency with the commonly used definition of extrinsic curvature in physics, namely $h\big(\up\nabla i_{\bar X}\up n i, \bar Y\big) = \up K i(\bar X, \bar Y)$ and $\up K i = \frac{1}{2}\mathfrak L_{\up n i}h$, where $\mathfrak L_{\up n i}$ represents the Lie derivative. 

\begin{figure}[htbp!]
	\centering
	\includegraphics[width=0.7\columnwidth]{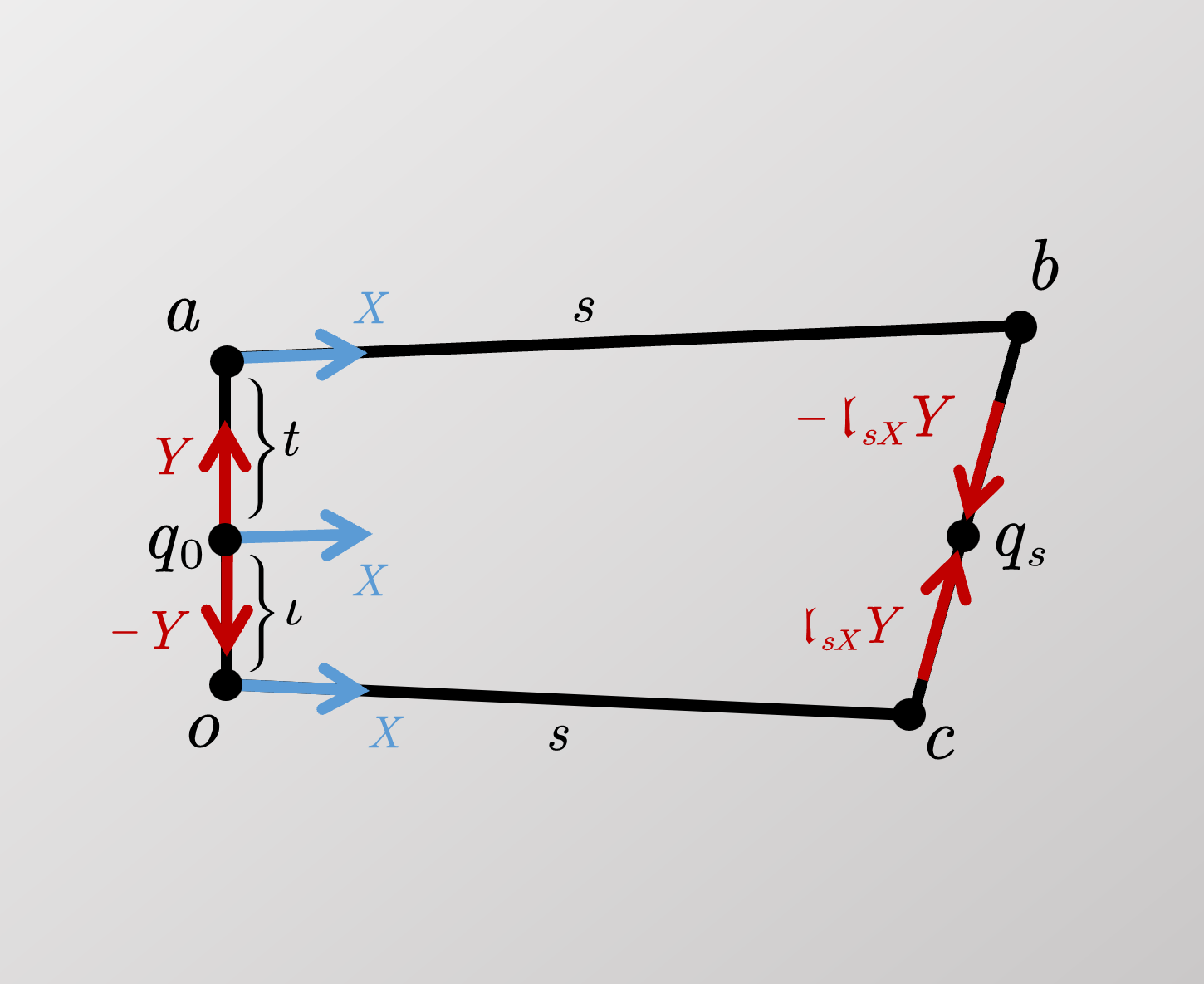}
	\caption{Defining curvature tensors by transport.}
	\label{Rq0}
\end{figure}

The curvature tensor $\up R i$ operates on any two tangent vector field $X$ and $Y$ within the same bulk, which results in a linear mapping $\up R i(X,Y)$,
\begin{eqnarray}
	\label{defineRi}
	\notag
	\up R i(X,Y) U&=&\up\nabla i_X\up \nabla i_Y U-\up\nabla i_Y\up \nabla i_X U-\up\nabla i_{[X,Y]}U\\
	&=&\lim_{s,t\to 0}\frac{\up\tau i_{sX}\up\tau i_{tY}U-\up\tau i_{t \vec Y}\up\tau i_{sX}U}{st}\,,
\end{eqnarray}
where the vector field $\vec  Y$ is defined as $\vec  Y=\mathfrak l_{sX}Y$,  with $\mathfrak l_{sX}$ representing Lie transport along $X$. Equation \eqref{defineRi} demonstrates the association of the curvature tensor with the closed loops (elements of the holonomy group) constructed by $X$ and $Y$, and $s=0$, $t=0$ marks the ``starting'' point  $q_0$ of the loop, which is also where $\up R i(X,Y)$ is evaluated at. In fact, the Ambrose-Singer theorem tells us that the curvature forms the holonomy algebra~\cite{kobayashi1996}. Therefore, we have the flexibility to choose equivalent elements from the holonomy group. We can therefore construct different  homotopic loops, as shown in Fig.~\ref{Rq0}, to extend Eq.~\eqref{defineRi} as
\begin{equation}
	\label{expanddefineRi}
	\up R i(X,Y)U = \lim_{s,\iota,t\to 0}\frac{\up\tau i_{-t\vec  Y}\up\tau i_{sX}\up\tau i_{t Y}U-\up\tau i_{\iota\vec  Y}\up\tau i_{sX}\up\tau i_{-\iota Y}U}{(t+\iota)s}\,,
\end{equation}
The parameter $t$ represents the positively-oriented integral curves of the vector $Y$, while $\iota$ is the parameter for the vector $Y$ in the negative direction, or equivalently, for $-Y$ in the positive direction, see e.g. Fig.~\ref{Rq0}. Equation \eqref{expanddefineRi} and \eqref{defineRi} are equivalent, which can be shown as 
\begin{align}
	\notag &
	\lim_{s,\iota,t\to 0}\frac{\tau_{-t\vec Y}\tau_{sX}\tau_{t Y}U-\tau_{\iota\vec Y}\tau_{sX}\tau_{-\iota Y}U}{(t+\iota)s}  \\
	\notag &
	=\lim_{s,\iota,t\to 0}\frac{\tau_{-t\vec Y}\tau_{sX}\tau_{t Y}\tau_{\iota Y}(\tau_{-\iota Y}U)-\tau_{\iota\vec Y}\tau_{sX}(\tau_{-\iota Y}U)}{(t+\iota)s}       \\
	\notag &
	=\lim_{s,\iota,t\to 0}\tau_{-t\vec Y}\frac{\tau_{sX}\left(\tau_{t Y}\tau_{\iota Y}\right)-\left(\tau_{t\vec Y}\tau_{\iota\vec Y}\right)\tau_{sX}}{(t+\iota)s}\left(\tau_{-\iota Y}U\right) \\
	\notag &
	=\lim_{s,\iota,t\to 0}\tau_{-t\vec Y}\frac{\tau_{sX}\tau_{(t+\iota) Y}-\tau_{(t+\iota)\vec Y}\tau_{sX}}{(t+\iota)s}\tau_{-\iota Y}U                                                                \\
	&= \lim_{s,\iota,t\to 0}\frac{\tau_{sX}\tau_{(t+\iota) Y}-\tau_{(t+\iota)\vec Y}\tau_{sX}}{(t+\iota)s}(U),
\end{align}
where the last line is the leading contribution in the limit $s,\iota,t\to 0$, which comes from the leading contribution from each of the three factors involving $\tau$ in the fourth line. In particular, the leading contribution of the first and the last factor is just the identity operator.  A change of variable from $(t+\iota)$ to $t$ (since both approach $0$) in the last line gives~\eqref{defineRi}. The advantage of equation~\eqref{expanddefineRi} is that it can be easily generalized to the case where $q_0$ is at the interface $\Sigma$. The curvature at $\Sigma$, denoted as $\up R\Sigma$, should encompass contributions from both $\up R 0$ and $\up R 1$. To define it, we attempt to construct all inequivalent loops $\mathscr O$ at $\Sigma$. In reality, there are only two independent cases, the first involves a loop entirely tangent to $\Sigma$, corresponding to the linear map $\up R\Sigma(\bar X, \bar Y)$. We will discuss this simple case later; for now, let us focus on the second possibility: a loop crosses $\Sigma$ and intersects it at two points, $q_0$ and $q_s$, corresponding to $\up R\Sigma(\bar X, \up n i)$. In this case, the loop can be seen as composed of two paths end to end,  $\up P 0$ and $\up P 1$, both connecting $q_0$ and $q_s$, i.e., $\mathscr O =\up P 0 - \up P 1$. To simultaneously include contributions from both bulks, we place these two paths inside $\up V0$ and $\up V1$, respectively. Therefore, for the component $\up R\Sigma(\bar X, \up n i)\bar U$, we define it as follows,
\begin{eqnarray}
	\label{addpathSigma}
	\notag
	&&\up R \Sigma(\bar X,\up n 0)\bar U\\
	\notag
	&&=-\up R \Sigma(\bar X,\up n 1)\bar U\\
	&&= \lim_{s,\iota,t\to 0^+}\frac{\up\tau0_{-t\vec  n^{[0]}}\up\tau0_{s\bar X}\up\tau0_{t n^{[0]}}\bar U-\up\tau1_{-\iota\vec  n^{[1]}}\up\tau1_{s\bar X}\up\tau1_{\iota n^{[1]}}\bar U}{(t+\iota)s},
\end{eqnarray}
 where $t,\iota$ now become infinitesimal parameters of the integral curves of $\up n 0,\up n 1$  respectively. We require $t,\iota$ to remain positive as they approach zero to ensure that the integral curves of $\up n i$ lie within $\up V i$ and do not cross $\Sigma$ into another bulk. A finite rescaling of these parameters changes the ``speed'' of this limiting process but does not change the result. 
 
\begin{figure}[t]
	\centering
	\includegraphics[width=0.7\columnwidth]{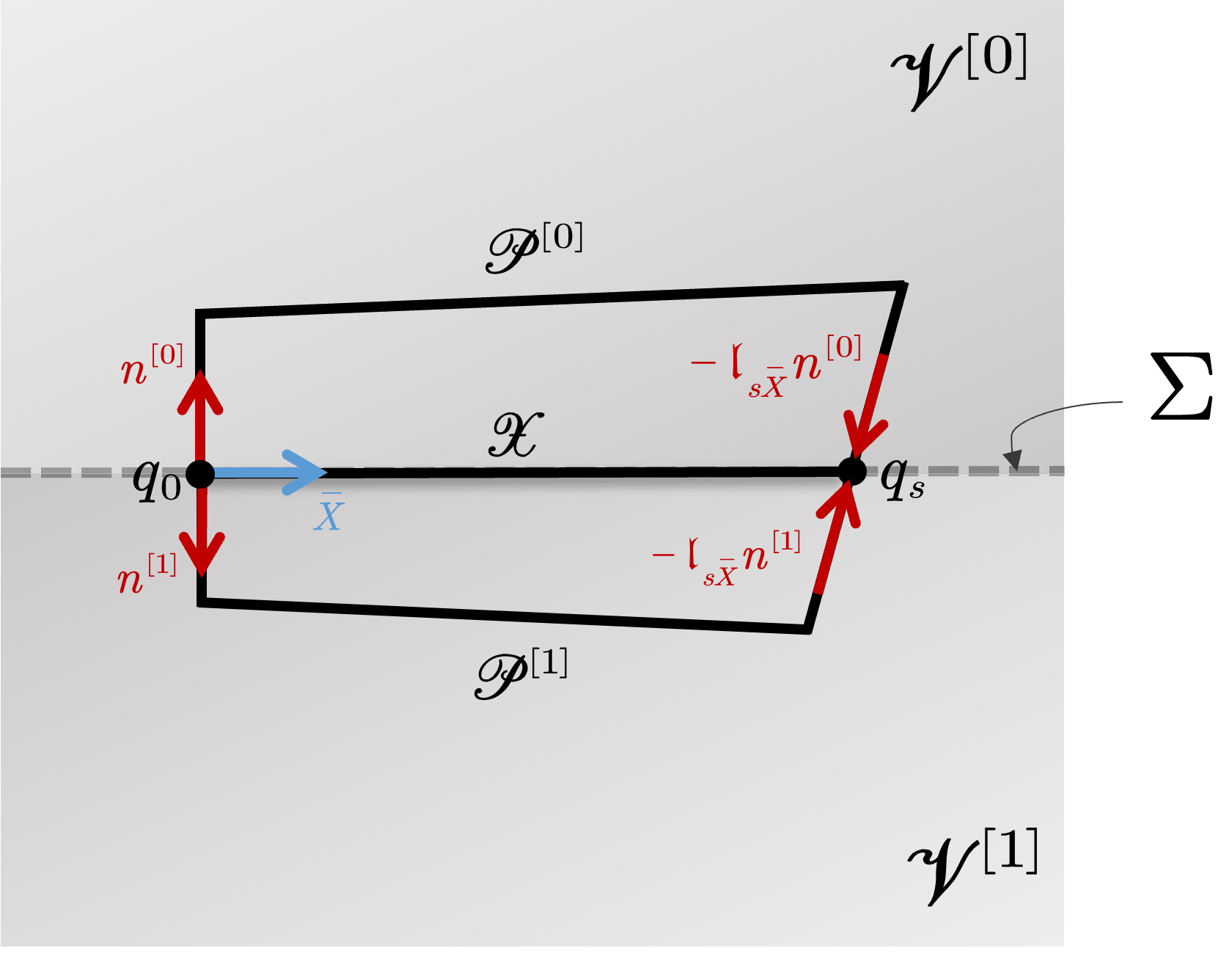}
	\caption{Geometric structure for gluing 2 bulks. For $i=0,1$, the path $\up P i$ is segmented into three parts: the section originating from $q_0$ is tangent to $\up n i$; the tangent vector of the middle section can be viewed as an continuation of $\bar X$, thus giving rise to a vector field associated with $\bar X$; and the final section merges into $q_s$. \label{2bulks}}
\end{figure}

 Definition \eqref{addpathSigma} is depicted in Fig.~\ref{2bulks}. At the initial point $q_0$, both $\up P 0$ and $\up P 1$ are tangent to the normal vector, ensuring a smooth joining of the two paths at this location. By virtue of $\mathfrak l_{s\bar X}\up n 0+\mathfrak l_{s\bar X}\up n 1=\mathfrak l_{s\bar X}\big(\up n 0+\up n 1\big)=0$, it is evident that the joining of $\up P 0$ and $\up P 1$ at the endpoint $q_s$ is also smooth. We introduce a path $\mathscr X$ that connects $q_0$ to $q_s$ and remains entirely tangent to $\Sigma$, representing the integral curve of the vector $\bar X$ at the point $q_0$. Subsequently, $\up P 0-\mathscr X$ forms a loop $\up O 0$, and $\up P 1-\mathscr X$ forms a loop $\up O 1$. It is obvious that the loop $\mathscr O$ is the combination of these two independent loops, $\mathscr O=\up O0-\up O 1$. We thus have
 \begin{widetext}
\begin{eqnarray}
	\label{decomRSigmaXn}
	\notag
	\up R \Sigma(\bar X,\up n 0)\bar U&=&\lim_{s,\iota,t\to 0^+}\frac{\up\tau0_{-t\vec  n^{[0]}}\up\tau0_{s\bar X}\up\tau0_{t \up n 0}\bar U-\up\tau0_{s\bar X}\bar U}{(\iota+t)s}+\frac{\up\tau1_{s\bar X}\bar U-\up\tau1_{-\iota\vec  n^{[1]}}\up\tau1_{s\bar X}\up\tau1_{\iota\up n 1}\bar U}{(\iota+t)s}+\frac{\up \tau0_{s\bar X}\bar U-\up\tau1_{s\bar X}\bar U}{(\iota+t)s}\\
	&=&\lim_{\iota,t\to 0^+}\frac{t}{t+\iota}\up R 0(\bar X,\up n 0)\bar U+\frac{\iota}{t+\iota}\up R 1(\bar X,\up n 0)\bar U+\frac{1}{t+\iota}\big(\up\II 1-\up\II 0\big)(\bar X,\bar U).
\end{eqnarray}
\end{widetext}
To proceed, we use a new parameter $\ell$ to denote the integral curve of the vector $\up n 0$ or $-\up n 1$ for any given positive $t$ and $\iota$, and the location of $\Sigma$ corresponds to $\ell=0$.  Let us define a distribution $\Delta^t_{-\iota}(\ell)$ such that, for $-\iota\leqslant\ell\leqslant t$, $\Delta^t_{-\iota}(\ell)=\frac{1}{t+\iota}$; otherwise, $\Delta^t_{-\iota}(\ell)=0$. We then have the integral
\begin{equation}
	\int \Delta^t_{-\iota}(\ell)\,\ed\ell=1\,,
\end{equation}
which will become the integral $\int \delta(\ell)\,\ed\ell=1$ when the limit of $\iota,t\to 0^+$ is taken. Thus we can say $\lim_{t,\iota\to 0^+}\frac{1}{t+\iota}=\delta(\ell)\big|_{\ell=0}$. In addition, as $t,\iota>0$ approach $0$ in different ways, the limit of $\frac{t}{t+\iota}$ will cover any value between $0$ and $1$, thus is indeterminate. This property is consistent with the uncertainty of $\Theta(\ell)\big|_{\ell=0}$. Hence, we might define $\lim_{t,\iota\to 0^+}\frac{t}{t+\iota}=\Theta(0)$. Then $\lim_{t,\iota\to 0^+}\frac{\iota}{t+\iota}=1-\Theta(0)$. Now we can express $\up R\Sigma (\bar X,\up n 0)\bar U$ as follows
\begin{eqnarray}
	\label{RXnU}
	\notag
	\up R \Sigma (\bar X,\up n 0)\bar U&=&\Theta(0)\cdot\up R 0(\bar X,\up n 0)\bar U\\
	\notag
	&&+\big(1-\Theta(0)\big)\cdot\up  R1(\bar X,\up n 0)\bar U\\
	\notag
	&&+\delta(0)\cdot\mathsf s\,\big(\up K 0+\up K 1\big)(\bar X,\bar U)\up n 0,\\
\end{eqnarray}
where Eq.~\eqref{defineKi} has been used. Applying the same method, we can define $\up R\Sigma(\bar X,\up n 0)\up n 0$. By substituting $\bar U$ in Eq.~\eqref{addpathSigma} with $\up n 0$, and following the calculation outlined in Eq.~\eqref{decomRSigmaXn}, it can be obtained that
\begin{eqnarray}
	\notag
	\up R \Sigma (\bar X,\up n 0)\up n 0&=&\Theta(0)\up R 0(\bar X,\up n 0)\up n 0\\
	\notag
	&&+\big(1-\Theta(0)\big)\up  R1(\bar X,\up n 0)\up n 0\\
	\notag
	&&-\delta(0)\big(\up\nabla0_{\bar X}n^{[0]}+\up\nabla1_{\bar X}\up n1\big).\\
\end{eqnarray}
In this manner, we have defined the components $\up R\Sigma(\bar X, \up n i)$ of the curvature tensor at $\Sigma$, by using the vector $\bar X$ tangent to $\Sigma$ and the normal vector $\up n i$ to construct a loop. 

Next, we define the remaining components $\up R\Sigma\big(\bar X,\bar Y\big)$, for which it is necessary to construct a loop entirely tangent to $\Sigma$. However, the situation becomes significantly different, as there is no natural way to decompose such a loop into two segments contained within $\up V 0$ and $\up V 1$ respectively. More specifically, let us discuss $\up R\Sigma\big(\bar X,\bar Y\big)\bar U$. Since $\up R\Sigma$ is linear, meaning that $\up R\Sigma(\bar X,\alpha\up n 0+\beta\bar Y)=\alpha\up R\Sigma(\bar X,\up n 0)+ \beta\up R\Sigma(\bar X,\bar Y)$, a natural definition is
\begin{eqnarray}
	\label{RSigmaXYU}
	\notag
	\up R\Sigma\big(\bar X,\bar Y\big)\bar U&=&\Theta(0)\up R0\big(\bar X,\bar Y\big)\bar U\\
	&&+\big(1-\Theta(0)\big)\up R1\big(\bar X,\bar Y\big)\bar U.
\end{eqnarray}
For the above definition, we can understand it as follows: instead of decomposing the loop, the vector $\bar U$ is now decomposed into two parts, $\bar U=\Theta(0)\bar U+\bigl(1-\Theta(0)\bigr)\bar U$. Subsequently, let $\Theta(0)\bar U$ undergo parallel translation induced by the connection of $\up V 0$ around the loop; and let $\bigl(1-\Theta(0)\bigr)\bar U$ undergo parallel translation induced by the connection of $\up V 1$. The definition of $\up R\Sigma(\bar X,\bar Y)\up n 0$ is entirely analogous
\begin{eqnarray}
	\label{RSigmaXYn}
	\notag
	\up R\Sigma\big(\bar X,\bar Y\big)\up n 0&=&\Theta(0)\up R 0\big(\bar X,\bar Y\big)\up n 0\\
	&&+\big(1-\Theta(0)\big)\up R1\big(\bar X,\bar Y\big)\up n 0.
\end{eqnarray}
We have now established the curvature tensor's definition at $\Sigma$. It is not hard to prove that this definition of $\up R\Sigma$ adheres to all the algebraic properties expected of a curvature tensor. From $\up R\Sigma$ the Ricci curvature tensor $\up\Ric\Sigma$ and the Ricci curvature scalar $\up{\tr R}\Sigma=R^{[\Sigma]}$ at $\Sigma$ can be calculated directly. In particular, $R^{[\Sigma]}$ reads
\begin{eqnarray}
	\label{LagrangianWithBoundary}
	\notag
	R^{[\Sigma]}&=&\Theta(0) R^{[0]}+\big(1-\Theta(0)\big) R^{[1]}\\
	&&-\delta(0)\cdot 2\mathsf s\,\big(\up{\tr K}0+\up{\tr K}1\big),
\end{eqnarray}
where the scalar $\up{\tr K}i=K^{[i]}$ represents the trace of the extrinsic curvature with respect to the metric $h$.
The Einstein tensor $\up G \Sigma$ at $\Sigma$ can be computed as 
\begin{eqnarray}
	\label{G}
	\notag
	\up G\Sigma&=&\Theta(0)\up G 0+\big(1-\Theta(0)\big)\up G 1\\
	\notag
	&&-\delta(0)\cdot\mathsf s\,\bigg\{\big(\up K 0+\up K 1\big)-\big(\up{\tr K}0+\up{\tr K}1\big)h\bigg\}.\\
\end{eqnarray} 
Since there may be a matter field energy-momentum tensor $\bar{\mathcal S}$ on $\Sigma$, the total energy-momentum tensor at $\Sigma$ can be expressed as 
\begin{eqnarray}
	\up T\Sigma=\Theta(0)\up T 0+\big(1-\Theta(0)\big)\up T 1+\delta(0)\bar {\mathcal S}.
\end{eqnarray}
Consequently, the Einstein field equations give rise to the junction condition~\eqref{jcfor2bulks}.

We have effectively re-established the junction condition. Throughout our derivation, we did not rely on the property $\delta(\ell)=\Theta'(\ell)$, which is different from the classical distribution method. Instead, we simply introduced $\Theta(0)$ and $\delta(0)$ just as symbols to replace the expressions $\lim_{t,\iota\to 0^+}\frac{t}{t+\iota}$ and $\lim_{t,\iota\to 0^+}\frac{1}{t+\iota}$. Furthermore, we have illustrated the geometric significance of the junction condition. 

In Appendix~\ref{sec3}, we meticulously clarified the indispensability of the continuity condition \eqref{dim2continu} for the junction condition when joining two bulks. Moreover, we explored the necessity of investigating the gluing of three (or more) bulks. 

\section{Junction conditions between three bulks\label{sec4}}
\begin{figure}[htbp!]
	\centering
	\includegraphics[width=0.7\columnwidth]{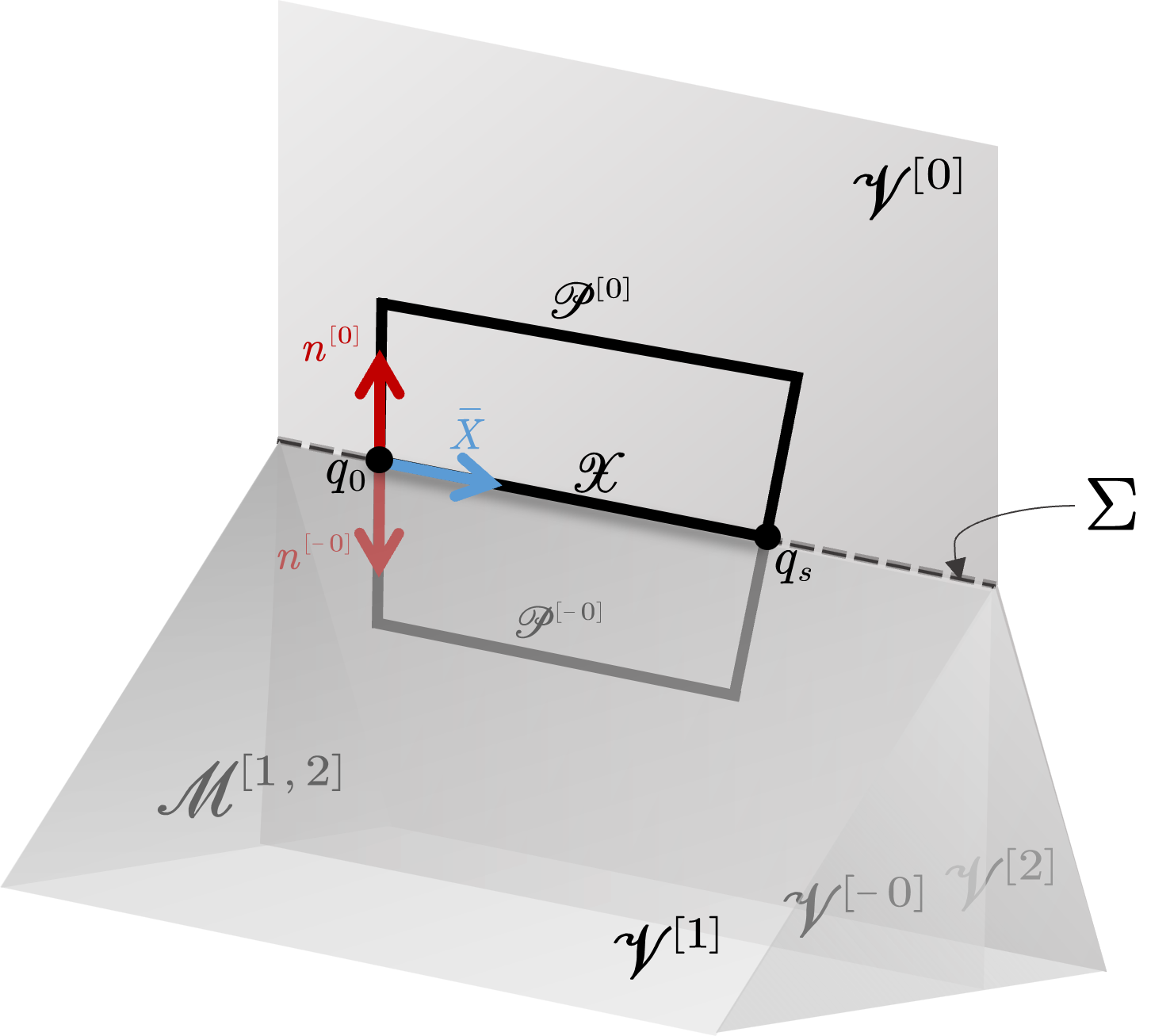}
	\caption{Geometric setting for the generalization of the junction condition for gluing three pages. Embed the booklet $\mathscr V$ into a $d+2$-dimensional manifold $\up M{1,2}$. Then, extending page $\up V 0$ along the direction of $-\up n 0$ results in an auxiliary bulk denoted as $\up V{-0}$. The geometry of this auxiliary bulk is entirely determined by $\up V 1$ and $\up V 2$.\label{ReverseExtendV0}}
\end{figure}

Before jumping into the general case of joining an arbitrary number of bulks along an interface, we first examine the simplest scenario of joining three bulks to minimize technical complexity. Nevertheless, the methodology developed in this section can be seamlessly applied to the general case. The geometry discussed in this section is shown in Fig.~\ref{ReverseExtendV0}, and to simplify the presentation we refer to these structure as booklet geometry, each joined bulk spacetime $\up V i$ as a page, and the entire joined spacetime $\mathscr V=\bigcup_i \up V i$ is referred to as the booklet. In this section, we assume all the normal vectors $\up ni$ are spacelike, namely $\mathsf s=\up g i(\up n i,\up n i)=1$ for $i=0,1,2$, for simplicity. We consider the natural generalization of \eqref{dim2continu} to the continuity condition:
\begin{equation}
	\label{dim3continu}
	\up n 0+\up n1+\up n2=0.
\end{equation}
 {More} sufficient and rigorous reasons for adopting Eq.~\eqref{dim3continu} will be provided in Sec.~\ref{subsec5A}. 

The general strategy of deriving the junction condition in this case goes as follows. We first arbitrarily select one of the pages, say, $\up V 0$, and choose the corresponding normal vector $\up n 0$. The Eq.~\eqref{dim3continu} means $-\up n 0=\up n 1+\up n 2$, which suggests to introduce an auxiliary bulk, denoted as $\up V{-0}$, whose  normal vector is $-\up n 0$. 
 In the following, we derive its geometric and physical properties directly from the corresponding attributes on $\up V 1$ and $\up V 2$. Obviously $\up V0$ and $\up V{-0}$ automatically satisfy \eqref{dim2continu} as $\up n 0 + (-\up n 0) = 0$. This enables us to apply the original Israel junction condition to deduce the condition among $\up V i, i=0,1,2$. Because of our choice, one of the pages appears more special compared to the other two pages. Therefore, we also need to confirm that if a different initial choice has been made, such as $\up V 1$ and $\up n 1$, or $\up V 2$ and $\up n 2$, the resulting junction condition will remain unchanged.

 In accordance with this perspective, we embed $\mathscr V$ into a $d+2$-dimensional manifold $\up M{1,2}$, then there should be $\mathscr T_\Sigma(\up M{1,2})\simeq\mathscr T(\Sigma)\oplus\Span\big\{\up n 1,\up n 2\big\}. $\footnote{In practice, since junction condition is a local property, it suffices to embed a small region of $\mathscr V$ containing $\Sigma$ into a small neighborhood of $\up M{1,2}$. In addition, we assume that Eq.~\eqref{dim3continu} is the sole constraint governing the relationship among the three normal vectors; therefore, any two of $\up n 0$, $\up n 1$ and $\up n 2$ are linearly independent. If this is not the case, all pages will be tangent to each other near $\Sigma$, and this represents a more degenerate case that will be excluded according to the discussion in Sec.~\ref{subsec5A} and Appendix~\ref{sec3}. } At this moment, $\up V 1$ and $\up V 2$ within $\up M{1,2}$ are regarded as submanifolds of codimension $1$, while $\Sigma$ is of codimension $2$. Subsequently, we introduce a metric $\up g{1,2}$ of $\up M{1,2}$ which satisfies
	\begin{eqnarray}
		\label{g12andg1andg2}
		\notag
		\up g{1,2}\big|_{\mathscr T_\Sigma(\up V i)}&=&\up g i, \qquad \forall \; i\in\big\{0,1,2\big\};\\
		&&\up g{1,2}\big|_{\mathscr T(\Sigma)}=h.
	\end{eqnarray}
In other words, both $\up g 1$ and $\up g 2$ are the induced metrics of $\up g{1,2}$. For the scenario involving only three pages, it can be proven that, under Eq.~\eqref{dim3continu}, the metric $\up g{1,2}$ satisfying Eq.~\eqref{g12andg1andg2} is unique. To illustrate this point, we introduce a function $\zeta$ on $\Sigma$,
\begin{equation}
	\label{gn1n2}
	\zeta=\up g{1,2}\big(\up n1,\up n2\big).
\end{equation}
Additionally, since $\up n 0 = -\big(\up n 1 + \up n2\big)$ and $\up n 0$ is normalized, we conclude
\begin{eqnarray}
	\notag
	1&=&\up g{1,2}(\up n0,\up n0)\\
	\notag
	&=&\up g{1,2}(\up n1,\up n1)+\up g{1,2}(\up n2,\up n2)+2\up g{1,2}(\up n1,\up n2)\\
	&=&2+2\zeta.
\end{eqnarray}
Namely $\zeta = -\frac{1}{2}$, indicating that $\up n 0-\up n 1,\quad \up n 1-\up n 2,\quad \up n 2-\up n 0$ form an equilateral triangle. But notice that if the booklet has more pages, the metric $\up g{1,2}$ cannot be uniquely determined. 

Using the metric $\up g{1,2}$, we can readily determine that the normal vector $\up N 1$ for $\up V 1$ in $\up M{1,2}$ is $\frac{\up n1+2\up n2}{\sqrt 3}$, the normal vector $\up N 2$ for $\up V 2$ is $-\frac{2\up n1 + \up n2}{\sqrt 3}$, and the normal vector $\up N 0$ for $\up V 0$ is $\frac{\up n1-\up n2}{\sqrt 3}$. Subsequently, by extending $\up V 0$ along the direction $\up n{-0}\coloneqq -\up n 0$, we obtain the bulk manifold referred to as $\up V{-0}$. The normal vector $\up N{-0}$ for $\up V{-0}$ in $\up M{1,2}$ is $-\frac{\up n1-\up n2}{\sqrt 3}$. Notably, the bulk $\up V{-0}$ and $\up V 0$ share a tangent space at $\Sigma$, signalling that $\up V{-0} \cup \up V 0$ possesses a well-defined differential structure at $\Sigma$. Furthermore, we have $\up g{-0} = \up g0$ at $\Sigma$. 

Currently, we only get the metric for $\up V{-0}$ at $\Sigma$, and the metrics at other locations are contingent upon the specific extension of $\up M{1,2}$ and $\up V{-0}$, that is, their second-order infinitesimal differential structure. Hence, we are unable to employ the metric for computing the connection on $\up V{-0}$. Nevertheless, the introduction of $\up V{-0}$ is intended to amalgamate the characteristics of $\up V1$ and $\up V2$ into a unified bulk. Thus we can directly derive the connection $\up \tau{-0}$ at $\Sigma$ for $\up V{-0}$, employing the connections $\up\tau i$ for $\up V i$, where $i=1,2$. To achieve this, drawing inspiration from Eq.~\eqref{IIi}, we define the extrinsic curvature $\up{\mathscr K}i$ of $\up V i$ in $\up M{1,2}$ near the junction (or the boundary of $\up V i$) parameterized by the limiting process $s\to 0$. To the leading $\mathcal O(s)$ order, we get
\begin{equation}
		\up \tau{1,2}_{s\bar X}\bar U=\up \tau i_{s\bar X}\bar U+s\up{\mathscr K}i(\bar X,\bar U)\up N i,
\end{equation}
where $\bar X, \bar U$ represent some vectors tangent to $\Sigma$ and $\forall [i]\in\big\{[-0],[1],[2]\big\}$  {and $\tau_{s\bar X}\bar U$ represents the vector $\bar U$ parallelly transported along the integral curve of $\bar X$ to the position $s\bar X$.}\footnote{This parameter $s$ should not be confused with the symbol $\mathsf s$ representing the norm of the vector $\up n i$.}

As $\Sigma$ is a common subspace of all pages $\up V i$, its tangent space $\mathscr T(\Sigma)$ is a common subspace of all $\mathscr T(\up V i)$ , the components $\up{\mathscr K}i(\bar X, \bar U)$ can be related among different pages. Consequently, we derive two independent vector equations:
\begin{eqnarray}
	\label{tau12=taui+Ki}
	\notag
	&&\up\tau{-0}_{s\bar X}\bar U-s\up{\mathscr K}{-0}(\bar X,\bar U)\dfrac{\up n1-\up n2}{\sqrt 3}\\
	\notag
	&&=\up \tau1_{s\bar X}\bar U+s\up{\mathscr K}1(\bar X,\bar U)\dfrac{\up n1+2\up n2}{\sqrt 3}\\
	&&=\up \tau2_{s\bar X}\bar U-s\up{\mathscr K}2(\bar X,\bar U)\dfrac{2\up n1+\up n2}{\sqrt 3},
\end{eqnarray}
where $\up\tau 1$ and $\up \tau 2$ are known, and our objective is to determine $\up\tau{-0}$. Note that the distinctions between the various $\up \tau i$ are solely attributed to the extrinsic curvature of $\Sigma$,
\begin{equation}
	\label{tauiXU=tauXU+KiXU}
	\up\tau i_{s\bar X}\bar U=\bar\tau_{s\bar X}\bar U+s\up K i(\bar X,\bar U)\up n i,
\end{equation}
for $[i]\in\big\{[-0],[1],[2]\big\}$. Substituting the above equation into Eq.~\eqref{tau12=taui+Ki}, using that $\up n 1$ and $\up n 2$ are linearly independent, we effectively obtain four independent scalar equations, 
	\begin{eqnarray}
		\label{(K+K)1}
		\notag
		&&\bigg(\up K{-0}-\frac{1}{\sqrt 3}\up{\mathscr K}{-0}\bigg)\up n 1+\bigg(\up K{-0}+\frac{1}{\sqrt 3}\up{\mathscr K}{-0}\bigg)\up n 2\\
		\notag
		&&=\bigg(\up K1+\frac{1}{\sqrt 3}\up{\mathscr K}1\bigg)\up n 1+\frac{2}{\sqrt 3}\up{\mathscr K}1\up n 2\\
		&&=-\frac{2}{\sqrt 3}\up{\mathscr K}2\up n 1+\bigg(\up K2-\frac{1}{\sqrt 3}\up{\mathscr K}2\bigg)\up n 2.
	\end{eqnarray}
All tensors in the above formula act on the direction $(\bar X,\bar U)$. With $\up K 1$ and $\up K 2$ being known, we can precisely determine the unknowns $\up{\mathscr K}1, \up{\mathscr K}2, \up{\mathscr K}{-0}$, $\up K{-0}$. First, the second equal sign yields
\begin{subequations}
	\begin{eqnarray}
		\up{\mathscr K}1(\bar X,\bar U)&=&\frac{\up K1+2\up K2}{\sqrt 3}(\bar X,\bar U),\\
		\up{\mathscr K}2(\bar X,\bar U)&=&-\frac{2\up K1+\up K2}{\sqrt 3}(\bar X,\bar U).
		\end{eqnarray}
\end{subequations}
Inserting the expression of $\up{\mathscr K}1$ into \eqref{(K+K)1}, the first equal sign yields
\begin{eqnarray}
	\label{scrKcalKXU}
	\notag
	\up{\mathscr K}{-0}(\bar X,\bar U)&=&\big(\up{\mathscr K}1+\up{\mathscr K}2\big)(\bar X,\bar U)\\
	&=&-\frac{\up K1-\up K2}{\sqrt 3}(\bar X,\bar U),
\end{eqnarray}
and
\begin{equation}
	\label{K-0}
	\up K{-0}=\up K1+\up K2.
\end{equation}
Notice that the components of $\up{\mathscr K}i$ with respect to $\up K 1,\up K 2$ are exactly equal to the components of $\up N i$ with respect to $\up n 1,\up n2$. Eq.\eqref{K-0} determines a part of the components of the connection for $\up V{-0}$ at $\Sigma$, specifically, $\up\tau{-0}_{s\bar X}\bar U$. First, we verify the compatibility of this part of the connection with the metric, implying that the inner product of two vectors $\bar U, \bar V$ remains unchanged under parallel translation along $\bar X$. Because of Eq.~\eqref{tauiXU=tauXU+KiXU} and considering the compatibility of $\bar\tau_{s\bar X}\bar U$ with $h$, the compatibility of $\up\tau{-0}_{s\bar X}\bar U$ with $\up g{-0}$ is satisfied regardless of the values of $\up K{-0}(\bar X,\bar U)$. Furthermore, building upon the fact that $\bar{\tau}_{s\bar X}\bar U$ is already torsion-free, the torsion-freeness of $\up \tau{-0}_{s\bar X}\bar U$ is equivalent to the symmetry $\up K{-0}(\bar X,\bar U)=\up K{-0}(\bar U,\bar X)$, as evident from Eq.~\eqref{K-0}, given that both $\up K 1$ and $\up K 2$ are symmetric tensors. 

We also need to further determine the other connection components. In fact, Eq.~\eqref{K-0}, along with the requirement for compatibility between the connection and the metric, can fully determine $\up \tau{-0}_{s\bar X}\up n{-0}$. To illustrate this, by employing $\up \tau{i}_{s\bar X}\up n{i}=\up n{i}-s\up\nabla{i}_{\bar X}\up n{i}$, and noting that $\up\nabla i_{\bar X}\up n i$ is always tangent to $\Sigma$, we can establish that
\begin{eqnarray}
	\notag
h\big(\up\nabla{-0}_{\bar X}\up n{-0},\bar Y\big)&=&\up K{-0}\big(\bar X,\bar Y\big)=\big(\up K1+\up K2\big)(\bar X,\bar Y)\\
&=&h\big(\up\nabla1_{\bar X}\up n1+\up\nabla2_{\bar X}\up n2,\bar Y\big)
\end{eqnarray}
holds for any $\bar Y$, then there must be
\begin{eqnarray}
	\notag
	\up\nabla{-0}_{\bar X}\up n{-0}&=&\up\nabla1_{\bar X}\up n{1}+\up\nabla2_{\bar X}\up n{2},\\
	 \up\tau{-0}_{s\bar X}\up n{-0}&=&\up\tau{1}_{s\bar X}\up n1+\up\tau{2}_{s\bar X}\up n2.
\end{eqnarray}
Now, using the above equation, we can directly observe the full compatibility of $\up\tau{-0}_{s\bar X}$ with the metric $\up g{-0}$. In addition, by
\begin{eqnarray}
	\notag
	&&\up \tau{-0}_{s\bar X}\up n{-0}+s\up{\mathscr K}{-0}\big(\bar X,\up n {-0}\big)\up N {-0}\\
	\notag
		&&=\up\tau{1,2}_{s\bar X}\up n{-0}=\up\tau{1,2}_{s\bar X}\up n1+\up\tau{1,2}_{s\bar X}\up n2\\
		\notag
		&&=\up\tau{1}_{s\bar X}\up n1+\up\tau{2}_{s\bar X}\up n2+s\up{\mathscr K}1\big(\bar X,\up n 1\big)\up N1\\
		&&\quad+s\up{\mathscr K}2\big(\bar X,\up n 2\big)\up N2,
\end{eqnarray}
while noting that $\up N{-0}=\up N1+\up N2$, we can obtain
\begin{equation}
	\label{scrK0=K1=K2}
	\up{\mathscr K}{-0}\big(\bar X,\up n {-0}\big)=\up{\mathscr K}1\big(\bar X,\up n 1\big)=\up{\mathscr K}2\big(\bar X,\up n 2\big).
\end{equation}
If we project $\up M{1,2}$ into the normal space, each page $\up V i$, where $[i]=[-0],[1],[2]$, is entirely characterized by the normal vector $\up n i$ at this stage. Consequently, the equal norms of all second fundamental forms $\mathscr K^{[i]}(\cdot,\up n i)\up N i$ imply the embeddings of all pages within $\up M{1,2}$ are identical. Moreover, by arbitrarily selecting a direction $\up n\beta=\beta_1\up n1+\beta_2\up n2$, an additional bulk $\up V\beta$ can be extended along it. A straightforward computation then establishes $\mathscr K^{[\beta]}(\cdot,\up n \beta)=\mathscr K^{[i]}(\cdot,\up n i)$ for $[i]=[-0],[1],[2]$. It is evident that the entire normal space exhibits rotational symmetry.

By utilizing $\up\tau{-0}_{s\bar X}\up n{-0}$ and imposing the torsion-freeness, we can determine $\up \tau{-0}_{\ell\up n{-0}}\bar X$. Finally, the only residual component is $\up\tau{-0}_{\ell\up n{-0}}\up n {-0}$, which not only depends on the specific extension of  $\up V{-0}$, but also encapsulates information related to $\up\tau{1,2}_{\ell\up n 1}\up n2$ (or $\up\tau{1,2}_{\ell\up n 2}\up n1$ equivalently). All these types of parallel translation cannot be determined solely based on the local geometry of $\up V1$ and $\up V2$ at $\Sigma$;  instead, they must rely on the specific construction of $\up M{1,2}$ and $\up V{-0}$ outside of $\Sigma$. Nevertheless, the component $\up\tau{-0}_{\ell\up n {-0}}\up n {-0}$ contributes nothing to the junction condition;  therefore, we can confidently disregard it. Utilizing the classical results derived for two pages, we have
\begin{eqnarray}
	\label{V0V-0fieldEq}
	\notag
	&&\Theta(0)\up G0+\big(1-\Theta(0)\big)\up G{-0}\\
	\notag
	&&\qquad -\up\delta0(0)\bigg\{\big(\up K0+\up K{-0}\big)-\big(\up{\tr K}0+\up{\tr K}{-0}\big)h\bigg\}\\
	&&=8\pi G\bigg\{\Theta(0)\up T0+\big(1-\Theta(0)\big)\up T{-0}+\up\delta 0(0)\bar {\mathcal S}\bigg\},
\end{eqnarray}
where $\up T{-0}$ is an effective energy-momentum tensor that we will introduce and elaborate on in Appendix~\ref{subsec4B}. In above equation, we have added the subscript $[0]$ to the Dirac function $\delta(\ell)$, indicating that $\up \delta0(\ell)$ is defined solely as a function on $\up V0\cup \up V{-0}$. This aligns with the following fact: the energy-momentum tensor $\bar{\mathcal S}$ on $\Sigma$ is an invariant, representing the density of matter within $\Sigma$. However, when observing this matter within different pages $\up V i$, the matter on $\Sigma$ exhibits varying densities, represented by different tensors $\up \delta i(\ell)\bar{\mathcal S}$.

Similar to \eqref{jcfor2bulks}, junction condition for $\up V0\cup \up V{-0}$ is
\begin{equation}
		\big(\up K 0+\up K {-0}\big)-\big(\up{\tr K}0+\up{\tr K}{-0}\big)h=-8\pi G\,\bar {\mathcal S}\ .
\end{equation} 
With the Eq.~\eqref{K-0}, the junction condition for joining three pages is thus given by
\begin{eqnarray}
	\label{jcfor3bulks}
	\notag
	&&\big(\up K 0+\up K 1+\up K2\big)-\big(\up{\tr K}0+\up{\tr K}{1}+\up{\tr K}{2}\big)h\\
	&&=-8\pi G\,\bar{\mathcal S}.
\end{eqnarray}
It can be seen that Eq.~\eqref{jcfor3bulks} possesses cyclic symmetry among the indices $[0],[1],[2]$. If we initially construct a $d+2$-dimensional manifold $\up M{0,2}$, whose local geometry is determined by $\up V0$ and $\up V 2$, independent of $\up V 1$, and then extend $\up V 1$ along the direction $\up n{-1}=\up n0+\up n2$, then we can redefine the page number $[i]$ in the cyclic manner $1\to 0\to 2\to 1$. It can be readily observed that the discussion above remains entirely the same, and we will once again obtain Eq.~\eqref{jcfor3bulks}. However, it is important to emphasize that Eq.~\eqref{V0V-0fieldEq} does not possess cyclic symmetry with respect to the indices $[i]$. It depends on our initial choices of $\up Vi$ and $\up ni$. It is interesting that different instances of Eq.~\eqref{V0V-0fieldEq} encapsulate the same junction condition. This implies that Eq.~\eqref{jcfor3bulks} is an intrinsic property, and we cannot simply view it as a consequence of the Israel junction condition for two pages.

From the previous discussion, we also see that the junction condition does not depend on how $\up M{1,2}$ is constructed or how $\up V{-0}$ is extended. 
This definition of the gluing process explained in the introduction is intrinsic, which indicates that our construction does not depend on an embedding manifold. Indeed, we can also arrive at the same conclusion by noticing that the resulting junction condition~\eqref{jcfor3bulks} does not depend on how this geometry is embedded (in the derivation by reverse extension) since the result only depends on the extrinsic curvatures within the various pages joining at the interface. It is also evident from the variational principle derivation of the junction condition that no embedding manifold is needed (see Sec.~\ref{subsec5B} for details). Therefore, the junction condition fundamentally stems from the tangent spaces of $\up V0, \up V1, \up V2$ at $\Sigma$, while $\up M{1,2}$ and $\up V{-0}$ merely serve as auxiliary manifolds and do not lead to any loss of generality.  

\section{Junction conditions between any number of bulks\label{sec5}}
\subsection{Method of reverse extension\label{subsec5A}}
We will employ the techniques developed in Sec.~\ref{sec4} to derive junction conditions in a general booklet geometry with an arbitrary number of pages. Along the discussion, we will highlight the pivotal role played by continuity condition \eqref{d+mdimcc}. Moreover, the  conclusions we will obtain in this section are independent of the specific gravitational model, which makes possible of straightforward derivations of junction conditions in alternative models, such as the dilaton gravity what we will discuss in the following section.

First, we must define the topology of the booklet $\mathscr V$, making it a topological space, which is fundamental for discussing continuity and thus particularly important. Recall that in order to glue all $\up V i$, we have provided the identification maps between the adjacent bulk boundaries $\psi_i:\partial\up V i\to\partial \up V{i+1}$, where $i=0,1,\cdots,m-1$. Since all the maps $\psi_i$ are diffeomorphisms, which necessarily preserve topology, all the topologies of $\partial \up V i$ are compatible and induce the topology of $\Sigma$. We define $\mathscr W\subset\mathscr V$ to be an open set of $\mathscr V$ if and only if for all $i$, $\up W i\coloneqq\mathscr W\cap\up V i$ is an open set of $\up V i$. It is easy to prove that this definition satisfies the three axioms of a topological space: $\varnothing$ and $\mathscr V$ are open sets; the intersection of a finite number of open sets is open; the union of any collection of open sets is open. In turn, if a collection of open sets $\up W i\subset\up V i$ have common parts $\up W i\cap\Sigma=\up W j\cap\Sigma,\;\forall i,j$, they can be pieced together to form an open set $\mathscr W$ of $\mathscr V$. From the definition, it is directly evident that the topology of $\mathscr V$ is compatible with the topologies of $\up V i$. Consequently, if a set $\up W i$ is an open set in $\up V i$ and $\up W i\cap\Sigma$ is the empty set, then $\up W i$ is also an open set in $\mathscr V$. When $m>1$, it is impossible to further define a differential structure at $\Sigma$ for $\mathscr V$, making it a differential manifold. Detailed mathematical proofs are placed in Appendix~\ref{AppenManifold}, see Proposition \ref{bookletmanifold}.

The first natural condition for gluing to get $\mathscr V$ is to require the existence of a metric $h$ on $\Sigma$ that is compatible with all bulk metrics $\up g i$,
\begin{equation}
	h=\up g i\big|_{\mathscr T(\Sigma)},\quad \forall i\in \big\{0,1,\cdots, m\big\}.
\end{equation}
 {The induced metric $h$ must be non-degenerate.}

As in the case of the Israel junction condition among 2 pages, we can embed the booklet $\mathscr V$ into an external manifold $\mathscr M$ of sufficient dimension, and straightforwardly generalizing the continuity condition of the normal vectors to $m+1$ pages 
\begin{equation}
	\label{d+mdimcc}
	\up n0+\up n1+\cdots+\up n m=0.
\end{equation}
In the following, we show the necessity of Eq.~\eqref{d+mdimcc} for the junction condition. Instead of directly assuming the condition above, we only require that each normal vector $\up n i$ is continuously distributed over the entire $\Sigma$.
 
Before that, let us first comment on the dimensions of the embedding space. First, according to Whitney's theorem, if we want to embed $\up V i$ entirely into $\mathscr M$, then in general, the dimension of $\mathscr M$ should be at least $2(d+1)$. However, the junction condition we consider is a local constraint that can be determined within open sets around $\Sigma$. Therefore, we do not need to embed the entire $\mathscr V$ into $\mathscr M$, we only need to locally embed an open set $\mathscr W\subset\mathscr V$ into $\mathscr M$. Each $\up W i = \mathscr W\cap\up V i$ can be sufficiently small, so as to be homeomorphic to an open set in $\mathbb R^{d+1}$ (this follows from the definition of a manifold). Thus, $\up W i$ can be embedded into any external manifold of dimension at least $d+1$. The interface $\Sigma$ can always be covered by some open sets $\Sigma= \bigcup_\alpha (\mathscr W_\alpha\cap\Sigma)$, where the index $\alpha$ distinguishes different open sets. Then, we can embed each $\mathscr W_\alpha$ into the external manifold separately and derive the junction conditions. The final form (see Eqs.~\eqref{betajunction},\eqref{d+mdimjunction} and \eqref{sijunction}) of the junction conditions depends only pointwise on local geometric quantities and is independent of the properties of the entire $\mathscr W_\alpha$. Therefore, within $\mathscr W_\alpha \cap \mathscr W_\gamma$, the junction conditions are compatible. This demonstrates that the method here is self-consistent and does not rely on the choice of the open cover $\bigcup_\alpha \mathscr W_\alpha$. In the following text, for notational simplicity, whenever it does not cause confusion, we denote the open set $\up W i=\mathscr W\cap\up V i$ as $\up V i$, denote $\mathscr W$ as $\mathscr V$; and denote $\mathscr W\cap\Sigma$ as $\Sigma$. 

Now, we embed $\mathscr V$ into a manifold $\mathscr M_{\mathrm L}$ of arbitrarily large dimension. Since $\Sigma$ is $d$-dimensional and there are $m+1$ normal vectors, we may assume $\dim\mathscr M_{\mathrm L}\geqslant d+m+1$. At each point, all normal vectors can generate a subspace $\mathscr N\coloneqq\Span\big\{\up n 0,\cdots,\up n m\big\}\subset \mathscr T_\Sigma(\mathscr M_{\mathrm L})$, whose dimension $\dim \mathscr N$ may vary from point to point, but there must be $\dim \mathscr N\leqslant m+1$. As long as $\Sigma$ is sufficiently small, we can choose a set of linearly independent bases $\big\{\mathcal E_0,\cdots,\mathcal E_m\big\}$ at each point, which generate an $m+1$-dimensional subspace $\mathscr E\subset \mathscr T_\Sigma(\mathscr M_{\mathrm L})$, ensuring that $\mathscr N\subset\mathscr E$ holds everywhere. This is feasible because all normal vectors are continuously distributed over $\Sigma$.

Next, we generate a $d+m+1$-dimensional submanifold $\mathscr M \subset \mathscr M_{\mathrm L}$ using $\mathscr T(\Sigma)\oplus \mathscr E$. This implies that the distribution of the subspace $\mathscr T(\Sigma)\oplus\mathscr E$ is integrable and must satisfy the prerequisites of Frobenius' theorem, i.e., $[\mathcal E_i, \mathcal E_j]$ is closed [$\mathrm{mod}\,\mathscr T(\Sigma)$]. Note that we actually only have the subspace distribution at $\Sigma$, which is the given initial condition. We can extend this subspace distribution in an appropriate manner to meet the prerequisites of Frobenius' theorem. Specifically, we can perform the following operation: let $\Sigma$ undergo a Lie translation along the integral curves of $\mathcal E_0$. As long as the translation parameter distance is sufficiently small, these integral curves do not intersect, thus forming a $d+1$-dimensional manifold $\mathscr M_0$. Then, smoothly extend $\mathcal E_1$ onto the manifold $\mathscr M_0$, and let $\mathscr M_0$ undergo a Lie translation along the integral curves of $\mathcal E_1$, thereby obtaining a $d+2$-dimensional manifold $\mathscr M_1$. By inducting this process, we eventually obtain a $d+m+1$-dimensional manifold $\mathscr M\coloneqq\mathscr M_m \subset \mathscr M_{\mathrm L}$. Discarding the redundant parts outside of $\mathscr M$, it can be observed that $\mathscr V$ can always be embedded into an external manifold $\mathscr M$ of dimension $d+m+1$. Furthermore, we establish the following simple proposition:
\begin{theorem}
	\label{theorem1}
	Assuming $\mathcal Q\subset\Sigma$ is a set of all points $q$ such that, in the tangent space $\mathscr T_q(\mathscr M)$, the vector set $\big\{\up n 0,\up n 1,\cdots,\up n m,\bar X\big\}$ is linearly independent for any $\bar X\in\mathscr T_q(\Sigma)$. Then, $\mathcal Q$ must be an open set.
\end{theorem}
\begin{proof}
	We provide the proof process briefly. Note that $\dim\mathscr M=d+m+1$. For any $q\in\mathcal Q$, consider a coordinate open set $\mathcal M_q\subset \mathscr M$ containing $q$. In $\mathcal M_q$, the coordinates of any point can be expressed as $(x^\mu, y^k)$ where $\mu=1,\cdots, d$ and $k=0,1,\cdots,m$. Moreover, for points within $\Sigma$,  {we require} $y^k=0$ for any $k$.  {Such a coordinate always exists; in fact, $y^k$ can be chosen as the parameters of the previously introduced basis $\mathcal{E}_k$.}
 Consequently, any vector $\bar X$ tangent to $\Sigma$ can be decomposed as $\bar X=\bar X^\mu\frac{\partial}{\partial x^\mu}$; and for each normal vector, there is $\up n i=(\up n i)^\mu\frac{\partial}{\partial x^\mu}+(\up n i)^k\frac{\partial}{\partial y^k}$. Then, $\big[(\up n i)^k\big]$ can be viewed as an $(m+1)\times(m+1)$-order matrix. This establishes a continuous mapping $\psi$ from $\mathcal M_q\cap\Sigma$ to the space of matrices. Within $\mathcal Q$, $\big\{\up n0,\cdots,\up n m,\frac{\partial}{\partial x^1},\cdots,\frac{\partial}{\partial x^d}\big\}$ is linearly independent, making $\big[(\up n i)^k\big]$ an invertible matrix, { since 
 \begin{eqnarray*}
	&&\sum_{j=0}^m\delta^i_j\,\up n j=\up n i=(\up n i)^\mu\frac{\partial}{\partial x^\mu}+(\up n i)^k\frac{\partial}{\partial y^k}\\
	&&=\bigg\{(\up n i)^\mu+(\up n i)^k\cdot y^\mu_k\bigg\}\frac{\partial}{\partial x^\mu}+\sum_{j=0}^m(\up n i)^k\cdot y_{kj}\,\up n j
 \end{eqnarray*}
 will yield $(\up n i)^k\cdot y_{kj}=\delta^i_j$ \footnote{Einstein's summation convention will be applied to tensor indices except for the page number $[i]$}. For points within $\Sigma-\mathcal Q$, $[(\up n i)^k\big]$ is not invertible either. Otherwise, let the inverse matrix be denoted as $(n^{-1})_{ik}$, then 
 \begin{eqnarray*}
	\sum_{i=0}^m(n^{-1})_{ik} \up n i=\sum_{i=0}^m(n^{-1})_{ik}\cdot (\up n i)^\mu\frac{\partial}{\partial x^\mu}+\frac{\partial}{\partial y^k}
\end{eqnarray*}
implies that linearly dependent $\up n i$ can generate linearly independent $\frac{\partial}{\partial y^k}$, which is impossible.} 
In the space of matrices, all invertible matrices form an open set $\mathcal N$, and we have $q\in\psi^{-1}(\mathcal N)\subset\mathcal Q$. Since $\psi$ is a continuous mapping, $\psi^{-1}(\mathcal N)$ is an open set. Therefore, $\mathcal Q$ as the union of all open sets $\psi^{-1}(\mathcal N)$, is also an open set.
\end{proof}
	If there exists such a non-empty open set $\mathcal Q$ as defined in proposition \ref{theorem1}, we can introduce an additional normal vector $\up n{m+1}$ that is continuously distributed over the entire $\Sigma$. Within $\mathcal Q$, let $\up n{m+1}$ be a linear combination of other normal vectors and a vector tangent to $\Sigma$; within the complement $\Sigma-\mathcal Q$, ensure that $\up n{m+1}$ is linearly independent of the other vectors. Starting from $\Sigma$ and along the direction of $\up n{m+1}$, we can extend a new page $\up V{m+1}$. At this moment, the total number of pages increases by $1$, equivalent to discussing the gluing of $m+2$ pages. Therefore, we can always assume that $\mathcal Q=\varnothing$. Physically, this process elucidates that the interaction among several independent bulk spacetimes at $\Sigma$ will inevitably generate a new bulk.  

Now, for any $p\in\Sigma$, the following linear combination necessarily exists,
\begin{equation}
	\label{sumalphaini0}
	\beta_0\up n 0+\beta_1\up n 1+\cdots+\beta_m\up nm=\bar w,
\end{equation}
where $\beta_i$ are not all $0$ and $\bar w$ is a vector tangent to $\Sigma$. Consequently, we can further reduce the redundant dimensions, setting $\mathscr M$ as a $d+m$ dimensional manifold since each point can have at most $m$ linearly independent normal vectors. However, at certain points, the constraints like Eq.~\eqref{sumalphaini0} may not be unique (up to equivalence). If there are additional constraints at point $p$, such as $\sum_{i=0}^m\alpha_i\up n i=\bar X$, where $\alpha_i$ and $\beta_i$ are not all equal, solving these two equations simultaneously will result in a new expression similar to Eq.~\eqref{sumalphaini0}, where some $\beta_i=0$, however. To explore this scenario, we define $m+1$ sets $\mathcal V_i$. Then $q\in\mathcal V_i$ is equivalent to the existence of a linear combination \eqref{sumalphaini0} at point $q$ with $\beta_i=0$. This definition implies that in $\mathcal V_i$, the $\up ni$ is redundant. When we remove $\up n i$, the remaining normal vectors are still linearly dependent [$\mathrm{mod}\,\mathscr T(\Sigma)$]. Nevertheless, we can confidently assert that the set $\mathcal V_i$ cannot contain any open subset; in other words, $\mathcal V_i$ can only consist of some boundary points, which we refer to as singular points. Otherwise, without loss of generality, let us consider that $\mathcal V_0$ contains an open subset. In that case, excluding the redundant $\up V0$ and only considering the joining of $\up V1$ to $\up Vm$ is sufficient to solve a junction condition, as it is merely a local property. However, the geometry of $\up V0$ may be incompatible with junction conditions generated by other pages.

Excluding all $\mathcal V_i$, we consider the remaining set $\mathcal W=\Sigma-\bigcup_{i=0}^m \mathcal V_i$. According to the definition, within $\mathcal W$, each normal vector $\up n i$ is linearly dependent on the other normal vectors [$\mathrm{mod}\,\mathscr T(\Sigma)$]. However, removing any $\up n i$, the remaining normal vectors are necessarily linearly independent [$\mathrm{mod}\,\mathscr T(\Sigma)$]. This indicates that, in $\mathcal W$, there exists only a unique constraint given by Eq.~\eqref{sumalphaini0} (up to equivalence), and $\beta_i\not=0$ for all $i$ and for all points. We immediately have the following proposition,
\begin{theorem}
	\label{theorem2}
	$\mathcal W$ is an open set of $\Sigma$.
\end{theorem}
\begin{proof}
	  Note that $\mathscr M$ is now $d+m$ dimensional, and $\mathcal W=\bigcap_{i=0}^m \left(\Sigma-\mathcal V_i\right)$. Within $\Sigma-\mathcal V_0$, the set $\big\{\up n 1, \cdots, \up n m, \bar X\big\}$ is linearly independent for any $\bar X\in\mathscr T(\Sigma)$; whereas within $\mathcal V_0$, they must be linearly dependent. Therefore, according to the conclusion of Proposition \ref{theorem1}, $\Sigma-\mathcal V_0$ must be an open set. Similarly, all $\Sigma-\mathcal V_i$ are open sets. $\mathcal W$ is the intersection of a finite number of open sets, and therefore is also an open set.
\end{proof}
$\mathcal W$ being an open set implies that $\bigcup_{i=0}^m\mathcal V_i$ is a closed set. Additionally, $\bigcup_{i=0}^m\mathcal V_i$ should not contain any open subset; otherwise, the joining of pages within this open set would be singular, and a junction condition would not be produced. Therefore, we observe that $\mathcal W$ is a dense open subset of $\Sigma$. Then we can establish the junction condition within $\mathcal W$. Based on the continuity of extrinsic curvature and other geometric quantities within $\Sigma$, it is reasonable to assume that the junction condition can also be extrapolated to singular points outside of $\mathcal W$. In this sense, in the subsequent discussions, there is no need to explicitly reference the open set $\mathcal W$, and we can assume that all $\beta_i\not=0$ holds across the entire $\Sigma$. Then, by dividing both sides of Eq.~\eqref{sumalphaini0} by $\beta_0$, or equivalently, redefining $\beta_0=1$, we can eliminate the formal degrees of freedom in Eq.~\eqref{sumalphaini0}. Now, Eq.~\eqref{sumalphaini0} is unique at every point. Furthermore, since all normal vectors are continuously distributed, each $\beta_i$ forms a continuous function on $\Sigma$; while $\bar w$ forms a continuous tangent vector field.

 We proceed to examine potential values for the vector field $\bar w$.  Within each $\up V i$, a metric $\up g i$ is present. Simultaneously, a metric $g$ can be established for the large manifold $\mathscr M$ at $\Sigma$. If we stipulate that the metric $g$ is compatible with each $\up g i$, namely, $g\big|_{\mathscr T_\Sigma(\up V i)}=\up g i$. Then, for any $\bar X$ tangent to $\Sigma$, 
 \begin{equation}
	0=\sum_{i=0}^m \beta_i\,\up g i\bigl(\up n i,\bar X\bigr)=g\bigg(\sum_{i=0}^m \beta_i\up n i,\bar X\bigg)=h(\bar w,\bar X).
 \end{equation}
 Because of the non-degeneracy of the induced metric $h$, we observe that $\bar w=0$ holds everywhere. Conversely, if $\bar w\not=0$ at certain points, certain indices $i$ must exist for which the metric $g$ is incompatible with $\up g i$. Consequently, we extend $\up V i$ along the direction of $-\up n i$ to obtain $\up V{-i}$. The metric $g$ on $\up V{-i}$ induces the metric $\up g {-i}$, and we have $\up g{-i}\not=\up g i$. Subsequently, the derivation in Appendix~\ref{sec3} tell us that in such cases, a well-defined junction condition does not exist. In the case of joining two pages, this physical requirement--- compatibility between the metrics $g$ and $\up g i$ --- reduces to the condition $\up g0\big|_\Sigma=\up g 1\big|_\Sigma$. As a result, Eq.~\eqref{sumalphaini0} can be reformulated as
\begin{equation}
	\label{n0=betaini}
	\up n 0+\beta_1\up n1+\cdots+\beta_m\up n m=0, \quad \forall\beta_i(q)\not=0,\; \forall q\in\Sigma\ .
\end{equation}
Unlike the continuity condition \eqref{d+mdimcc}, the above expression merely states the fact that the normal vectors must be linearly dependent. Becuase of the arbitrariness in the choices of all $\beta_i$, Eq.~\eqref{n0=betaini} essentially does not presuppose any specific relationship among the normal vectors.

The preceding lengthy explanation actually illustrates how $\mathscr V$ is permitted to be embedded in a higher-dimensional spacetime $\mathscr M$. Singular points are possible, and the ``embedding'' degenerates at these singular points. This could potentially create a misconception that the junction condition is determined by this specific embedding method. In reality, the junction condition can only be dictated by the matter on $\Sigma$ and represent an intrinsic property of $\mathscr V$. So, in actuality, it works the other way around—the embedding method is determined by the junction condition. Therefore, we must emphasize that ``embedding'' serves as an auxiliary tool. In practice, there is no observer capable of observing $\mathscr V$ from the perspective of $\mathscr M$. For observers within $\mathscr V$, they can solely perceive the junction condition and remain unaware of Eq.~\eqref{n0=betaini} or the continuity condition \eqref{d+mdimcc}.

\begin{figure}[htbp!]
	\centering
	\includegraphics[width=0.7\columnwidth]{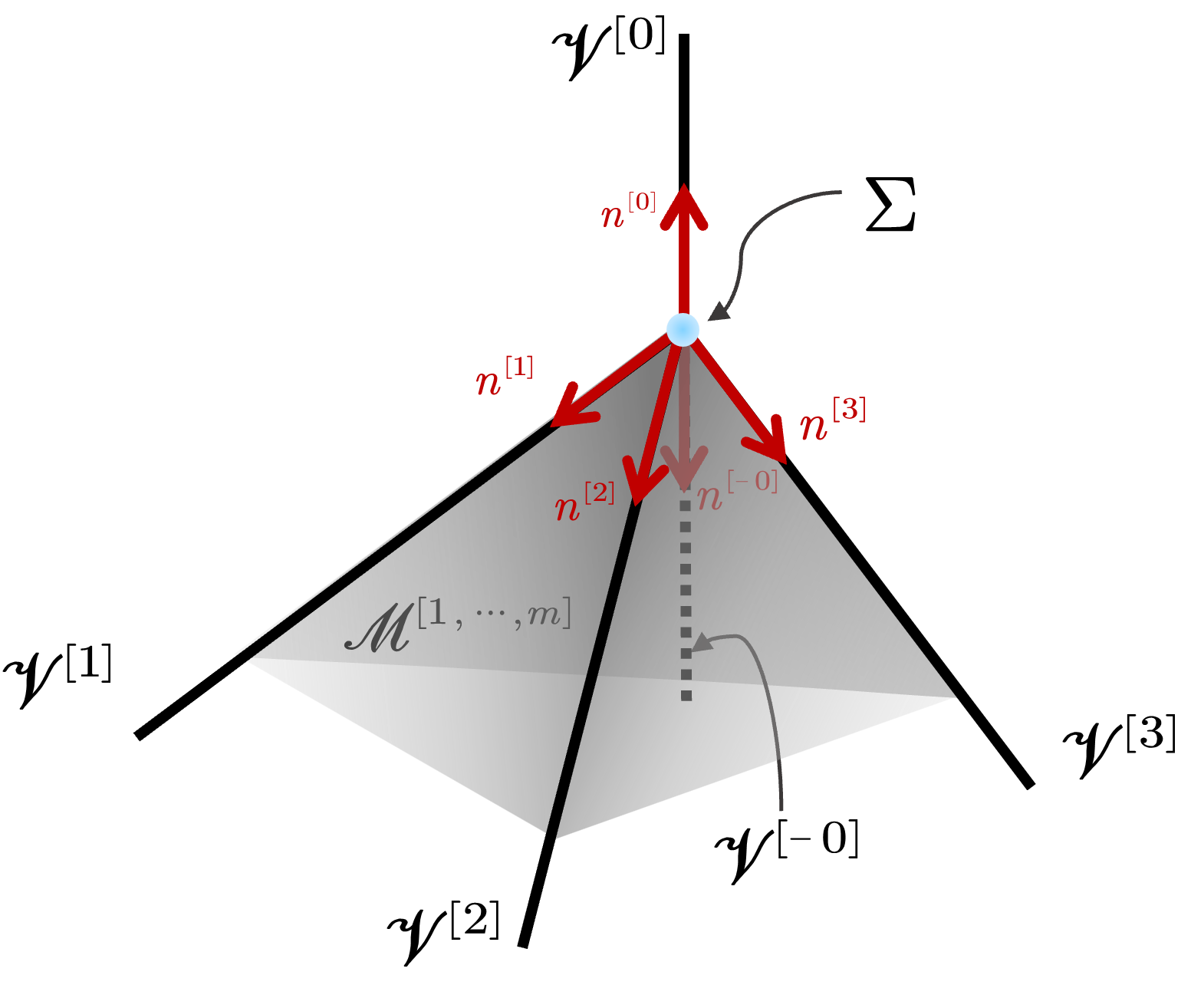}
	\caption{The geometric configuration of $\up M{1,\cdots,m}$ and the reverse extension of page $\up V0$.\label{4bulks}}
\end{figure}

By Eq.~\eqref{n0=betaini}, we can see that the normal space generated by $\up n 0,\cdots,\up n m$ is of dimension $m$, and therefore it can be represented as $\mathscr N\coloneqq\Span\big\{\up n 1,\cdots,\up n m\big\}$. It is easy to prove
\begin{eqnarray}
	\mathscr T_\Sigma(\mathscr M)=\mathscr T(\Sigma)\oplus \mathscr N\ .
\end{eqnarray} 
This equation is necessary for defining the second fundamental form of $\Sigma$ in $\mathscr M$. Following the approach in Sec.~\ref{sec4}, we contemplate the construction of the $d+m$ dimensional manifold $\up M{1,\cdots,m}$, which possesses a geometric structure that smoothly continues from $\up V1, \up V2,\cdots,\up Vm$, but has no direct connection to $\up V0$. The Fig.~\ref{4bulks} can conceptually illustrate the specific construction of $\up M{1,\cdots,m}$. Consequently, the metric $\up g{1,\cdots,m}$ of $\up M{1,\cdots,m}$ at the location of $\Sigma$ is a combination of $\up g1,\cdots,\up gm$, or conversely, $\up gi$ is the induced metric of $\up g{1,\cdots,m}$ on $\up V i$. For the sake of simplicity, we will use the abbreviation $g$ in place of $\up g{1,\cdots,m}$. To fully define $g$, a set of functions $\zeta^{ij}$ must also be provided,
\begin{equation}
	\label{zetaij}
	g(\up n i,\up n j)=\zeta^{ij},\quad \forall i,j\in \big\{1,2,\cdots,m\big\}.
\end{equation}
Specifically, when $i=j$, $\zeta^{ij}=\pm 1$. In general, not all normal vectors need to have the same sign. That is, some $\up n i$ can be timelike, while other normal vectors $\up n j$ can be spacelike. In this case, different pages will have different causal structures. Specifically, if the metric $h$ is timelike and there exists a timelike normal vector $\up n k$, this would imply that the metric $\up g k$ has two independent time directions, which is physically impossible. Therefore, we naturally assume that all normal vectors have a common sign $\mathsf s=\up g i(\up n i,\up n i)$, hence $\zeta^{ii}=\mathsf s$. It should be emphasized that if we only focus on the junction conditions, the coexistence of both timelike and spacelike normal vectors is entirely permissible. We will briefly discuss this at the end. 
The normalization of $\up n0$ will yield a constraint:
\begin{equation}
	\label{betazeta}
	\mathsf s=g(\up n0,\up n0)=\sum_{i,j}\zeta^{ij}\beta_i\beta_j,
\end{equation}
all $\zeta^{ij}$ has $m(m-1)/2$ degrees of freedom, so except for the case of $m+1=3$ pages, the above constraint cannot uniquely determine $\zeta^{ij}$, and thus $g$. Note that $g$ does not necessarily have to be Lorentzian, meaning its signature is not necessarily $(1,d+m-1)$. Therefore, even if the normal space is spanned by space-like vectors $\up n i$, namely $\mathsf s=1$, there may still exist time-like or light-like vectors within it.

We define the second fundamental form $\up\SF i$ for $\up V i$ within the manifold $\up M{1,\cdots,m}$. For vectors $X$ and $Y$ tangent to $\up V i$,
\begin{eqnarray}
	\label{SF}
	\notag
	\up\SF i(X,Y)&=&\up\nabla{1,\cdots,m}_XY-\up\nabla i_XY\\
	&=&-\mathsf s\sum_{j=1}^m{\mathscr K}^{[i]}_{[j]}(X,Y)\up n j.
\end{eqnarray}
Unlike the case with three pages, within the manifold $\up M{1,\cdots,m}$, the direction of normal vectors orthogonal to $\up V i$ is not unique. Therefore, when we expand $\up\SF i(X, Y)$ in the basis $\up n j,\; j=1,\cdots,m$, the double-indexed ${\mathscr K}^{[i]}_{[j]}(X,Y)$ play the role of extrinsic curvature. Because of the orthogonality of $\up\SF i(X, Y)$ to $\up V i$, we encounter a constraint
\begin{equation}
	\label{gSFini}
	0=-\mathsf s\,g\big(\up n i,\up\SF i(X,Y)\big)=\sum_{j=1}^m{\mathscr K}^{[i]}_{[j]}(X,Y)\zeta^{ij}.
\end{equation}
Note that the second fundamental form of $\Sigma$ within $\up M{1,\cdots,m}$ can also be defined, it provides a link to all the pages $\up V i$,
\begin{eqnarray}
	\label{SFiplusKi}
	\notag
	&&\up\nabla{1,\cdots,m}_{\bar X}\bar Y-\bar\nabla_{\bar X}\bar Y
	\\
	\notag
	&&=\up\nabla{1,\cdots,m}_{\bar X}\bar Y-\up\nabla i_{\bar X}\bar Y+\up\nabla i_{\bar X}\bar Y-\bar\nabla_{\bar X}\bar  Y \\
	\notag
	&&=\up \SF i(\bar X,\bar Y)+\up \II i(\bar X,\bar Y)\\
	\notag
	&&=-\mathsf s\,\up K i(\bar X,\bar Y)\up n i-\mathsf s\sum_{j=1}^m\up{\mathscr K} i_{[j]}(\bar X,\bar Y)\up n j\\
	&&=-\mathsf s\sum_{j=1}^m \big(\up{\mathscr K}i_{[j]}+\delta^i_j\up K i\big)(\bar X,\bar Y)\up n j.
\end{eqnarray}
The above equation is valid for any $i\in\big\{1,2,\cdots,m\big\}$, and the first row not contain $[i]$, which leads to $m-1$ independent vector equations or $m(m-1)$ independent scalar equations, which can be represented more explicitly as
\begin{equation}
	\label{scrK+deltaK}
	\big(\up{\mathscr K}i_{[k]}+\delta^i_k\up K i\big)(\bar X,\bar Y)=\big(\up{\mathscr K}j_{[k]}+\delta^j_k\up K j\big)(\bar X,\bar Y),
\end{equation}
for all $i,j,k\in\big\{1,2,\cdots,m\big\}$. Along with Eq.~\eqref{gSFini} providing $m$ constraints,  we discern that all $\mathscr K^{[i]}_{[j]}(\bar X,\bar Y)$ can be fully determined, and their expressions can be explicitly specified. Adding $\mathsf s\,\up K j(\bar X,\bar Y)$ to both sides of Eq.~\eqref{gSFini}, noting $\mathsf s=\zeta^{ii}$, and employing Eq.~\eqref{scrK+deltaK}, while fixing one index $[i]$,
\begin{eqnarray}
	\label{K=scrK+deltaK}
	\notag
	\up K j(\bar X,\bar Y)&=&\mathsf s\sum_{k=1}^m\big({\mathscr K}^{[j]}_{[k]}+\delta^j_k\up K j\big)(\bar X,\bar Y)\zeta^{jk}\\
	&=&\mathsf s\sum_{k=1}^m\big({\mathscr K}^{[i]}_{[k]}+\delta^i_k\up K i\big)(\bar X,\bar Y)\zeta^{jk}.
\end{eqnarray}
Furthermore, note that $\zeta^{jk}$ represents the component coefficients of the induce metric $g\big|_{\mathscr N}$ of metric $g$ in the normal space. The normal space $\mathscr N$ and $\mathscr T(\Sigma)$ are orthogonal, indicating that $g\big|_{\mathscr N}$ is non-degenerate. Otherwise, there would exist a vector $N\in\Span\big\{\up n1,\cdots,\up nm\big\}$ that is orthogonal to the entire $\mathscr T(\mathscr M)$, which contradicts the non-degeneracy of $g$. As a result, matrix $\zeta^{jk}$ is inherently invertible, and its inverse matrix is denoted as $(\zeta^{-1})_{jk}$. We can proceed to solve $\mathscr K^{[i]}_{[j]}(\bar X,\bar Y)$,
\begin{equation}
	\label{scrKij}
	{\mathscr K}^{[i]}_{[j]}(\bar X,\bar Y)=-\delta^i_j\up K i(\bar X,\bar Y)+\mathsf s\sum_{k=1}^m(\zeta^{-1})_{jk}\up Kk(\bar X,\bar Y),
\end{equation}
$\forall i\in\big\{1,\cdots,m\big\}$. Substituting the above equation into Eq.~\eqref{SFiplusKi}, we can observe that 
\begin{equation}
	\label{SecondFundforSigma}
	\big(\up \II i+\up \SF i\big)(\bar X,\bar Y)=-\sum_{j,k}(\zeta^{-1})_{jk}\up Kk(\bar X,\bar Y)\up nj
\end{equation}
is an invariant for all pages. This invariant demonstrates that the components of the connection $\up\nabla{1,\cdots,m}_{\bar X}\bar Y$  are entirely determined by the corresponding components of the connections of each $\up V i$, namely $\up\nabla i_{\bar X}\bar Y$. Consequently, the torsion of the connection and its compatibility with the metric are both evident.

Now, within $\up M{1,\cdots,m}$, we extend $\up V0$ in the direction along $\up n{-0}\coloneqq-\up n0$, creating an auxiliary bulk $\up V{-0}$ whose geometry is induced by the geometry of $\up M{1,\cdots,m}$. Consequently, the extrinsic curvature $\up K{-0}$ of $\Sigma$ within $\up V{-0}$ and the second fundamental form $\up\SF{-0}$ of $\up V{-0}$ within $\up M{1,\cdots,m}$ are entirely determined by the geometry of the other $\up Vi$. Since Eq.~\eqref{SecondFundforSigma} is also applicable to $\up V{-0}$,  and note that $\up\SF{-0}$ is orthogonal to $\up n{-0}$, it follows that
\begin{eqnarray}
	\label{K-0tobetaiKi}
	\notag
	\up K{-0}(\bar X,\bar Y)&=&-g\bigg(\up n{-0},\big(\up \II{-0}+\up\SF{-0}\big)(\bar X,\bar Y) \bigg)\\
	\notag
	&=&\sum_{j,k}(\zeta^{-1})_{jk}\up Kk(\bar X,\bar Y)g\big(\up n{-0},\up n j\big)\\
	\notag
	&=&\sum_{i,j,k} (\zeta^{-1})_{jk}\zeta^{ij}\beta_i\up Kk(\bar X,\bar Y)\\
	&=&\sum_{i=1}^m \beta_i\up K i(\bar X,\bar Y),
\end{eqnarray}
namely,
\begin{equation}
	\label{K0=betaiKi}
	\up K{-0}= \beta_1\up K 1+\beta_2\up K2+\cdots+\beta_m\up Km.
\end{equation}
This is a significant conclusion that is independent of gravitational models. If we were only to aim at proving the above equation, we could use another method by extending the normal vector $\up n{-0}$ into a vector field on the bulk $\up V{-0}$ and then utilize $\up K{-0}=\frac{1}{2}\mathfrak L_{\up n{-0}}h$ for the proof. However,  on one hand, we hope to manifest that Eq.~\eqref{K0=betaiKi} possesses a localized characteristic that remains unaffected by the choice of extension method of $\up V{-0}$ and $\up n{-0}$; On the other hand, we hope to explore the comprehensive geometric structure concerning the embedding of $\up Vi$ into the manifold $\up M{1,\cdots,m}$. When we combine Eq.~\eqref{K0=betaiKi} with \eqref{SecondFundforSigma}, 
\begin{equation}
	\mathscr K^{[-0]}_{[i]}(\bar X,\bar Y)=\sum_{j=1}^m\big(\mathsf s\,(\zeta^{-1})_{ij}-\beta_i\beta_j\big)\up Kj(\bar X,\bar Y).
\end{equation}
We can readily confirm that $\mathscr K^{[-0]}_{[i]}(\bar X,\bar Y)$ also complies with the orthogonal constraint condition \eqref{gSFini}.

By utilizing Eq.~\eqref{K0=betaiKi}, we can not only determine all the connection components of $\up V{-0}$ of the form $\up\nabla{-0}_{\bar X}\bar Y$, but also establish $\up\nabla{-0}_{\bar X} \up n{-0}$, and further applying the torsion-free condition to determine $\up\nabla{-0}_{\up n{-0}} \bar X$. Note that $\up\nabla{-0}_{\bar X} \up n{-0}$ is a vector tangent to $\Sigma$,
\begin{eqnarray}
	\notag
	h\big(\up\nabla{-0}_{\bar X}\up n{-0},\bar Y\big)&=&\up K{-0}(\bar X,\bar Y)=\sum_{i=1}^m \beta_i\up K i(\bar X,\bar Y)\\
	&=&h\bigg(\sum_{i=1}^m \beta_i\up \nabla i_{\bar X}\up n i,\bar Y\bigg)
\end{eqnarray}
holds for any $\bar Y$. Therefore, the non-degeneracy of the metric implies
\begin{equation}
	\label{nabla0n0}
	\up\nabla{-0}_{\bar X}\up n{-0}=\sum_{i=1}^m\beta_i\up \nabla i_{\bar X}\up n i.
\end{equation}
On the other hand, since
\begin{eqnarray}
	\label{nabla1tomn0}
	\notag
	&&\up\nabla{-0}_{\bar X}\up n{-0}+\up\SF{-0}(\bar X,\up n{-0})\\
	\notag
	&&=\up\nabla{1,\cdots,m}_{\bar X}\up n{-0}\\
	\notag
	&&=\sum_{i=1}^m\beta_i\up\nabla{1,\cdots,m}_{\bar X}\up n i+\mathrm d\beta_i(\bar X)\up n i\\
	&&=\sum_{i=1}^m\beta_i\up\nabla i_{\bar X}\up n i+\mathrm d\beta_i(\bar X)\up n i+\beta_i\up\SF i(\bar X,\up n i),\qquad
\end{eqnarray}
where the components on both sides of the equation that are tangential and orthogonal to $\Sigma$ are exactly equal, respectively. This not only reproduces Eq.~\eqref{nabla0n0}, but we  also get
\begin{subequations}
	\label{SF0=biSFi+dbi}
	\begin{eqnarray}
		\label{SF0=biSFi+dbi1}
		\up\SF{-0}(\bar X,\up n{-0})&=&\sum_{i=1}^m\mathrm d\beta_i(\bar X)\up n i+\beta_i\up\SF i(\bar X,\up n i);\qquad \;\\
		\label{SF0=biSFi+dbi2}
		\mathscr K^{[-0]}_{[j]}(\bar X,\up n{-0})&=&-\mathsf s\,\mathrm d\beta_j(\bar X)+\sum_{k=1}^m\beta_k\mathscr K^{[k]}_{[j]}(\bar X,\up n k),\qquad\;  
	\end{eqnarray}
\end{subequations}
where $\forall j\in\big\{1,2,\cdots,m\big\}$. This equation extends Eq.~\eqref{scrK0=K1=K2} to the case of $m+1$ pages. We can observe that $\up\SF{-0}(\up n{-0},\cdot)$ contains the twist terms $\mathrm d\beta_i$, which indicates that $\up\SF{-0}(\up n{-0},\cdot)$ depends not only on the local values of the normal vector $\up n{-0}$ but also on how $\up n{-0}$ varies along $\Sigma$.

In Eq.~\eqref{gSFini}, taking the direction $(\bar X,\up n i)$, yields the constraint condition for $\up\SF i(\bar X,\up n i)$,
\begin{equation}
	\label{gSFiXni0}
	0=-\mathsf s\,g\big(\up n i,\up\SF i(\bar X,\up n i)\big)=\sum_{j=1}^m\zeta^{ij}\mathscr K^{[i]}_{[j]}(\bar X,\up n i).
\end{equation}
Specially, $g\big(\up n{-0},\up\SF{-0}(\bar X,\up n{-0})\big)$ will introduce constraints,
\begin{eqnarray}
	\label{offshell}
	\notag
	0&=&-g\big(\up n{-0},\up\SF{-0}(\bar X,\up n {-0})\big)\\
	\notag
	&=&\mathsf s\sum_{i,j}\beta_i\zeta^{ij}\mathscr K^{[-0]}_{[j]}(\bar X,\up n {-0})\\
	\notag
	&=&\mathsf s\sum_{i,j,k}\beta_i\beta_k\zeta^{ij}\mathscr K^{[k]}_{[j]}(\bar X,\up n k)-\sum_{i,j}\beta_i\zeta^{ij}(\bar X\beta_j)\\
	\notag
	&=&\mathsf s\sum_{i,j,k}\beta_i\beta_k\zeta^{ij}\mathscr K^{[k]}_{[j]}(\bar X,\up n k)+\frac{1}{2}\sum_{i,j}\beta_i\beta_j(\bar X\zeta^{ij})\\
	\notag
	&=&\frac{1}{2}\sum_{i,j}\beta_i\beta_j\bigg(\bar X\zeta^{ij}+\mathsf s\sum_{k=1}^m\zeta^{ik}\mathscr K^{[j]}_{[k]}(\bar X,\up n j)\\
	&&\hspace{0.39\columnwidth}+\zeta^{jk}\mathscr K^{[i]}_{[k]}(\bar X,\up n i)\bigg),
\end{eqnarray}
where the fourth equal sign makes use of $\bar X\big(\sum_{i,j}\zeta^{ij}\beta_i\beta_j\big)=\bar X(1)=0$. It is imperative to emphasize that $\beta_i$ should be considered as free variables here, rather than fixed parameters. Actually, we can disentangle the relationship between $\up n{-0}$ and $\up n 0$. We begin by arbitrarily selecting a set of functions $\beta_i$, without satisfying the constraint \eqref{n0=betaini} necessarily. Subsequently, we define $\up n{-0}\coloneqq \beta_1\up n 1+\cdots+\beta_m\up n m$ in reverse. It is evident that $\up n{-0}$ and $\up n 0$ are independent, and $\up n{-0}$ does not require normalization. Then, We extend along the direction of  $\up n{-0}$ to obtain $\up V{-0}$. However, $\up n{-0}$ cannot be null, which is equivalent to $\sum_{i,j}\zeta^{ij}\beta_i\beta_j\not=0$. Otherwise, the metric on $\up V{-0}$ degenerates, leading to undefined connections and extrinsic curvatures. Additionally, we impose the condition $\sum_{i,j}\zeta^{ij}\beta_i\beta_j$ to be constant to ensure the validity of the derivation process in Eq.~\eqref{offshell}. This requirement does not introduce extra constraints since Eq.~\eqref{offshell} holds pointwise. This type of extension $\up V{-0}$ where $\up n{-0}\not=-\up n0$ is referred to as ``off-shell'', while the reverse is termed ``on-shell''. Certainly, for the off-shell $\up V{-0}$, according to the definition of the second fundamental form, $\up \SF{-0}$ should still remain orthogonal to $\up n{-0}$, that is, Eq.~\eqref{offshell} still applies. Fixing an index $[k]$, let $\beta_i=\delta_{ik}$. This leads to $\beta_i\beta_j=\delta_{ik}\delta_{jk}$, and due to $\bar X\zeta^{kk}=\bar X(1)=0$, Eq.~\eqref{gSFiXni0} arises again. It can be seen that Eq.~\eqref{gSFiXni0} is merely a specific instance of  Eq.~\eqref{offshell}. 

If we denote the vector $(\beta_1,\cdots,\beta_m)$ as $\beta$, then $\beta_i\beta_j$ correspond to the matrix elements of $\beta\otimes\beta$. Since Eq.~\eqref{offshell} is a linear equation of $\beta\otimes\beta$,  the number of independent constraints that Eq.~\eqref{offshell} can provide should be equal to the dimension of the vector space generated by all $\beta\otimes\beta$, which can be proven to be $m(m+1)/2$. In fact, all $\beta_i\beta_j$ have $m(m+1)/2$ components, hence the number of degrees of freedom $\leqslant m(m+1)/2$. To prove that equality is always achieved, we can directly construct $m(m+1)/2$ vectors $\beta$ such that all $\beta\otimes\beta$ are linearly independent.  Firstly, note that the symmetric matrix $\zeta$ is a non-degenerate metric. Choose $m$ vectors $\beta^{(a)},\; 1\leqslant a\leqslant m$, forming an orthonormal basis with respect to $\zeta$, i.e., $\zeta(\beta^{(a)},\beta^{(b)})=\pm \delta^{ab}$. Then, define $m(m-1)/2$ vectors $\beta^{(bc)}=\beta^{(b)}+\kappa\beta^{(c)}$, where $1\leqslant b<c\leqslant m$, and $0<\kappa<1$ is a constant. We have $\zeta(\beta^{(bc)},\beta^{(bc)})=\pm 1\pm \kappa^2\not=0$. It is evident that all $\zeta(\beta,\beta)$ are non-zero constants, and all $\beta^{(a)}\otimes \beta^{(a)}, \beta^{(bc)}\otimes \beta^{(bc)}$ are linearly independent. Therefore, the vector space generated by all $\beta\otimes\beta$ is precisely the space of $m+1$-order symmetric matrices. Consequently, each summation component in Eq.~\eqref{offshell} must be zero, i.e., 
\begin{equation}
	\label{constrainKXn}
	\sum_{k=1}^m\zeta^{ik}\mathscr K^{[j]}_{[k]}(\cdot,\up n j)+\zeta^{jk}\mathscr K^{[i]}_{[k]}(\cdot,\up n i)=-\mathsf s\,\mathrm d\zeta^{ij},
\end{equation}
$\forall i,j\in\big\{1,2,\cdots,m\big\}$, which includes Eq.~\eqref{gSFiXni0} as a special case. Using the constraint \eqref{constrainKXn}, we can prove the compatibility of the connections $\up\nabla{1,\cdots,m}_{\bar X}$ with the metric $g$. We only need to show that $\bar X\cdot g\left(\up n i,\up n j\right)-g\left(\up\nabla{1,\cdots,m}_{\bar X}\up n i,\up n j\right)-g\left(\up n i,\up\nabla{1,\cdots,m}_{\bar X}\up n j\right)=0$. Notice that the first term is $\bar X\cdot\zeta^{ij}$, and using the definition of the second fundamental form $\up \SF i$ we can rewrite this equation into Eq.\eqref{constrainKXn}.

Even though Eq.~\eqref{constrainKXn} provides $m(m+1)/2$ constraints, it is still insufficient to determine the $m^2$ components of $\mathscr K^{[i]}_{[j]}(\bar X, \up ni)$. Thus, the components of the connection $\up \nabla{1,\cdots,m}_{\bar X}\up n i$ cannot be determined. For the case of three pages, where $m+1=3$, the $\mathscr K^{[i]}_{[j]}(\bar X, \up n i)$ is left with only $m^2-m(m+1)/2=1$ degree of freedom, which is the norm of the extrinsic curvature for all bulks, as in Eq.~\eqref{scrK0=K1=K2}.

After establishing the local geometries of $\up M{1,\cdots,m}$ and $\up V{-0}$ at $\Sigma$, the junction condition emerges as a straightforward consequence. By applying the two pages junction condition for $\up V0\cup \up V{-0}$ and incorporating Eq.~\eqref{K0=betaiKi}, the junction condition for $m+1$ pages will be derived as
\begin{eqnarray}
	\label{betajunction}
	\notag
	&&\big(\up K0+\beta_1\up K1+\cdots+\beta_m\up K m\big)\\
	&&\quad-\big(\up{\tr K}0+\cdots+\beta_m\up{\tr K}m\big)h=-8\pi G\mathsf s\,\bar{\mathcal S}.\quad
\end{eqnarray}
However, we promptly recognize that the equation above lacks cyclic symmetry with respect to the index $[i]$. If we perform a reverse extension to $\up V1$, we obtain
\begin{equation}
	\up n{-1}\coloneqq -\up n 1=\frac{1}{\beta_1}\up n0+\frac{\beta_2}{\beta_1}\up n2+\cdots+\frac{\beta_m}{\beta_1}\up nm,
\end{equation}
which leads to
\begin{eqnarray}
	\notag
	&&\big(\up K0+\beta_1\up K1+\cdots+\beta_m\up K m\big)\\
	&&\quad -\big(\up{\tr K}0+\cdots+\beta_m\up{\tr K}m\big)h=-8\pi G\mathsf s\, \beta_1\bar{\mathcal S}.\qquad
\end{eqnarray}
If one seeks a self-consistent junction condition that is independent of the choice of $\up V i$, it is necessary to set $\beta_1=\beta_2=\cdots=\beta_m=1$. Namely, the continuity condition \eqref{d+mdimcc} holds. At this time, the junction condition takes the form of 
\begin{equation}
	\label{d+mdimjunction}
	\sum_{i=0}^m \up K i-\up{\tr K}i\,h=-8\pi G\mathsf s\,\bar{\mathcal S}.
\end{equation}
The significance of the continuity condition \eqref{d+mdimcc} for the junction condition becomes apparent. This also demonstrates that the junction condition is not governed by the geometry of the embedding manifold $\mathscr M$; instead, it is entirely determined by the matter distribution $\bar{\mathcal S}$ on $\Sigma$ and the geometries of all $\up Vi$. Conversely, the junction condition dictates how the booklet $\mathscr V$ is embedded within $\mathscr M$. While there can be multiple choices for $\mathscr M$, Eq.~\eqref{d+mdimcc} must always be satisfied. Therefore, we emphasize that Eqs.~\eqref{d+mdimcc} and \eqref{d+mdimjunction} are closely related. Equation \eqref{d+mdimjunction} is determined by matter and is observed by an internal observer within $\mathscr V$, while Eq.~\eqref{d+mdimcc} represents the external manifestation when embedded in the higher dimensional spacetime $\mathscr M$. In this sense, we say that the junction condition and the continuity condition for the normal vector are dual to each other.

Now we can briefly discuss the case where some of the normal vectors are timelike and some are spacelike. Thus, the previously uniform sign $\mathsf s$ needs to be replaced with a page-dependent $\up{\mathsf s}i=\up g i(\up n i,\up n i)$. All the above discussions are similar, and the continuity condition for the normal vectors are modified to
\begin{eqnarray}
	\up{\mathsf s}0\up n 0+\up{\mathsf s}1\up n 1+\cdots+\up{\mathsf s}m\up n m=0\ .
\end{eqnarray}
It is easy to see that $\up{\mathsf s}i$ plays the role of $\beta_i$ in Eq.\eqref{n0=betaini}, thus satisfying the constraint $\sum_{i,j}\zeta^{ij}\up{\mathsf s}i\up{\mathsf s}j=\up{\mathsf s}0$. This condition will revert to $\up{\mathsf s}0=\up{\mathsf s}1=\mathsf s$ in the 2-page scenario. At the same time, the junction conditions become
\begin{eqnarray}
	\label{sijunction}
	\sum_{i=0}^m \up{\mathsf s}i\bigg\{\up K i-\up{\tr K}i\,h\bigg\}=-8\pi G\,\bar{\mathcal S}.
\end{eqnarray}

\subsection{Method of variation of the action\label{subsec5B}}
For the case of joining 2 pages, we can integrate Eq.\eqref{LagrangianWithBoundary}, the terms proportional to $\Theta$ and $1-\Theta$ will yield the Hilbert bulk actions for $\up V0$ and $\up V1$, respectively. The terms proportional to the $\delta$ function will produce the Gibbons-Hawking-York (GHY) boundary terms\cite{GH1977,Y1972} on $\partial\up V0$ and $\partial\up V1$. Therefore, the action $I_{\mathscr V}$ in the booklet is the direct sum of the Hilbert actions $I_{\up V i},\,i=0,1$ for each page, each with the GHY boundary term. Varying this action can concurrently yield both the Einstein field equations and the Israel junction condition\cite{Moss2005}. This can be straightforwardly extended to the case of $m+1$ pages
\begin{eqnarray}
	\label{sumaction}
	I_{\mathscr V}=I_{\up V 0}+I_{\up V 1}+\cdots+I_{\up V m}\ .
\end{eqnarray}
The dynamical variables of $I_{\mathscr V}$ should consist of the metrics of each page, but all metrics at $\Sigma$ should satisfy the consistency constraint $\up g i\big|_{\mathscr T(\Sigma)}=h$. The Hilbert action on each page will independently produce the Einstein field equations within this bulk, while the consistency at the boundary requires the summation of the boundary contributions of the action, ultimately resulting in the multi-page junction condition. This line of derivation is entirely independent of the reverse extension method, which reflects the fact that the junction conditions are intrinsic physical constraints of the booklet, determined by the matter on the interface. Therefore, deriving the junction conditions using the action variation method can serve as a cross-check for the reverse extension method.  The action on the booklet can be written as
\begin{eqnarray}
	\label{IscrV}
	\notag
	I_{\mathscr V}&=&\frac{1}{16\pi G}\int_{\mathscr V} \sum_{i=0}^m\big(\up\Theta i\, R^{[i]}-\up\delta i(\ell)\cdot 2\mathsf s\,K^{[i]}\big)\up\varepsilon i\\
	&=&\frac{1}{16\pi G}\sum_{i=0}^m\int_{\up Vi} R^{[i]}\,\up\varepsilon i-2\mathsf s\oint_{\Sigma}K^{[i]}\,\bar\varepsilon_h.
\end{eqnarray}
Here, $R^{[i]}=\mathrm{tr}_{\up g i}\up\Ric i,\; K^{[i]}=\up{\tr K}i$; and $\up \varepsilon i,\bar\varepsilon_h$ represent the volume elements of $\up V i,\Sigma$ respectively. It is necessary to provide an elaborate explanation of the various symbols involved in Eq.~\eqref{IscrV}. We define $\up\Theta i$ as a distribution on the booklet $\mathscr V$, serving as a natural generalization of the Heaviside distribution $\Theta(\ell)$. Within $\up V k-\Sigma$, $\up\Theta i=\delta^{ik}$; at $\Sigma$, the value of $\up\Theta i\big|_\Sigma$ becomes indistinct, which can be considered as any number between $0$ and $1$. However, the following property should be satisfied,
\begin{equation}
	\label{Theta}
	\sum_{k=0}^m\up\Theta k=1,\qquad \mathfrak L_{\up n i}\up\Theta i\big|_\Sigma=\up\delta i(0).
\end{equation}
Relative to $\up\Theta i$, $\up \delta i(\ell)$ is merely a distribution defined within $\up V i$ (and its reverse continuation), where $\ell$ is the parameter of the integral curves of $\up n i$. This is because, according to the details of our established method of reverse extension, $\up\delta i(\ell)$ only plays a role in $\up V i\cup\up V {-i}$. So, to allow the second equality in Eq.~\eqref{IscrV} to hold, we only need to formally establish the following formula concerning $\up\delta i(\ell)$,
\begin{equation}
	\int_{\up V j}\up \delta i(\ell)\,\up\varepsilon i =\delta^i_j\oint_{\Sigma}\bar\varepsilon_h,\; \int_{\mathscr V}\up \delta i(\ell)\,\up\varepsilon i=\oint_{\Sigma}\bar\varepsilon_h.
\end{equation}
We can formally simplify the above expression as $\bar\varepsilon_h=\up \delta i(\ell)\up \varepsilon i$, which is sufficient for deriving the properties of $\up \delta i(\ell)$.

The second equation of \eqref{Theta} is, in fact, an expression of Stokes theorem. In the preceding geometric derivation, this property is not necessary. However, when computing variations of the action, if our sole concern is the junction condition, utilizing this property allows us to swiftly extract the boundary terms (as the junction condition necessarily only involves the part that contributes to the boundary $\Sigma$ in the variation process), while other terms can be disregarded.

For ease of computation, we select a coordinate system for $\up V i$ near $\Sigma$. Let $x^{\bar \mu}=(x^0, x^1, \cdots, x^{d-1})$ be any set of internal coordinates of $\Sigma$, together with the parameter $\ell$ along with $\up ni$, they form Gaussian coordinates $x^\mu=(x^0, \cdots, x^{d-1},\ell)$. However, $\up ni$ is only a local vector at $\Sigma$. We further extend $\up ni$ to a vector field defined over an open set that contains $\Sigma$, with $\ell$ serving as the parameter of the integral curves of this vector field. Furthermore, setting $\ell=\text{const}$ defines a hypersurface $\Sigma_\ell$. We extend $\up ni$ such that it is orthogonal to $\Sigma_\ell$ everywhere and $\mathfrak l_{\ell\up n i}\Sigma=\Sigma_\ell$. It can be proved that such an extension inevitably exists and is unique within a small enough open set. At this point, the integral curves of $\up ni$ are generally not geodesics. The induced metric on $\Sigma_\ell$ can still be denoted as $h$. Since the variation calculation is similar for each page $\up V i$, we can omit the specific index $[i]$. In this setup, $g_{\ell\ell}\equiv 1$ and $g_{\bar \mu\ell}\equiv 0$.  We can choose $\delta g_{\mu\nu}$ without yielding $\delta n^\mu$, and this behavior will not impact the results at the boundary. Hence, only the components $\delta g_{\bar\mu\bar\nu}=\delta h_{\bar\mu\bar\nu}$ has significance.

During the variation, on the boundary, $\delta h_{\bar\mu\bar\nu}$ and $\partial_\ell \delta h_{\bar\mu\bar\nu}$ should be treated as independent dynamical variables. The latter is exactly equal to $2\delta K_{\bar\mu\bar\nu}$. This is because $\ell$ is not an internal coordinate tangent to $\Sigma$, so $\partial_\ell$ does not yield a total derivative term on $\Sigma$. Alternatively, $\partial_\ell\delta(\ell)$ is ill-defined. Additionally, $h_{\bar\mu\bar\nu}$ represents the intrinsic metric of $\Sigma$, and $K_{\bar\mu\bar\nu}$ quantifies how $\Sigma$ is embedded in $\up V i$, so the independence of the two is reasonable. However, in practice, the outcome of the variation does not depend on $\delta K_{\bar\mu\bar\nu}$. 

The variation of the boundary Lagrangian yields
\begin{equation}
	\delta\big(2K\bar\varepsilon_h\big)=\bigg\{\big(2K_{\bar\mu\bar\nu}-K h_{\bar\mu\bar\nu}\big)\delta h^{\bar\mu\bar\nu}+2h^{\bar\mu\bar\nu}\delta K_{\bar\mu\bar\nu}\bigg\}\bar\varepsilon_h.
\end{equation}
As for the Lagrangian of the bulk part, $\Theta(\ell) R\,\varepsilon$, our focus should be on terms that contain $\partial_\ell$ within the Ricci scalar $R$. For terms of the form $\Theta(\ell)\partial_\ell$, by employing the method of integration by parts, the action of $\partial_\ell$ will be transferred to $\Theta(\ell)$, resulting in the appearance of terms containing $\partial_\ell\Theta(\ell)=\delta(\ell)$. This process is equivalent to, using Stokes' theorem to transform total derivative terms into boundary terms. Consequently, 
\begin{equation}
	\label{partialell}
	\delta\big(\Theta(\ell) R\,\varepsilon\big)=
	\delta(\ell)\bigg\{\mathsf s\,\delta{\Gamma^{\bar\nu}}_{\ell\bar\nu}-h^{\bar\rho\bar\sigma}\delta{\Gamma^\ell}_{\bar\rho\bar\sigma}\bigg\}\varepsilon+\text{others},
\end{equation}
where “others” represents terms that do not contribute to the boundary variation. Using
\begin{eqnarray}
	\notag
	{\Gamma^\ell}_{\bar\rho\bar\sigma}&=&\mathsf s\,g\big(\nabla_{\partial_{\bar\rho}}\partial_{\bar\sigma},\partial_\ell\big)=-\mathsf s\,g\big(\partial_{\bar\sigma},\nabla_{\partial_{\bar\rho}}\partial_\ell\big)\\
	\notag
	&=&-\mathsf s\,g\big(\partial_{\bar\sigma},\nabla_{\partial_\ell}\partial_{\bar\rho}\big)=-\mathsf s\,K_{\bar\rho\bar\sigma},\\
	{\Gamma^{\bar\nu}}_{\ell\bar\nu}&=&h^{\bar\mu\bar\nu}g\big(\nabla_{\partial_\ell}\partial_{\bar\nu},\partial_{\bar\mu}\big)=h^{\bar\mu\bar\nu}K_{\bar\nu\bar\mu}=K,
\end{eqnarray}
we get
\begin{equation}
	\label{deltaThetaR}
	\delta\big(\Theta(\ell) R\,\varepsilon-2\mathsf s\,K\bar\varepsilon_h\big)=-\mathsf s\big(K_{\bar\mu\bar\nu}-K h_{\bar\mu\bar\nu}\big)\bar\varepsilon_h\delta h^{\bar\mu\bar\nu}+\text{others}.
\end{equation}
The results here are consistent with those presented in the reference~\cite{Moss2005}. Restoring the index $[i]$ and summing over it, we rediscover Eq.~\eqref{d+mdimjunction}, which we recast as
\begin{equation}
	\sum_{i=0}^m K^{[i]}_{\bar\mu\bar\nu}-K^{[i]} h_{\bar\mu\bar\nu}=-8\pi G\cdot\mathsf s\,\bar S_{\bar\mu\bar\nu},
\end{equation}
where $\bar S_{\bar\mu\bar\nu}$ represents the components of the energy-momentum tensor $\bar {\mathcal S}$ in the coordinate basis $x^{\bar\mu}$.

The process of deriving the junction condition through variations of the action may seem formally simpler. Nevertheless, this approach obscures the geometric essence of the junction condition.  

 {Based on the previous discussion, we also see that the junction condition does not depend on how $\up M{1,\cdots,m}$ is constructed or how $\up V{-0}$ is extended. First, it is clear that the definition of the gluing process presented in the introduction is intrinsic, which indicates that our construction does not depend on the embedding manifold. Indeed, we can also arrive at the same conclusion by noticing that the resulting junction condition~\eqref{jcfor3bulks} does not depend on how this geometry is embedded (in the derivation by reverse extension) since the result only depends on the extrinsic curvatures in the directions of the various pages joining at the interface. It is also evident from the variational principle derivation of the junction condition that no embedding manifold is needed. Therefore, the junction condition fundamentally stems from the tangent spaces of $\up V0, \up V1, \cdots, \up V m$ at $\Sigma$, while $\up M{1,\cdots,m}$ and $\up V{-0}$ merely serve as auxiliary tools to get the final result. 
}

\section{Junction conditions in dilaton gravity \label{sec6}}
Dilaton gravity, such as 2d JT gravity, plays important role in recent breakthrough of the black hole information paradox in low dimension. It is therefore interesting to consider dilation gravity theories in general dimensions. In this section, we will only consider the scenario where all normal vectors are spacelike, thus fixing $\mathsf s=1$. Results for timelike normal vectors can be derived similarly.
\subsection{Equations of motion and change of frames\label{subset6A}}
We consider the following action for a general dilaton gravity theory in $d+1$ spacetime dimension
\begin{eqnarray}
	\label{dilatongravity}
	\notag
	I_{\up V i}=&&\frac{1}{16\pi \GN}\int_{\up V i} \bigg\{(\up\Phi i)^2 R^{[i]}\\
	&&+\lambda\,\big(\partial\up\Phi i\big)^2-\up{\mathscr U} i\big((\up\Phi i)^2\big)\bigg\}\up\varepsilon i,\quad
\end{eqnarray}
where $\up\varepsilon i$ is the $d+1$ dimensional volume element of page $\up V i$, and the dimensionless scalar field $(\up\Phi i)^2$ is identified as the dilaton. The $\up{\mathscr U} i\big((\up\Phi i)^2\big)$ denotes a scalar potential. Note that the potential possesses a dimensionality of $\text{Length}^{-2}$. Consequently, the form of the function $\up{\mathscr U}i$ depends on the characteristic length $\up L i$ of $\up V i$ (e.g. if $\up V i$ corresponds to an Anti-de Sitter (AdS) spacetime then $\up L i$ is the $\AdS$ radius). Given that different pages may have distinct characteristic lengths, we use the page number $[i]$ to distinguish the potential functions across the pages. 

This action is not the most general action for dilaton gravity; we choose this form simply because when $d=1$ it take the usual form of $2$D dilaton  gravity, see e.g.~\cite{almheiri2015,Maldacena:2019cbz,GRUMILLER2002327,Turiaci2021,Maeda2000,Henz2013}. We will work with this general action and focus on its interesting special case of JT gravity in the companion papers~\cite{SPL,SPL2}. Notice that although this action~\eqref{dilatongravity} is not the most general one, it has direct connections with many other known dilaton gravity models. For example, turning off the potential term $\up{\mathscr U}i\big((\up\Phi i)^2\big)$ in~\eqref{dilatongravity} gives the Brans-Dicke gravity action~\cite{BransDick1961}. In addition, the more general action of dilaton gravity can be expressed as~\cite{Philippe2010, Gasperini2001,Damour2002},
\begin{equation}
	\label{generaldilaton}
	I_{\text{DG}}=\frac{1}{16\pi \GN}\int \bigg\{e^{-2\mathsf \psi(\phi)} R+\mathsf Z(\phi)\big(\partial\phi\big)^2-\mathsf V(\phi)\bigg\}\varepsilon,
\end{equation}
where we omit the page number $[i]$ for the sake of convenience. It is straightforward to show that if we redefine the dilaton as $\Phi=\mathscr F(\phi)$, which satisfies 
\begin{equation}
	\mathsf Z(\phi)=\lambda\,\mathscr F^\prime(\phi)^2,\quad \mathscr U(\Phi)=\mathsf V\bigl(\mathscr F^{-1}(\Phi)\bigr)=\mathsf V(\phi),
\end{equation}
our action~\eqref{dilatongravity} reduces to a special case of~\eqref{generaldilaton} with  \footnote{Notice that as long as $\mathsf Z\not=0$, it is ensured that $\mathscr F^\prime\not=0$, thereby rendering the inverse function $\mathscr F^{-1}$ well-defined.}
\begin{equation}
	\psi(\phi)=-\ln \big|\mathscr F(\phi)\big|.
\end{equation}

We are interested in the dilaton gravity which couples to extra matter fields, which means
\begin{equation}
I=\sum_i I_{\up V i}+\int_{\up V i} \mathcal L^{[i]}_\text{matt} \,\up\varepsilon i.
\end{equation}
The equations of motion of the metric $g^{\mu\nu}$ and the scalar field $\Phi$ are
\begin{subequations}
	\begin{eqnarray}
		\label{vartophi}
		&&\Phi R-\lambda\Box\Phi-\Phi\mathscr U^\prime(\Phi^2)=4\pi \GN\,\Xi,\\
		\notag
		\label{vartog}
		&&\Phi^2\left(R_{\mu\nu}-\frac{1}{2}g_{\mu\nu}R\right)+\lambda \left(\partial_\mu\Phi\partial_\nu\Phi-\frac{1}{2}g_{\mu\nu}(\partial\Phi)^2\right)\qquad\\
		\notag
		&&\quad-\left(\nabla_\mu\partial_\nu\Phi^2-g_{\mu\nu}\Box\Phi^2\right)+\frac{1}{2}g_{\mu\nu}\,\mathscr U\left(\Phi^2\right)\\
		&&=8\pi \GN\, T_{\mu\nu}.
	\end{eqnarray}
\end{subequations}
where the scalar source strength $\Xi$ is 
\begin{equation}
	\Xi=-2\frac{\delta \mathcal L_\text{matt}}{\delta\Phi}.
\end{equation}
In the sake of generality, we assume the matter fields also couple to the dilaton, if this is not the case we simply set $\Xi=0$ and all the following discussions apply directly.

For the 2-dimensional case, the kinetic term $\lambda (\partial\Phi)^2$ can be eliminated \cite{almheiri2015} by a Weyl transformation on the metric,
\begin{equation}
	\label{metricWeyl}
	    \widetilde g_{\mu\nu}=\Phi^{2\alpha} g_{\mu\nu},\quad \widetilde g^{\mu\nu}= \Phi^{-2\alpha} g^{\mu\nu},
\end{equation}
while the form of other terms remains unchanged, where $\alpha$ is a dimensionless constant. We refer to $g_{\mu\nu}$ as the original metric and $\widetilde g_{\mu\nu}$ as the conformal metric.  {Notice that we do not assume the theory to be invariant under Weyl transformation, the above is simply a change of variables, or a change of frame.} 

In cases higher than $2$ dimensions, this approach can still eliminate the kinetic term but cannot keep the form of $\Phi^2R$ invariant (unless the $\Phi$ field undergoes redefinition). Now, let us examine the specific results for the $d+1$ dimensional case. The conformal factor $\Phi^{2\alpha}$ leads to the original connection being incompatible with the conformal metric, requiring an additional contortion tensor to fix this, 
\begin{equation}
	\label{contortion}
	{\widetilde\Gamma^\rho}_{\;\mu\nu}-{\Gamma^\rho}_{\mu\nu}= \alpha\Phi^{-1}\left(\delta^\rho_{\nu}\partial_\mu\Phi  +\delta^\rho_{\mu}\partial_\nu \Phi -g_{\mu\nu}\partial^\rho\Phi  \right).
\end{equation}
The Riemann curvature, Ricci tensor and Ricci scalar of the conformal metric can be expressed as
\begin{widetext}
\begin{subequations}
	\label{conformalRcurvature}
	\begin{eqnarray}
		\widetilde R^\rho_{\;\;\mu\gamma\nu}-{R^\rho}_{\mu\gamma\nu}
		&=&\nabla_\gamma\bigl(\,{\widetilde\Gamma^\rho}_{\;\mu\nu}-{\Gamma^\rho}_{\mu\nu}\bigr)+\alpha^2\Phi^{-2}\big(\delta^\rho_\gamma\pd_\nu\Phi\pd_\mu\Phi+\delta^\rho_\nu g_{\gamma\mu}(\pd\Phi)^2+g_{\mu\nu}\pd_\gamma\Phi\pd^\rho\Phi\big)-\big(\nu\leftrightarrow \gamma\big),\\
		\widetilde R_{\mu\nu}-R_{\mu\nu}&=&\alpha^2(d-1)\,\Phi^{-2}\big(\partial_\mu\Phi\partial_\nu\Phi-g_{\mu\nu}(\partial\Phi)^2\big) -\alpha(d-1)\,\nabla_\mu(\Phi^{-1}\partial_\nu\Phi)-\alpha g_{\mu\nu}\nabla_\rho(\Phi^{-1}\partial^\rho\Phi),\\
		\Phi^{2\alpha}\widetilde R-R&=&-\alpha^2d(d-1)\,\Phi^{-2}(\pd\Phi)^2-2\alpha d\,\nabla_\rho\big(\Phi^{-1}\pd^\rho\Phi\big).
	\end{eqnarray}
\end{subequations}
\end{widetext}
The spacetime measure is affected by the change~\eqref{metricWeyl}, while its orientation  remains unchanged, this leads to 
\begin{equation}
	\label{conformalvolume}
	\widetilde \varepsilon=\Phi^{\alpha(d+1)}\varepsilon.
\end{equation}
To demonstrate this equation, let us arbitrarily select a tetrad composed of $d+1$ orthonormal $1$-forms, and represent the volume element as their wedge product. Following the Weyl transformation, the normalization is redefined, with each member of the tetrad magnified by a factor of $\Phi^\alpha$, leading to~\eqref{conformalvolume}.

Based on these data, we obtain the Lagrangian after the Weyl transformation
\begin{eqnarray}
	\label{L-conformalL}
	\notag
	16\pi \GN\,\mathcal L\varepsilon&=&\bigg\{\Phi^2R+\lambda(\pd\Phi)^2-\mathscr U(\Phi^2)\bigg\}\varepsilon\\
	&=&\alpha d\,\nabla_\rho(\pd^\rho\Phi^2)\varepsilon
	+16\pi \GN\,\widetilde{\mathcal L}\,\widetilde\varepsilon,
\end{eqnarray}
where
\begin{eqnarray}
	\label{conformalL}
	\notag
	16\pi \GN\,\widetilde{\mathcal L}&=&\Phi^{2-\alpha(d-1)}\widetilde R+\big(\lambda+\alpha^2d(d-1)\\
	&&-4\alpha d\big)\Phi^{-\alpha(d-1)}(\widetilde\partial\Phi)^2-\widetilde{\mathscr U}(\Phi^2)\,,
\end{eqnarray}
and $\widetilde{\mathscr U}(\Phi^2)=\Phi^{-\alpha(d+1)}\mathscr U(\Phi^2)$. We further redefine the dilaton $\widetilde\Phi^2=\Phi^{2-\alpha(d-1)}$, and accordingly rewrite the potential  $\widetilde{\mathsf U}(\widetilde\Phi^2)=\widetilde{\mathscr U}(\Phi^2)$. The Lagrangian becomes  
\begin{eqnarray}
	\label{redefdilaton}
	\notag
	16\pi \GN\,\widetilde{\mathcal L}&=&\widetilde\Phi^2\widetilde R+\frac{4\big(\lambda+\alpha^2d(d-1)-4\alpha d\big)}{\bigl(\alpha(d-1)-2\bigr)^2}(\widetilde\partial\widetilde\Phi)^2\\
	&&-\widetilde{\mathsf U}\bigl(\widetilde \Phi^2\bigr),\qquad
\end{eqnarray}
provided $\alpha\not=\frac{2}{d-1}$. The conformal Lagrangian $\widetilde{\mathcal L}\widetilde\varepsilon$ differs from the original one by only a boundary term, and thus they necessarily result in equivalent dynamics. 
It is clear from~\eqref{redefdilaton} that if we set
\begin{equation}
		\label{lambda}
		\lambda+\alpha^2d(d-1)-4\alpha d=0,
\end{equation}
namely,
\begin{equation}
	\alpha=
	\begin{cases}
		\lambda/4, &d=1\\
		\dfrac{2}{d-1}\bigg(1\pm\sqrt{1-\dfrac{d-1}{d}\dfrac{\lambda}{4}}\,\bigg),&d>1
	\end{cases}
	~,
\end{equation}
the kinetic term of the dilaton can be eliminated, then the conformal Lagrangian takes the form $\Phi^{2-\alpha(d-1)}\widetilde R-\widetilde {\mathscr U}(\Phi^2)$. Obviously, when $d=1$, we have $\widetilde\Phi=\Phi$ regardless of the value of $\alpha$. When $d > 1$, if we choose $\alpha=\frac{2}{d-1}$, the $\widetilde R$ and $\Phi$ decouple. We therefore refer to this decoupled action as being in the Einstein frame; in contrast, the action \eqref{dilatongravity} is refered to be in the Jordan frame. 

 {The change of variables~\eqref{metricWeyl} induces a change in the matter Lagrangian due to its coupling with gravity. We use variables with a tilde to represent the Lagrangian after the change}
\begin{equation}
	\widetilde{\mathcal L}_{\text{matt}}(\widetilde g, \Phi)\widetilde \varepsilon=\mathcal L_{\text{matt}}(g,\Phi)\varepsilon\ .
\end{equation}
The following computation 
\begin{subequations}
	\begin{eqnarray}
		\notag
		\widetilde \Xi \widetilde\varepsilon&=&-2\frac{\delta \mathcal {\widetilde L}_\text{matt}}{\delta\Phi}\widetilde\varepsilon=-2\frac{\delta\mathcal L_\text{matt}}{\delta\Phi}\varepsilon-2\frac{\delta g^{\mu\nu}}{\delta\Phi}\frac{\delta\big( \mathcal L_\text{matt}\varepsilon\big)}{\delta g^{\mu\nu}}\\
		&= &\big(\Xi+2\alpha\Phi^{-1} {T^\mu}_\mu\big)\varepsilon\\
		\notag
		\widetilde T_{\mu\nu}\widetilde\varepsilon&=&-2\frac{\delta \big(\mathcal {\widetilde L}_\text{matt}\widetilde\varepsilon\big)}{\delta\widetilde g^{\mu\nu}}=-2\Phi^{2\alpha}\frac{\delta \left(\mathcal L_\text{matt}\varepsilon\right)}{\delta g^{\mu\nu}}\\
		&=& \Phi^{2\alpha} T_{\mu\nu}\varepsilon,
	\end{eqnarray}
\end{subequations}
leads to
\begin{eqnarray}
	\label{conformalT}
	\notag
	\widetilde \Xi&=&\Phi^{-\alpha(d+1)}\left(\Xi+2\alpha\Phi^{-1}{T^\mu}_\mu\right),\\
	\widetilde T_{\mu\nu}&=&\Phi^{-\alpha(d-1)}T_{\mu\nu}.
\end{eqnarray}

The equations of motion in the transformed frame are
\begin{widetext}
\begin{subequations}
	\label{conformalmotion}
	\begin{eqnarray}
		\label{conformalvartophi}
		\notag
		&&\big(2-\alpha(d-1)\big)\Phi^{1-\alpha(d-1)}\widetilde R -\big(\lambda+\alpha^2d(d-1)-4\alpha d\big)\cdot\Phi^{-\alpha(d-1)-1}\\
		&&\hspace{4.5cm}\times\bigg(2\Phi\widetilde{\Box}\Phi-\alpha(d-1)\big(\widetilde\partial\Phi\big)^2\bigg)-2\Phi\widetilde{\mathscr U}^{\,\prime}(\Phi^2)=8\pi \GN\, \widetilde\Xi,\\
		\label{conformalvartog}
		\notag
		&&\Phi^{2-\alpha(d-1)}\left(\widetilde R_{\mu\nu}-\frac{1}{2}\widetilde g_{\mu\nu}\widetilde R\right)-\big(\widetilde\nabla_\mu\partial_\nu-\widetilde g_{\mu\nu}\widetilde \Box\big)\Phi^{2-\alpha(d-1)}+\big(\lambda+\alpha^2d(d-1)-4\alpha d\big)\cdot\Phi^{-\alpha(d-1)}\\
		&&\hspace{4.5cm}\times\bigg(\widetilde\partial_\mu\Phi\widetilde\partial_\nu\Phi-\frac{1}{2}\widetilde g_{\mu\nu}\big(\widetilde\partial\Phi\big)^2\bigg)+\frac{1}{2}\widetilde g_{\mu\nu}\,\widetilde{\mathscr U}(\Phi^2)=8\pi \GN\, \widetilde T_{\mu\nu}.
	\end{eqnarray}
\end{subequations}
\end{widetext}
It is clear from the above equations that when $\alpha=\frac{2}{d-1}$ the dilaton does not directly couple to the curvature and can be absorbed in to the stress energy tensor, so that~\eqref{conformalvartog} reduces to the Einstein equation. Moreover, the equations of motion~\eqref{vartophi} and \eqref{vartog}, as well as their conformal counterparts, \eqref{conformalvartophi} and \eqref{conformalvartog}, are equivalent. This is nothing surprising; the variations of the actions $\delta\widetilde{\mathcal L}$ and $\delta \mathcal L$ are related through a Jacobian  $\frac{\partial (\widetilde g,\Phi)}{\partial(g,\Phi)}$ that is invertible due to  $\det\frac{\partial (\widetilde g,\Phi)}{\partial(g,\Phi)}= \Phi^{2\alpha}\not=0$. Therefore, the two sets of equations of motion must simultaneously equal zero. Indeed, we can confirm their equivalence by explicit calculations with the help of Eqs.~\eqref{conformalRcurvature} and \eqref{conformalT}.

\subsection{Junction conditions for dilaton gravity\label{subsec6B}}
In this section we derive the multiway junction condition in dilaton gravity. Let us consider $m+1$ pages, denoted as $\up V i$, intersecting at $\Sigma$ whose tangent space is denoted by $\mathscr T(\Sigma)$; each of these pages hosts a dilaton field $\up \Phi i$ that are required to satisfy the following continuity condition
\begin{equation}
	\label{continfordilaton}	
	\bar\Phi\coloneqq\up\Phi 0\big|_\Sigma=\up\Phi 1\big|_\Sigma=\cdots=\up \Phi m\big|_\Sigma.
\end{equation}
Each $\up\Phi i$ can generate a $1$-form within the cotangent space of $\up V i$, denoted as $\ed\up\Phi i$. In accordance with Eq.~\eqref{continfordilaton}, these $1$-forms also satisfy the continuity condition
\begin{equation}
	\label{continfordPhi}	
	\ed\bar\Phi=\ed\up\Phi 0\big|_{\mathscr T(\Sigma)}=\cdots=\ed\up \Phi m\big|_{\mathscr T(\Sigma)},
\end{equation}
where $\big|_{\mathscr T(\Sigma)}$ denotes the restriction of the operation domain of $1$-forms to $\mathscr T(\Sigma)$. We further have $\ed\bar\Phi(\up n 0)=\ed\bar\Phi(\up n 1)=\cdots=\ed\bar\Phi(\up n m)=0$ to enable $\ed\bar\Phi$ to act on all tangent spaces $\mathscr T_\Sigma(\up V i)$, then 
\begin{equation}
	\label{compofdPhii}
	\ed\up\Phi i=\ed\bar\Phi+\big(\mathfrak L_{\up  n i}\up\Phi i\big)\,\up{n^*}i,
\end{equation}
where $\up{n^*}i=\up g i(\up n i,\cdot)$ is a $1$-form within the cotangent space of $\up V i$. For a vector $\bar X$ tangent to $\Sigma$, we have $\up{n^*}i(\bar X)=0$; however $\up{n^*}i(\up n i)=1$. 

Based on the above results, we introduce a notation $\ed\up\Phi\Sigma$, similar to the $\up R\Sigma$ notation used earlier
\begin{eqnarray}
	\label{dPhiSigma}
	\notag
	\ed\up\Phi\Sigma&=&\sum_{i=0}^m\up\Theta i(0)\ed\up\Phi i\\
	&=&\ed\bar\Phi +\sum_{i=0}^m\up\Theta i(0)\big(\mathfrak L_{\up  n i}\up\Phi i\big)\,\up{n^*}i.
\end{eqnarray}

Next, our goal is to derive the junction condition for dilaton gravity by examining the singularities at $\Sigma$ of the field equations \eqref{vartophi} and \eqref{vartog}. First notice that terms such as $\up{\mathscr U}i\big((\up\Phi i)^2\big)$, $\frac{1}{2}\up g i_{\mu\nu}\,\up{\mathscr U}i\big((\up\Phi i)^2\big)$ and  $\up\Phi i\mathscr U^{[i]\prime}\big((\up\Phi i)^2\big)$ do not exhibit singularity proportional to $\delta(\ell)$ at $\Sigma$, since the functions $\up{\mathscr U}i$ and $\mathscr U^{[i]\prime}$ does not depend on the derivatives of $\up\Phi i$. Next we consider the
behavior of terms with derivatives in the action and the motion equations, such as $\partial_\mu\up\Phi i\partial_\nu\up\Phi i$ and $(\partial\up\Phi i)^2$, near $\Sigma$. Concerning $\partial_\mu\up\Phi i\partial_\nu\up\Phi i$, it represents the components of the tensor $\ed\up\Phi i\otimes\ed\up\Phi i$ in the coordinate basis. Eq.\eqref{continfordPhi} has already guaranteed the continuity of $\ed\up\Phi i$, and the tensor product operation is a localized algebraic multiplication. Consequently, the tensor product of $\ed\up\Phi i$ with itself will also satisfy continuity at  $\Sigma$. We use the distribution $\up\Theta i$ to concatenate all $\ed\up\Phi i\otimes\ed\up\Phi i$ at $\Sigma$, obtaining the tensor $\big(\ed\Phi\otimes\ed\Phi\big)^{[\Sigma]}$, 
\begin{equation}
	\big(\ed\Phi\otimes\ed\Phi\big)^{[\Sigma]}=\sum_{i=0}^m\up\Theta i(0)\ed\up\Phi i\otimes \ed\up\Phi i.
\end{equation}
In a similar manner, we can represent $(\partial\up\Phi i)^2$ as $\up{g^*}i\big(\ed\up\Phi i,\ed\up\Phi i\big)$, signifying the inner product of the $1$-form $\ed\up\Phi i$ with itself under the bivector $\up {g^*}i$ (which is the inverse of $\up gi$). Just like the tensor product operation, $\up {g^*}i$ is a localized multiplication and thus $(\partial\up\Phi i)^2$ will exhibit continuity at $\Sigma$. Thus, we have demonstrated that the term $\lambda\big(\partial_\mu\up\Phi i\partial_\nu\up\Phi i-\frac{1}{2}\up g i_{\mu\nu}(\partial\up\Phi i)^2\big)$ within the field equations does not contribute to the junction condition.

The remaining terms in the motion equations are $\up\nabla i_\mu\partial_\nu(\up\Phi i)^2$, $\up \Box i\up\Phi i$ and $\up\Box i(\up\Phi i)^2$. These wave terms, similar to the curvature tensor, involve second-order derivatives of the dynamical variables, and thus, will inevitably contribute nontrivially to the junction condition. For convenience, we directly consider such a scalar field $\up\psi i=\mathscr F(\up\Phi i)$ instead of dilaton $\up\Phi i$, where the continuous function $\mathscr F$ depends pointwise on the values of $\up\Phi i$. For instance, $\mathscr F$ is a power function of $\up\Phi i$, which is the case we will employ. Immediately, we can deduce that at $\Sigma$,
\begin{equation}
	\label{dFdilaton}
	\ed\up\psi i=\mathscr F^\prime (\up\Phi i)\,\ed\up\Phi i=\mathscr F^\prime(\bar\Phi)\,\ed\up\Phi i\ .
\end{equation}
Therefore, the continuity condition \eqref{continfordPhi} for $\ed\up\Phi i$ can be used to derive the continuity condition for $\ed\up\psi i$. Adopting the method introduced in Eq.~\eqref{addpathSigma}, we establish a tensor at the location of $\Sigma$ denoted as $\big(\nabla\ed\psi\big)^{[\Sigma]}$. This tensor must account for contributions from all $\up\nabla i\ed\up\psi i$ and will additionally generate terms that are proportional to $\delta(\ell)$.

\subsubsection{Junction condition from reverse extension}
Repeating the process of Sec.~\ref{subsec5A}, we smoothly extend the dilatons across the entire $\up M{1,\cdots, m}$, resulting in a scalar field denoted as $\up\Phi{1,\cdots,m}$; then we restrict this field to $\up V{-0}$, yielding the $\up\Phi{-0}$. By applying the continuity condition \eqref{d+mdimcc} of normal vectors,
\begin{eqnarray}
	\notag
	\ed\up\Phi{-0}(\up n{-0})&=&\ed\up\Phi{1,\cdots,m}(\up n{-0})\\
	\notag
	&=&\sum_{i=1}^m\ed\up\Phi{1,\cdots,m}(\up n i)=\sum_{i=1}^m\ed\up \Phi i(\up n i),\\
\end{eqnarray}
equivalently,
\begin{equation}
	\label{Ln0Phi0}
	\mathfrak L_{\up n{-0}}\up\Phi{-0}=\mathfrak L_{\up n1}\up\Phi1+\mathfrak L_{\up n2}\up\Phi2+\cdots+\mathfrak L_{\up n m}\up\Phi m.
\end{equation}
Thanks to the continuity condition \eqref{continfordilaton} of dilatons, Eq.~\eqref{dFdilaton} holds true for any $[i]\in\big\{[-0],[1],\cdots,[m]\big\}$, directly we have
\begin{equation}
	\ed\up\psi{-0}(\up n{-0})=\sum_{i=1}^m\ed\up \psi i(\up n i)=\sum_{i=1}^m \mathfrak L_{\up n i}\up\psi i,
\end{equation}
 while for vectors $\bar X$ tangent to $\Sigma$, $\ed\up\psi{-0}(\bar X)=\ed\bar\psi(\bar X)$, where $\bar\psi=\mathscr F(\bar\Phi)$. Therefore,
 \begin{eqnarray}
	\notag
 	\ed\up\psi{-0}&=&\ed\bar\psi+\bigg(\sum_{i=1}^m \mathfrak L_{\up n i}\up\psi i\bigg)\up{n^*}{-0}\\
	&=&\ed\bar\psi-\bigg(\sum_{i=1}^m \mathfrak L_{\up n i}\up\psi i\bigg)\up{n^*}0.
 \end{eqnarray}
 Taking $i=0$ in Eq.~\eqref{compofdPhii}, we can compare $\ed\up\psi 0$ and $\ed\up\psi{-0}$,
 \begin{eqnarray}
	\notag
 	\ed\up\psi 0-\ed\up\psi{-0}&=&\bigg(\sum_{i=0}^m \mathfrak L_{\up n i}\up\psi i\bigg)\up{n^*}0\\
	&=&\mathscr F^\prime(\bar\Phi)\bigg(\sum_{i=0}^m \mathfrak L_{\up n i}\up\Phi i\bigg)\up{n^*}0.\quad
 \end{eqnarray} 
 
 At this point, we possess all the necessary information to define $\big(\nabla\ed\psi\big)^{[\Sigma]}$ on $\up V{-0}\cup \up V0$. Let us begin with the component in the normal direction, denoted as $\big(\nabla_{\up n0}\ed\psi\big)^{[\Sigma]}$,
 \begin{widetext}
 	\begin{eqnarray}
 		\label{nablan0dpsiSigma}
 	\notag
 	\big(\nabla_{\up n0}\ed\psi\big)^{[\Sigma]}&=&\lim_{s,\iota\to 0^+}\frac{\tau^{[0]*}_{s\up n0}\ed\up\psi 0-\tau^{[-0]*}_{-\iota\up n0}\ed\up\psi{-0}}{s+\iota}\\
 	\notag
 	&=&\lim_{s,\iota\to 0^+}\bigg\{\frac{s}{s+\iota}\frac{\tau^{[0]*}_{s\up n0}\ed\up\psi 0-\ed\up\psi 0}{s}
	 +\frac{\iota}{s+\iota}\frac{\ed\up \psi{-0}-\tau^{[-0]*}_{-\iota\up n0}\ed\up\psi{-0}}{\iota}+\frac{1}{s+\iota}\big(\ed\up\psi 0-\ed\up\psi{-0}\big)\bigg\}\\
 	&=&\up\Theta0(0) \up\nabla 0_{\up n0}\ed\up\psi 0+\big(1-\up\Theta0(0)\big)\up\nabla {-0}_{\up n0}\ed\up\psi {-0}+\up\delta0(0)\mathscr F^\prime(\bar\Phi)\bigg(\sum_{i=0}^m \mathfrak L_{\up n i}\up\Phi i\bigg)\up{n^*}0.
 \end{eqnarray}
 Here, the parallel translation of the $1$-forms is defined as $\tau^{[0]*}_{s\up n0}\ed\up\psi0 (X)=\ed\up\psi0\big(\tau^{[0]}_{s\up n0}X\big)$ with $X\in\mathscr T_\Sigma(\up V 0)$, namely, pulling back $\ed\up\psi0$ from the location corresponding to parameter $s$ to $\Sigma$ along the $-\up n0$ direction. Regarding the components in the direction $\bar X$ tangent to $\Sigma$, denoted as $\big(\nabla_{\bar X}\ed\psi\big)^{[\Sigma]}$, we can define it directly as
\end{widetext}

 \begin{equation}
 	\label{nablaXdpsiSigma}
 	\big(\nabla_{\bar X}\ed\psi\big)^{[\Sigma]}=\up\Theta0(0) \up\nabla 0_{\bar X}\ed\up\psi 0+\big(1-\up\Theta0(0)\big)\up\nabla {-0}_{\bar X}\ed\up\psi {-0},
 \end{equation}
 which ensures the linearity of $\big(\nabla\ed\psi\big)^{[\Sigma]}$. Upon contracting $\big(\nabla\ed\psi\big)^{[\Sigma]}$ with the metric, we arrive at
 \begin{eqnarray}
 	\label{BoxpsiSigma}
 	\notag
 	\big(\Box \psi\big)^{[\Sigma]}&=&\mathrm{tr}_{\up g 0}\big(\nabla\ed\psi\big)^{[\Sigma]}\\
	 \notag
 	&=&\up\Theta0(0)\up\Box 0\up\psi 0+\big(1-\up\Theta0(0)\big)\up\Box{-0}\up\psi{-0}\\
	&&+\up\delta0(0)\,\mathscr F^\prime(\bar\Phi)\sum_{i=0}^m \mathfrak L_{\up n i}\up\Phi i\ .
 \end{eqnarray}
 
 The field equation \eqref{vartophi} includes the term $\lambda\big(\Box \Phi\big)^{[\Sigma]}$. In this case, we choose $\mathscr F(\bar\Phi)=\bar\Phi$, with which the singular term will be $\up\delta 0(\ell)\sum_{i=0}^m \mathfrak L_{\up n i}\up\Phi i$. Eq.\eqref{vartog} includes the term  $\big(\nabla_\mu\pd_\nu\Phi^2-g_{\mu\nu}\Box\Phi^2\big)^{[\Sigma]}$, hence we choose $\mathscr F(\bar\Phi)=\bar\Phi^2$ and with this choice, the singular term will manifest as $-2\up\delta 0(\ell)\bar\Phi\,h_{\mu\nu}\sum_{i=0}^m \mathfrak L_{\up n i}\up\Phi i $.
 
 Regarding the matter part in the field equations, the energy-momentum tensor $\bar{\mathcal S}\,\bar\varepsilon_h$ on $\Sigma$ manifests itself as $\up\delta i(\ell)\bar{\mathcal S}\,\up\varepsilon i$ contained in $\up T i$. Similarly, if the matter Lagrangian on $\Sigma$ contains dilaton, it will also generate a scalar source strength $\bar\varrho\,\bar\varepsilon_h$, which manifests itself as $\up\delta i(\ell)\bar\varrho\,\up\varepsilon i$  in $\Xi$. Therefore, Eqs.~\eqref{vartophi}\eqref{vartog}  yield the following junction condition for dilaton gravity 
 \begin{subequations}
	\label{dilatonjunction}
	\begin{eqnarray}
		\label{junctionpartPhi}
		&&\sum_{i=0}^m 2\bar\Phi\,\up{\tr K}i+\lambda\mathfrak L_{\up n i}\up\Phi i=-4\pi \GN\, \bar\varrho, \\
		\label{junctionpartK}
		\notag
		&&\sum_{i=0}^m \bar\Phi^2\,\big(\up K i-\up{\tr K}i\, h\big)-2\bar\Phi \mathfrak L_{\up n i}\up\Phi i\, h=-8\pi \GN \bar{\mathcal S}.\\
	\end{eqnarray}
 \end{subequations}
 Obviously, the above equation exhibits cyclic symmetry with respect to the page numbers $[i]$, so as expected, the junction condition does not depend on our choice of which page to be reversely extended.
 
 \subsubsection{General prescription to derive junction conditions in other theories}
 Now we can summarize the general principles for finding junction conditions in various gravitational models, such as those involving additional scalar, vector, or tensor fields besides a metirc. When gluing together $m+1$ pages,  the dynamical variables $\up Q i$ must adhere to continuity conditions. Generally, $\up Q i$ takes the form of tensors of various orders, and the continuity conditions require that the projection of $\up Q i$ onto $\Sigma$ results in a consistent tensor $\bar Q$ independent of page numbers $[i]$. Furthermore, the continuity conditions necessitate the existence of a smooth extension $Q^{[1,\cdots,m]}$ within the external manifold $\mathscr M$ that is compatible with all $\up Q i$. Then, terms such as $\up \nabla i\up Q i$ or $\up \nabla i\up \nabla i\up Q i$ present in the field equations may lead to singularities proportional to $\delta(\ell)$. By employing the method of reverse extension, $\up Q i$ will adhere to an equation similar to \eqref{Ln0Phi0}; when combined with the model-independent equation \eqref{K0=betaiKi}, we will arrive at the junction condition. In particular, for gravity models with torsion, such as Einstein-Cartan gravity, teleparallel gravity, according to the definition $\up\nabla i_XY-\up\nabla i_YX-[X,Y]$ of torsion, it is sufficient to consider asymmetric extrinsic curvatures or second fundamental forms. The derivation in Sec.~\ref{subsec5A} requires no modification; Eq.~\eqref{K0=betaiKi} remains valid. However, these results are contingent on the assumption that the connection and metric are compatible. For gravity theories that deviate from this compatibility, such as Palatini gravity, $\up \nabla i_{\bar X}\bar Y-\bar{\nabla}_{\bar X}\bar Y$ might not be orthogonal to $\Sigma$. In such cases, the normal vector $\up n i$ should be replaced by a suitable vector linearly independent of the tangent space $\mathscr T(\Sigma)$, still denoted as $\up n i$, ensuring the invariance of the extrinsic curvature definition. The continuity conditions are expressed more generally by the Eq.~\eqref{sumalphaini0}.  Additional constraints are required, as dictated by the specific premises of the model. 
 
\subsubsection{Junction condition from variation of the action}
In the rest of this subsection, we will construct GHY-like boundary terms for dilaton gravity and, using the action variation approach, rederive the junction condition. Starting from the action \eqref{dilatongravity}, when compared with Eq.~\eqref{LagrangianWithBoundary}, it is evident that $(\up\Phi i)^2 R^{[i]}$ results in a boundary term of $-2\up\delta i(\ell)\,\bar\Phi^2 K^{[i]}$. Furthermore, Eq.~\eqref{dPhiSigma} ensures that $\lambda(\partial\up\Phi i)^2$ does not generate terms proportional to $\up\delta i(\ell)$; and $\up{\mathscr U}i\big((\up\Phi i)^2\big)$ cannot induce boundary terms either. Thus, the action for the entire booklet spacetime $\mathscr V$ takes the following form
\begin{eqnarray}
	\notag
	16\pi\GN\,I_{\mathscr V}&=&\int_{\mathscr V}\sum_{i=0}^m \up\Theta i\bigg\{(\up\Phi i)^2 R^{[i]}+\lambda(\partial\up\Phi i)^2\\
	\notag
	&&-\up{\mathscr U}i\big((\up\Phi i)^2\big)\bigg\}\up\varepsilon i-2\up\delta i(\ell)\bar\Phi^2 K^{[i]}\up\varepsilon i\\
	\notag
	&=&\sum_{i=0}^m\int_{\up Vi} \bigg\{(\up\Phi i)^2 R^{[i]}+\lambda(\partial\up\Phi i)^2\\
	\notag
	&&-\up{\mathscr U}i\big((\up\Phi i)^2\big)\bigg\}\up\varepsilon i-2\oint_\Sigma\bar\Phi^2 K^{[i]}\bar\varepsilon_h.\\
\end{eqnarray} 
Similar to the approach in Sec.~\ref{subsec5B}, we adopt a Gaussian coordinate, denoted by $x^\mu = (x^{\bar\mu}, \ell)$, where $x^{\bar\mu}$ represents the internal coordinates of $\Sigma$, and $\ell = \text{const}$ characterizes a family of hypersurfaces $\Sigma_\ell$. We extend the normal vector $\up n i$ to a vector field within $\up V i$ in such a way that it is orthogonal to $\Sigma_\ell$ everywhere and satisfies $\mathfrak l_{\ell\up ni}\Sigma=\Sigma_\ell$. For now, we omit the page numbers $[i]$ wherever this does not raise any confusion. 
We respectively express the bulk and boundary Lagrangian as $16\pi\GN\,\mathcal L=\Phi^2 R+\lambda(\partial\Phi)^2-\mathscr U\big(\Phi^2\big)$  and $16\pi\GN\,\bar {\mathcal L}_\Sigma=-2\bar\Phi^2 K$. 

Varying with respect to $h_{\bar\mu\bar\nu}$, the boundary Lagrangian yields
\begin{eqnarray}
	\notag
	&&16\pi\GN\,\delta_h\big(\bar {\mathcal L}_\Sigma\,\bar\varepsilon_h\big)\\
	&&=-\bar\Phi^2\bigg\{\big(2K_{\bar\mu\bar\nu}-K h_{\bar\mu\bar\nu}\big)\delta h^{\bar\mu\bar\nu}+2h^{\bar\mu\bar\nu}\delta K_{\bar\mu\bar\nu}\bigg\}\bar\varepsilon_h.\qquad
\end{eqnarray}
The contribution from the bulk Lagrangian $\mathcal L$ comes from those components depending on $\partial_\ell$. In our case, in addition to Ricci curvature, the only relevant term in the Lagrangian $\mathcal L$ is $\lambda (\pd_\ell\Phi)^2$, which however is independent of $h_{\bar\mu\bar\nu}$ and thus do not contribute. Therefore 
\begin{eqnarray}
	\label{partialelldeltah}
	\notag
	&&16\pi\GN\delta_h\big(\Theta(\ell)\mathcal L\varepsilon\big)\\
	\notag
	&&=-\big(\delta(\ell)\Phi^2+\Theta(\ell)\pd_\ell\Phi^2\big)\big(h^{\bar\rho\bar\sigma}\delta{\Gamma^\ell}_{\bar\rho\bar\sigma}
    -\delta{\Gamma^{\bar\nu}}_{\ell\bar\nu}\big)\varepsilon\\
	\notag
	&&\quad-\Theta(\ell) \bigg(h^{\bar\rho\bar\sigma}\pd_{\bar\nu}\Phi^2\delta{\Gamma^{\bar\nu}}_{\bar\rho\bar\sigma}+\pd_{\bar\nu}\Phi^2\delta{\Gamma^{\bar\nu}}_{\ell\ell}\\
	&&\quad-h^{\bar\rho\bar\sigma}\pd_{\bar\sigma}\Phi^2\delta{\Gamma^{\bar\nu}}_{\bar\rho\bar\nu}-h^{\bar\rho\bar\sigma}\pd_{\bar\sigma}\Phi^2\delta{\Gamma^{\ell}}_{\bar\rho\ell}\bigg)\varepsilon+\text{others},\qquad\quad
\end{eqnarray}
where ``others'' represents terms that do not contribute to the boundary variation. Comparing with Eq.~\eqref{partialell}, we observe several additional terms. These arise as corrections due to the coupling between $\Phi$ and $R$. The above equation only involves variations $\delta {\Gamma^\rho}_{\mu\nu}$ of the Christoffel symbols. By using the definition of the Christoffel symbols ${\Gamma^\rho}_{\mu\nu}=\frac{1}{2}g^{\rho\gamma}\big(\partial_\mu g_{\gamma\nu}+\partial_\nu g_{\gamma\mu}-\partial_\gamma g_{\mu\nu}\big)$, we can see that only 
\begin{eqnarray}
	\notag
	\delta {\Gamma^\ell}_{\bar\mu\bar\nu}&=&-\frac{1}{2}\partial_\ell\,\delta h_{\bar\mu\bar\nu}+\text{others},\\
	\delta {\Gamma^{\bar\mu}}_{\ell\bar\nu}&=&\frac{1}{2}h^{\bar\mu\bar\gamma}\partial_\ell\,\delta h_{\bar\gamma\bar\nu}+\text{others},
\end{eqnarray}
will involve $\partial_\ell \delta h_{\bar\mu\bar\nu}$. In Eq.~\eqref{partialelldeltah}, we can further neglect terms that are not relevant,
\begin{eqnarray}
	\notag
	&&16\pi \GN\,\delta_h\big(\Theta(\ell)\mathcal L\varepsilon\big)\\
	\notag
	&&=-\delta(\ell)\Phi^2\big(h^{\bar\rho\bar\sigma}\delta{\Gamma^\ell}_{\bar\rho\bar\sigma}-\delta{\Gamma^{\bar\nu}}_{\ell\bar\nu}\big)\varepsilon\\
	&&\quad+\delta(\ell)\pd_\ell\Phi^2h_{\bar\mu\bar\nu}\delta h^{\bar\mu\bar\nu}\,\varepsilon+\text{others}.\quad
\end{eqnarray}
Finally we obtain
\begin{eqnarray}
	\label{deltahforbulkplusbdy}
	\notag
	&&16\pi\GN\,\delta_h\big(\Theta(\ell)\mathcal L\varepsilon+\bar {\mathcal L}_\Sigma\bar\varepsilon_h\big)\\
	\notag
	&&=-\bigg\{\bar\Phi^2\big(K_{\bar\mu\bar\nu}-K h_{\bar\mu\bar\nu}\big)-2\bar\Phi\partial_\ell\Phi\,h_{\bar\mu\bar\nu}\bigg\}\delta h^{\bar\mu\bar\nu}\,\bar\varepsilon_h\\
	&&\quad+\text{others}.
\end{eqnarray}

Next, when we vary the action with respect to $\Phi$, the boundary Lagrangian yields
\begin{equation}
	16\pi\GN\,\delta_\Phi\big(\bar{\mathcal L}_\Sigma\bar\varepsilon_h\big)=-4\bar\Phi K\delta\Phi\,\bar\varepsilon_h.
\end{equation}
For the variation of the bulk Lagrangian, we only need to consider the terms proportional to $\partial_\ell\,\delta \Phi$,
\begin{equation}
	16\pi\GN\,\delta_\Phi\big(\Theta(\ell)\mathcal L\varepsilon\big)=-2\lambda\delta(\ell)\partial_\ell\Phi\,\varepsilon\delta\Phi+\text{others}.
\end{equation}
Finally we have
\begin{eqnarray}
	\label{deltaPhiforbulkplusbdy}
	\notag
	&&16\pi \GN\,\delta_\Phi\big(\Theta(\ell)\mathcal L\varepsilon+\bar {\mathcal L}_\Sigma\bar\varepsilon_h\big)\\
	&&=-2\bigg\{2\bar\Phi K+\lambda\,\partial_\ell\Phi\bigg\}\bar\varepsilon_h\,\delta\Phi+\text{others}.\quad  
\end{eqnarray}
Restoring the page number $[i]$ and summing over $i$, we once again obtain the junction condition \eqref{dilatonjunction},
\begin{subequations}
	\begin{eqnarray}
		\label{dilatensjunction1}
		&&\sum_{i=0}^m 2\bar\Phi K^{[i]}+\lambda\,\partial_\ell\Phi^{[i]}=-4\pi \GN\,\bar\varrho,\\
		\label{dilatensjunction2}
		\notag
		&&\sum_{i=0}^m \bar\Phi^2\big(K^{[i]}_{\bar\mu\bar\nu}-K^{[i]} h_{\bar\mu\bar\nu}\big)-2\bar\Phi\partial_\ell\Phi^{[i]}\,h_{\bar\mu\bar\nu}=-8\pi \GN\,\bar S_{\bar\mu\bar\nu},\\
	\end{eqnarray}
\end{subequations}
where $\bar\varrho$ and $\bar S_{\bar\mu\bar\nu}$ represent the scalar source strength and energy-momentum tensor on $\Sigma$, respectively.
Comparing $\bar{\mathcal L}_\Sigma$ with the familiar GHY boundary term~\cite{GH1977,Y1972}, we see that the dilaton couples with the extrinsic curvature, which consequently results in the  $\Phi^2\chi$ coupling between the interface tension $\chi$ and the dilaton. 

\subsection{Equivalence of the junction conditions in different frames\label{subsec6C}}
The equations of motion have been shown to be equivalent under the change of frame~\eqref{metricWeyl}. In this section, we check  the equivalence of the junction conditions before and after a transformation, which is less straightforward compared to the invariance of the equations of motion. This is because, as indicated by Eq.~\eqref{L-conformalL}, the conformal Lagrangian $\widetilde{\mathcal L}\,\widetilde\varepsilon$, differs from $\mathcal L\varepsilon$ by a boundary term, $\alpha d\,\nabla_\rho\partial^\rho\Phi^2$, which may have an impact on the junction condition. To address this issue, we first use Stokes theorem to rewrite this boundary term
\begin{equation}
	\int\nabla_\rho\partial^\rho\Phi^2\,\varepsilon=-\oint_\Sigma\partial_\ell\Phi^2\,\bar\varepsilon_h.
\end{equation}
As per Eq.~\eqref{conformalL}, in the conformal Lagrangian $16\pi\GN\,\widetilde {\mathcal L}$, only $\Phi^{2-\alpha(d-1)}\widetilde R$ will results in a GHY-like boundary term at $\Sigma$, which is $-2\bar\Phi^{2-\alpha(d-1)}\widetilde K$. Recalling that the GHY-like boundary term generated by the original Lagrangian, $16\pi\GN\,\mathcal L$, is $-2\bar\Phi^2 K$. Therefore, to show the equivalence of the junction condition before and after a change of frame~\eqref{metricWeyl}, we need to demonstrate 
\begin{equation}
	\label{bdyjunctionequiv}
	\delta\big(2\bar\Phi^{2-\alpha(d-1)}\widetilde K\,\bar\varepsilon_{\widetilde h}-2\bar\Phi^2K\bar\varepsilon_h\big)=\delta\big(\alpha d\, \partial_\ell\Phi^2\,\bar\varepsilon_h\big),
\end{equation} 
where $\bar\varepsilon_{\widetilde h}$ is the volume element of the conformal metric $\widetilde h$.

We start from the continuity condition. Since the~\eqref{metricWeyl} preserves orthogonality, the conformal induce metric is $\widetilde h=\bar \Phi^{2\alpha}h$ and the conformal normal vector is $\up{\widetilde n}i=\bar\Phi^{-\alpha}\up n i$. The continuity condition of normal vectors is clearly preserved
\begin{equation}
	\up{\widetilde n}0+\up{\widetilde n}1+\cdots+\up{\widetilde n}m=\bar\Phi^{-\alpha}\big(\up n 0+\up n1+\cdots+\up nm\big)=0,
\end{equation}
which is attributed to the continuity condition of the dilaton. Then, we can directly deduce
\begin{equation}
	\label{LtildentildePhi}
	\mathfrak L_{\up{\widetilde n}i}\up\psi i=\bar\Phi^{-\alpha}\mathfrak L_{\up ni}\up\psi i=\mathscr F^\prime(\bar\Phi)\bar\Phi^{-\alpha}\mathfrak L_{\up ni}\up\Phi i,
\end{equation}
where $\up\psi i=\mathscr F(\up\Phi i)$. Now we can perform a direct computation of the conformal extrinsic curvature. Note that we have extended $\up n i$ to become a vector field over $\up V i$,
\begin{eqnarray}
	\notag
		&&2\up{\widetilde{\mathcal K}}i(\bar X,\bar Y)\\
		\notag
		&&=\mathfrak L_{\up{\widetilde n}i}\widetilde h(\bar X,\bar Y)\\
		\notag
		&&=\mathfrak L_{\up{\widetilde n}i}\big(\widetilde h(\bar X,\bar Y)\big)-\widetilde h(\mathfrak L_{\up{\widetilde n}i}\bar X,\bar Y)-\widetilde h(\bar X,\mathfrak L_{\up{\widetilde n}i}\bar Y)\\
	\notag
	&&=\bar\Phi^\alpha\mathfrak L_{\up n i}\big(h(\bar X,\bar Y)\big)+2\alpha\bar\Phi^{\alpha-1}\big(\mathfrak L_{\up n i}\up\Phi i\big)h(\bar X,\bar Y)\\
	\notag
	&&\quad-\bar\Phi^\alpha h(\mathfrak L_{\up n i}\bar X,\bar Y)-\bar\Phi^\alpha h(\bar X,\mathfrak L_{\up n i}\bar Y)\\
	&&=2\bar\Phi^\alpha\up K i(\bar X,\bar Y)+2\alpha\bar\Phi^{\alpha-1}\big(\mathfrak L_{\up n i}\up\Phi i\big)h(\bar X,\bar Y),\qquad
\end{eqnarray}
which means
\begin{equation}
	\label{conformalKi}
	\up{\widetilde{\mathcal K}}i=\bar\Phi^\alpha\up K i+\alpha\bar\Phi^{\alpha-1}\big(\mathfrak L_{\up n i}\up\Phi i\big)h.
\end{equation}
Eq.\eqref{conformalT} provides the forms of the conformal energy-momentum tensor ${\widetilde {\mathcal T}}^{[i]}$ and the conformal scalar source strength $\widetilde{\Xi}^{[i]}$. Since the form of $\widetilde h=\bar\Phi^{2\alpha}h$ is the same as that in Eq.~\eqref{metricWeyl}, while $\widetilde h$ is a $d$ dimensinal metric and $\widetilde g^{[i]}$ is $d+1$ dimensional, it is sufficient to perform a substitution $d+1\to d$ in Eq.~\eqref{conformalT}, and we directly get forms of the conformal energy-momentum tensor $\widetilde{\mathcal S}$ and the conformal scalar source strength $\widetilde\varrho$ within $\Sigma$,
\begin{equation}
	\label{conformalSandrho}
	\widetilde {\mathcal S} =\bar\Phi^{-\alpha(d-2)}\bar {\mathcal S},
	\quad \widetilde{\varrho}=\bar\Phi^{-\alpha d}\big(\bar\varrho+2\alpha\bar\Phi^{-1}\mathrm{tr}_h\mathcal{\bar S}\big).
\end{equation}
The volume element $\bar \varepsilon_{\widetilde h}$ is obtained through the wedge product of $d$ orthonormal $1$-forms. Thus, performing a substitution $d+1\to d$ in Eq.~\eqref{conformalvolume} provides that
\begin{equation}
	\bar \varepsilon_{\widetilde h}=\bar\Phi^{\alpha d}\bar\varepsilon_h.
\end{equation}
Utilizing $\bar\varepsilon_{\widetilde h}=\up\delta i(\widetilde\ell)\up{\widetilde\varepsilon}i$ and Eq.~\eqref{conformalvolume}, or by direct dimensional analysis, we obtain the behavior of the Dirac function under~\eqref{metricWeyl},
\begin{equation}
	\up\delta i(\widetilde\ell)=\Phi^{-\alpha}\up\delta i(\ell).
\end{equation}

Now that we have assembled all the required information, the behavior of the junction condition under the Weyl transformation can be computed explicitly. Taking the trace of both sides of Eq.~\eqref{conformalKi}, a straightforward calculation reveals
\begin{eqnarray}
	\notag
	&&2\bar \Phi^{2-\alpha(d-1)}\big(\mathrm{tr}_{\widetilde h}\widetilde {\mathcal K}^{[i]}\big)\,\bar\varepsilon_{\widetilde h}-2\bar\Phi^2\big(\mathrm{tr}_h{\mathcal K}^{[i]}\big)\bar\varepsilon_h\\
	&&\qquad =\alpha d\, \mathfrak L_{\up n i}(\up\Phi i)^2\,\bar\varepsilon_h,
\end{eqnarray}
which is a stronger conclusion than Eq.~\eqref{bdyjunctionequiv}, so the validity of the latter follows. We have thus proven the equivalence of the junction conditions under a Weyl transformation. More directly, we can confirm this by analyzing the field equation \eqref{conformalvartophi} and \eqref{conformalvartog}. It is evident that only the curvature terms, the matter terms $\widetilde {\mathcal T}^{[i]}$ and $\widetilde \Xi^{[i]}$, and terms such as $\widetilde\Box^{[i]}\up\Phi i$ and $\big(\up{\widetilde\nabla}i_\mu\partial_\nu-\up{\widetilde g}i_{\mu\nu}\up{\widetilde \Box}i\big)(\up\Phi i)^{2-\alpha(d-1)}$ will exhibit singularities proportional to $\delta(\widetilde\ell)$. Utilizing Eqs.~\eqref{nablan0dpsiSigma}\eqref{nablaXdpsiSigma}\eqref{BoxpsiSigma} while setting $\mathscr F(\bar\Phi)=\bar\Phi$ and $\mathscr F(\bar\Phi)=\bar\Phi^{2-\alpha(d-1)}$ respectively, we can then derive the junction conditions after the change of frame~\eqref{metricWeyl}
\begin{subequations}
	\begin{eqnarray}
		\notag
		 &&\bar\Phi^{-\alpha(d-1)}\sum_{i=0}^m\bigg\{\big(2-\alpha(d-1)\big)\bar\Phi\,\mathrm{tr}_{\widetilde h}{\widetilde {\mathcal K}}^{[i]}+\big(\lambda\\
		 &&\quad +\alpha^2d(d-1)-4\alpha d\big)\mathfrak L_{\widetilde n^{[i]}}\up\Phi i\bigg\}=-4\pi \GN\,\widetilde \varrho,\qquad \quad \\
		 \notag
		 &&\sum_{i=0}^m\bar\Phi^{2-\alpha(d-1)}\bigg(\widetilde{\mathcal K}^{[i]}-\mathrm{tr}_{\widetilde h}{\widetilde {\mathcal K}}^{[i]}\,\widetilde h\bigg)\\
		 &&\quad  -\mathfrak L_{\up{\widetilde n} i}(\up\Phi i)^{2-\alpha(d-1)}\,\widetilde h=-8\pi \GN\,\widetilde{\mathcal S}.
	\end{eqnarray}
\end{subequations}
Substituting Eqs.~\eqref{conformalKi} and \eqref{conformalSandrho} into the condition above, we  reproduce the original junction condition \eqref{dilatonjunction}. Specifically, for $d=1$, we observe that the junction condition after the change of frame~\eqref{metricWeyl} retains the exact same form as the original one.

\subsection{Tension along the interface\label{sec7}}
We introduce a tension term on the interface $\Sigma$, denoted as $\chi$. To isolate the effect of the matter fields on the interface, we do not include matter fields in the bulks. There are two potential forms for the tension term
\begin{subequations}
	\begin{eqnarray}
		\label{tensionaction}
		I_{\text{decoup}}&=&-\frac{1}{8\pi\GN}\int_\Sigma \chi\bar\epsilon_h,\\
	\label{dilatontensionaction}
		I_{\text{coup}}&=&-\frac{1}{8\pi\GN}\int_\Sigma \bar \Phi^2\chi\bar\epsilon_h. 
	\end{eqnarray} 
\end{subequations}
In the first scenario, tension and the dilaton remain decoupled. This suggests that tension, regarded as a vacuum energy density within $\Sigma$, encompasses contributions not only from the extrinsic curvature but also from the dilaton. Conversely, in the second scenario, tension becomes coupled with the dilaton, indicating that the origin of tension lies exclusively in the ``stretching'' of $\Sigma$ by each of the page $\up V i$, which is to say, the contribution from the extrinsic curvature of $\Sigma$ embedded in $\up V i$. In fact, it is easy to see that $I_{\text{decoup}}$ can be transformed into $I_{\text{coup}}$ through the change of frame~\eqref{metricWeyl} with $\alpha=-2/d$, and vice versa. We will separately solve the junction conditions for these two distinct scenarios.

\subsubsection{The case of decoupled dilaton}
Taking the variation of Lagrangian in \eqref{tensionaction} will yield
\begin{equation}
	16\pi \GN\, \delta\big(\bar{\mathcal L}_{\text{decoup}}\bar\varepsilon_h\big)=-2\delta\big(\chi\bar\varepsilon_h\big)=\chi h_{\bar\mu\bar\nu}\bar\varepsilon_h\delta h^{\bar\mu\bar\nu}.
\end{equation}
Therefore, the energy-momentum tensor $\bar S_{\bar\mu\bar\nu}$ and the scalar source strength $\bar\varrho$ generated by the tension are given by
\begin{equation}
	\bar S_{\bar\mu\bar\nu}=-\frac{1}{8\pi\GN}\chi h_{\bar\mu\bar\nu},\qquad \bar\varrho=0.
\end{equation}
Substituting the above results into Eqs.~\eqref{dilatensjunction1} and \eqref{dilatensjunction2}, the junction condition becomes
\begin{eqnarray}
	\label{tensionjunction}
	\notag
	&&\bar\Phi^2\sum_{i=0}^m K^{[i]}_{\bar\mu\bar\nu}=-\frac{\lambda}{\lambda(d-1)-4d}\,\chi h_{\bar\mu\bar\nu},\\
	&&\sum_{i=0}^m\pd_\ell(\up\Phi i)^2
	=\frac{4d}{\lambda(d-1)-4d}\,\chi.
\end{eqnarray}
Taking the trace on both sides of the first equation,
\begin{equation}
	\sum_{i=0}^m \bar\Phi^2K^{[i]}=-\frac{\lambda d}{\lambda(d-1)-4d}\chi,
\end{equation}
we observe that the tension is proportional to $\bar\Phi^2K^{[i]}$. In particular, for the JT gravity where $d=1$ and $\lambda=0$, the dilaton is non-zero (and even approach infinity) on the boundary. Consequently, 
the junction conditions imply that the sum of all extrinsic curvatures equals zero,
\begin{equation}
	\sum_{i=0}^m K^{[i]}_{\bar\mu\bar\nu}=0, \quad\sum_{i=0}^m K^{[i]}=0, \quad \sum_{i=0}^m\pd_\ell(\up\Phi i)^2=-\chi.
\end{equation}

\subsubsection{The case of coupled dilaton}
Varying the Lagrangian~\eqref{dilatontensionaction},
\begin{eqnarray}
	\notag
	16\pi \GN \delta\big(\mathcal L_{\text{coup}}\bar\varepsilon_h\big)&=&-2\delta\big(\bar\Phi^2\chi\bar\varepsilon_h\big)\\
	&=& \bar\Phi^2\chi h_{\bar\mu\bar\nu}\delta h^{\bar\mu\bar\nu}\, \bar\varepsilon_h-4\bar\Phi\chi\delta\bar\Phi\,\bar\varepsilon_h,\qquad\;
\end{eqnarray}
leads to the energy-momentum tensor and the scalar source strength 
\begin{equation}
		\bar S_{\bar\mu\bar\nu}=-\frac{1}{8\pi\GN}\bar\Phi^2\chi h_{\bar\mu\bar\nu},\quad \bar\varrho=\frac{1}{2\pi\GN}\bar\Phi\chi.
\end{equation}
The junction conditions \eqref{dilatensjunction1} and \eqref{dilatensjunction2} yields
\begin{equation}
	\sum_{i=0}^m\pd_\ell(\up\Phi i)^2=\frac{4}{\lambda(d-1)-4d}\bar\Phi^2 \chi,
\end{equation}
and
\begin{eqnarray}
	\label{sumKipropchih}
	\notag
	&&\sum_{i=0}^mK^{[i]}_{\bar\mu\bar\nu}=-\frac{\lambda-4}{\lambda(d-1)-4d}\chi h_{\bar\mu\bar\nu},\\
	&&\sum_{i=0}^mK^{[i]}=-\frac{(\lambda-4)d}{\lambda(d-1)-4d}\chi.
\end{eqnarray}
We observe that the tension $\chi$ can be completely determined by the extrinsic curvature of $\Sigma$ in all pages, independent of the dilaton $\bar\Phi^2$ on $\Sigma$. The tension on $\Sigma$ then gives a constraint on the variation of the dilaton along the normal vector direction, namely, $\partial_\ell(\up\Phi i)^2$.

Specifically, setting $d=1$ and $\lambda=0$ leads to a solution that is applicable to JT gravity. In this scenario, $\Sigma$ is one-dimensional, and we only need to select one internal coordinate, denoted as $u$, which induces a metric component $h_{uu}$. Then we have $K^{[i]}_{uu}=K^{[i]}h_{uu}$.   The $u$ coordinate can always be reparameterized so that $h_{uu}$ is constant. The junction conditions then reduce to
\begin{subequations}
	\label{JTjunction}
	\begin{eqnarray}
		\label{JTjunctionK}
		&&\sum_{i=0}^mK^{[i]}=-\chi,\\
		\label{JTjunctionPhi}
		&&\sum_{i=0}^m\partial_\ell(\up\Phi i)^2=-\bar\Phi^2\chi.
	\end{eqnarray}
\end{subequations}
We refer to Eq.~\eqref{JTjunctionK} as the junction condition for the extrinsic curvature and Eq.~\eqref{JTjunctionPhi} as the junction condition for the dilaton. In the companion papers~\cite{SPL,SPL2}, we show how these two constraints determine the solution of JT gravity on the pages that can be glued together. 

\subsection{A simple example of gluing AdS$_{d+1}$ spacetime}
Now we consider the case where each $\up V i$ to be glued together is an $\AdS_{d+1}$ spacetime with AdS radii  $\up Li$ respectively, whose Ricci curvature is
\begin{equation}
	R^{[i]}=-\frac{d(d+1)}{ (\up L i)^2},\quad 
	R^{[i]}_{\mu\nu}=-\frac{d}{ (\up L i)^2}g^{[i]}_{\mu\nu}.
\end{equation}
Contracting the Gauss equation~\eqref{GaussEq}, we get
\begin{equation}
	\label{GaussforAdS}
	R^{[i]}-2R^{[i]}_{\ell\ell}=\bar R+K^{[i]}_{\bar\mu\bar\nu}(K^{[i]})^{\bar\mu\bar\nu}-(K^{[i]})^2,
\end{equation}
for all $i\in\big\{0,1,\cdots,m\big\}$, which results in $m+1$ independent equations,
\begin{equation}
	\bar R+\frac{d(d-1)}{(\up L i)^2}=(K^{[i]})^2-K^{[i]}_{\bar\mu\bar\nu}(K^{[i]})^{\bar\mu\bar\nu}.
\end{equation}
Upon eliminating $\bar R$, we get $m$ independent constraints. On the other hand, we can redefine $\frac{(\lambda-4)d}{\lambda(d-1)-4d}\chi\to\chi$ in Eq.~\eqref{sumKipropchih}, which is equivalent to setting $\lambda=0$. This results in
\begin{equation}
	\label{lamdbais0sumKi}
	\sum_{i=0}^m K^{[i]}_{\bar\mu\bar\nu}=-\frac{1}{d}\chi h_{\bar\mu\bar\nu},\qquad \sum_{i=0}^m K^{[i]}=-\chi,
\end{equation}
which provides an additional $d(d+1)/2$ independent constraints. However, the $m+1$ unknown extrinsic curvatures $K^{[i]}_{\bar\mu\bar\nu}$ together have a total of $(m+1)d(d+1)/2$ components. Therefore, there are $m\big(d(d+1)/2-1\big)$ degrees of freedom that cannot be determined. We can decompose the extrinsic curvature $K^{[i]}_{\bar\mu\bar\nu}$ into two parts,
\begin{equation}
	\label{sumKi+chi=0}
	K^{[i]}_{\bar\mu\bar\nu}=\frac{1}{d}K^{[i]} h_{\bar\mu\bar\nu}+\up\xi i_{\bar\mu\bar\nu},\quad h^{\bar\mu\bar\nu}\up\xi i_{\bar\mu\bar\nu}=0.
\end{equation}
Eq.~\eqref{lamdbais0sumKi} immediately imposes constraints on $\up\xi i_{\bar\mu\bar\nu}$, then we have
\begin{equation}
	\label{sumxitensor0}
	\sum_{j=0}^m \up\xi j_{\bar\mu\bar\nu}=0,
\end{equation}
and Eq.~\eqref{GaussforAdS} can be reformulated as
\begin{equation}
	\label{K2-Xi2}
	\bar R+\frac{d(d-1)}{(\up L i)^2}=\frac{d-1}{d}(K^{[i]})^2-\xi^{[i]}_{\bar\mu\bar\nu}(\xi^{[i]})^{\bar\mu\bar\nu}.
\end{equation}
Additional constraints can also be obtained by employing the Codazzi equation
\begin{equation}
	-R^{[i]}_{\ell\bar\sigma\bar\mu\bar\nu}=\bar\nabla_{\bar\mu} K^{[i]}_{\bar\nu\bar\sigma}-\bar\nabla_{\bar\nu}K^{[i]}_{\bar\mu\bar\sigma},
\end{equation}
$\forall i\in\big\{0,1,\cdots,m\big\}$. For AdS spacetime, we have $R^{[i]}_{\ell\bar\sigma\bar\mu\bar\nu}=0$, which implies
\begin{equation}
	\label{nablaKxi}
	\bar\nabla^{\bar\nu}\up\xi i_{\bar\nu\bar\mu}=\bigg(1-\frac{1}{d}\bigg)\partial_{\bar\mu} K^{[i]}.
\end{equation}
However, these constraints involve covariant derivatives, which means they are not algebraic equations that should depend solely on the local value of $\up\xi i_{\bar\mu\bar\nu}$ at a single point.

Therefore the constraints are not enough to determine $K^{[i]}_{\bar\mu\bar\nu}$ unless in specific scenarios. For example, when $d=1$ it is clear that $\up\xi i_{\bar\mu\bar\nu}=0$; or when $m=1$, $\up\xi 0_{\bar\mu\bar\nu}+\up\xi 1_{\bar\mu\bar\nu}=0$ is given. In the latter case, Eq.~\eqref{K2-Xi2} shows that $(K^{[1]})^2-(K^{[0]})^2$ is independent of $\up\xi 0_{\bar\mu\bar\nu}$ and $\up\xi 1_{\bar\mu\bar\nu}$, which agree with the computation in reference~\cite{kawamoto2023}. For others instances where $d>1$ and $m>1$, the form of $\up\xi i_{\bar\mu\bar\nu}$ depends on the embedding of $\Sigma$ in each $\up V i$. For large values of $d$ and $m$, comprehending the bulk gluing conditions is a non-trivial endeavor. In the companion papers~\cite{SPL,SPL2} of this series, we will outline the gluing conditions for various scenarios within the specific context of $d=1$ and arbitrary $m$. This will show that, even in the lowest dimension, the problem is highly non-trivial.

Nevertheless, we will explore relatively simple scenarios in the following. For example, in the case with $d>1$ when all $\up \xi i_{\bar\mu\bar\nu}=0$, all $K^{[i]}$ and $\bar R$ become constants due to Eq.~\eqref{nablaKxi} and~\eqref{K2-Xi2}. Then Eq.~\eqref{K2-Xi2} provides us with $m$ independent algebraic equations
\begin{equation}
	\label{Ki-K0=Li-L0}
	(K^{[i]})^2-(K^{[0]})^2=\frac{d^2}{(\up L i)^2}-\frac{d^2}{(\up L 0)^2},
\end{equation}
for all $i\in\big\{1,\cdots,m\big\}$. By combining Eq.~\eqref{lamdbais0sumKi}, we can determine all $K^{[i]}$, expressing them in terms of $\chi$ and all $\up Li$. It is important to highlight that $\bar R$ can also be represented as a function of the tension and all AdS radii, $\bar R\big(\chi,\up L0,\cdots,\up L m\big)$. Hence,
\begin{equation}
	K^{[i]}= \pm\sqrt{\frac{d}{d-1}\bar R+\frac{d^2}{(\up L i)^2}}.
\end{equation}
The fact that $\bar R$ is constant implies that $\Sigma$ is a space of constant curvature.  

Cases where $\up \xi i_{\bar\mu\bar\nu}=0$ for all pages do exist. One example is simply the AdS spacetime. In the Poincar\'e patch of each page, the metric $\up g i$ can be put into the form
\begin{equation}
	\up g i=\frac{(\up L i)^2}{z^2}\big(-\ed t^2+\cdots+\ed z^2\big),\quad z>0.
\end{equation}
Embedding the hypersurface $\Sigma$ into $\up V i$ as $\up z i=\up C i>0$, $\Sigma$ then divides the patch of $\AdS_{d+1}$ spacetime into two regions: $z>\up C i$ and $0<z<\up C i$. The normal vector can be represented as $\pm\frac{\up C i}{\up L i}\partial_z$, pointing in the directions of decreasing and increasing conformal factor $\frac{1}{z^2}$, respectively. Now, the extrinsic curvature is given by
\begin{equation}
	\label{AdSd+1Ki}
	K^{[i]}_{\bar\mu\bar\nu}=\frac{1}{2}\mathfrak L_{\up n i}h=\mp \frac{1}{\up L i}h_{\bar\mu\bar\nu},\quad K^{[i]}=\mp \frac{d}{\up L i}.
\end{equation}
It is clear that $\xi^{[i]}_{\bar\mu\bar\nu}=0$. Moreover, if we select each page $\up V i$ to be the part $z\geqslant\up C i$, then $K^{[i]}<0$. The other choice results in $K^{[i]}$ with opposite signs. By making the coordinate functions, other than $z$, independent of the page number $[i]$, the degrees of freedom associated with coordinate transformations can be eliminated. In this case, the continuity of $h$ across the interface $\Sigma$ requires $\up C i/\up L i$ to be a constant independent of $[i]$. At this point, $h$ is a flat metric, we have $\bar R=0$. In this case, the junction condition yields
\begin{equation}
	\label{chiequaldsum1/Li}
	\chi=d\bigg(\frac{1}{\up L0}+\frac{1}{\up L1}+\cdots+\frac{1}{\up Lm}\bigg).
\end{equation}

\section{Discussion}\label{cmts}

\subsection{Spacetime ``Feynman diagrams'' and web geometries}

\begin{figure}[hbt]
  \includegraphics[width=\columnwidth]{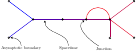}
  \caption{An illustration of the web geometry obtained from gluing nine bulks along four interfaces. It is evident that this resembles a Feynman diagram where each spacetime is analogous to a propagator, a junction between two bulks is analogous to a mass insertion, an $n$-way junction is analogous to an $n$-vertex, and an asymptotic boundary is analogous to an external particle.}
  \label{webs}
\end{figure}

The junction conditions between multiple bulks make it possible to glue up a web of spacetimes in a way similar with how we build up a Feynman diagram.
Indeed, each spacetime to be glued can be considered as the analog of a propagator in a Feynman diagram,  junctions between two bulks can be considered as the analog of a ``mass" term, or a $2$-valent vertex, and the multiway junction an analog of higher-valent vertices in the Feynman diagrams. Then each glued geometry looks like a Feynman diagram, as illustrated in Fig.~\ref{webs}, and the gravitational path integral consisting of a sum over different geometries for a given set of asymptotic boundaries and integrating over the physical degrees of freedom, which is analogous to summing over Feynman diagrams and integrating over the internal momenta for a given set of external legs.  

The junction condition can be considered as the momenta conservation condition. This indeed immediately indicates that the path integral over the curvature is similar with the integration over momenta in Feynman rules as we expected, and the curvatures are the analog of momenta in Feynman diagrams. Then the junction condition~\eqref{d+mdimjunction}, which can be written into a $delta$ function, is analogous to the momenta conservation $delta$ function in Feynman rules. In particular, the stress-energy tensor $\bar{\mathcal S}$ on the interface is analogous to the momenta carried by  (and inserted into the process) the vertices in the Feynman diagram; the case with vanishing $\bar{\mathcal S}$ correspond to normal vertices that does not carry momentum while non-vanishing $\bar{\mathcal S}$ corresponds to momentum carrying vertices. In the case of dilaton-gravity, the extra condition from the dilaton~\eqref{dilatonjunction} can be considered as conservation of momentum in an extra dimension as in the familiar dimensional reduction of field theories and the Feynman diagrams therein. To make this analogy more explicit, we notice that utilizing the Hamiltonian formalism of general relativity allows us to establish the relationship between the extrinsic curvature and canonical momenta \cite{Jha2023,Steven2019} in $\up V i$ corresponding to the metric $h_{\bar\mu\bar\nu}$ of $\Sigma$, 
\begin{equation}
    \up\pi i_{\bar\mu\bar\nu}=\sqrt{|\det h|}\bigl(K^{[i]}_{\bar\mu\bar\nu}-K^{[i]}h_{\bar\mu\bar\nu}\bigr).
	\label{cp1}
\end{equation}
Therefore, the junction condition can be formulated as $\sum_i\up\pi i_{\bar\mu\bar\nu}=-8\pi G\mathsf s\sqrt{|\det h|}\bar S_{\bar\mu\bar\nu}$, which illuminates the analogy between curvature and momenta.\footnote{However, it is crucial to acknowledge that, our discussion in the main text is more general; $\up n i$ may not necessarily be timelike in our discussion, although Eq.~\eqref{cp1} assumes it being timelike so that the Hamiltonian formalism is applicable.}

Let us emphasize that this is only an analog and the real gravitational path integral is much more complicated than Feynman integrals. For example, in the Feynman diagram the momentum entering the vertex is the same as that carried by the propagator; on the other hand, the analogous conservation in the glued geometry is among the extrinsic curvatures ``going into" the junction, which is different from the curvatures in the bulk. However, there is not a contradiction since we can solve the field equation of motion from the boundary into the bulk with the specified boundary condition. This is just one example demonstrating that the gravitational analysis is more complicated than Feynman diagrams. It is therefore interesting to understand in details about possible deeper connections between them, probably via the worldline formalism of quantum field theory~\cite{Bern:1990cu,Bern:1991aq,Strassler:1992zr}. We will defer this analysis to later publications. 

\subsection{Application to JT gravity}
A direct application of our junction conditions is to the Jackiw-Teitelboim (JT) gravity~\cite{JACKIW1985343,TEITELBOIM198341,Maldacena:2016upp, Harlow:2018tqv,Yang:2018gdb,Maldacena:2019cbz,kolchmeyer2023,PhysRevD.99.046016,PhysRevD.101.106023,blommaert2019,grumiller2021,mertens2023} that greatly deepens our understanding of the blackhole information paradox via its connection to the Sachdev-Ye-Kitaev (SYK) model~\cite{SY1993,Maldacena:2016hyu,Maldacena:2016upp,Kitaev:2017awl}. In the companion papers~\cite{SPL,SPL2}, we will apply the junction conditions to JT gravity, classify the dilaton solution into three types, and get allowed dilaton profiles in the glued booklet geometry. We will show that indeed multi-boundary geometries can be constructed in JT gravity subject to the junction condition we derived, and the junction condition imposes some rather lenient requirements on the geometric parameters of the pages. Furthermore the condition imposes some selection rules on the allowed dilaton configurations in different pages. Moreover, our construction applies to higher dimensional gravitational theories as well, it is therefore relevant in the study of high-dimensional disordered models that extend the JT/SYK story to high dimensions~\cite{Gu:2016oyy,Berkooz:2016cvq,Peng:2018zap,Chang:2021fmd,Chang:2021wbx,Peng4d}. 

The generalization of the junction condition to glue an arbitrary number of pages generates numerous possibilities, significantly expanding the degrees of freedom in constructing physical models, which allows for direct manipulation of fundamental properties such as the topology and geometry of spacetime to meet various desired characteristics. In addition, it is interesting to consider junction conditions in the context of higher-spin gravity as a future direction. 
 
\subsection{Summary of the paper}

Let us summarize what we have achieved in this paper. We first rederive the Israel junction condition for a booklet geometry $\mathscr V$ formed by gluing two pages along an interface $\Sigma$. This is done slight different from classical treatment; we provide an explicit definition of the curvature tensor at the interface $\Sigma$ and the derivation is done  in the geometric framework. For the 3-way junction, we embed the three-pages booklet into a larger $(d+2)$-dimensional spacetime $\mathscr M$  and select an arbitrary page, such as $\up V 0$. Then we extend $\up V 0$ across the $\Sigma$ along the direction $-\up n 0$ to get $\up V{-0}$. The junction condition for $\up V 0\cup\up V{-0}$ establishes the criteria for gluing the three pages. Importantly, we prove the independence of this result on the choice of $\up V 0$. This is the method of reverse extension.  Next, employing the same method of reverse extension we explore the junction condition for gluing an arbitrary number of pages. Our discussion, characterized by considerable generality, elucidates the close relation between the junction condition and the continuity condition of the normal vector. In Appendix~\ref{sec3}, we delve into the significance of the continuity condition \eqref{dim2continu} of the normal vectors, highlighting its pivotal role in the junction condition when dealing with two pages, and naturally leads to its necessity of joining three pages. 

Next, we explore dilaton gravity. Employing the method of reverse extension, we establish definitions for certain geometric quantities associated with the dilaton at $\Sigma$. Utilizing these definitions, we deduce the junction condition for dilaton gravity. In addition, we show that the same result can be obtained from variation of the action of the theory under consideration. We then show the equivalence between the junction condition before and after the change of frame~\eqref{metricWeyl}. We then solve the junction condition when tension is introduced on the  interface in both scenarios decoupled and coupled to the dilaton for some simple example. 

Let us finally comment on the relation with the gravitational path integral. In the recent discussion of multi-boundary geometries, in particular the replica wormhole geometry appearing in the computation of the entanglement entropy of a field theory coupled with gravity, the gravity region has Euclidean signature. 
On the other hand, ultimately we would like to have better understandings of gravity in Lorentzian signature. Therefore one interesting question is to construct similar multi-boundary geometries in Lorentzian signature, and gluing by appropriate junction condition discussed in this paper is one way to construct such geometries. In principle, the gravitational path integral should include all possible topologies and geometrical fluctuations, so there is no reason to exclude multiway junction geometries from contributing to the gravitational path integral. This is also what we learned from the recent development of the gravitational path integral where connected geometries, such as double-cone geometries, as well as other off-shell configurations are required in the gravitational path integral to reproduce the holographic dual field theory results (see e.g.~\cite{Saad:2019lba}). 
Therefore we think the multi-boundary geometry constructed in this way is relevant and should be included in the gravitational path integral.

\begin{acknowledgments}
	CP thanks Song He, Jieqiang Wu and Zhenbin Yang for helpful discussions on related topics. CP is supported by NSFC NO.~12175237, and NO.~12247103 the Fundamental Research Funds for the Central Universities, and funds from the Chinese Academy of Sciences. JS and LL are supported by the NSFC NO.~11973014.
\end{acknowledgments}

\appendix

\section{Necessity of the continuity condition for two bulks\label{sec3}}
The continuity condition \eqref{dim2continu} has been noted to imply the  {consistency} of the ``differential structure,'' specifically, $\mathscr T_\Sigma(\up V 0)=\mathscr T_\Sigma(\up V 1)$, as well as the continuity of the metric $\up g0\big|_\Sigma=\up g 1\big|_\Sigma$. In other references~\cite{PhysRevD.106.064007,Brassel2023,Deruelle2008junction,Aviles2020,PhysRevD.67.024030,PhysRevD.103.104069,Poisson2004}, a more prevalent convention concerning the orientation of the normal vector is to identify $\up n 0$ and $\up n 1$, namely $\up n 0=\up n1$. This convention seems reasonable in the context of gluing two bulks together. However, when we attempt to glue 3 or more bulks along $\Sigma$, this convention becomes challenging to generalize. For example, if we were to assume $\up n 0=\up n 1=\up n 2$, the presence of the additional bulk $\up V 2$ would unavoidably overlap with either $\up V 1$ or $\up V 0$ within an infinitesimal vicinity of $\Sigma$. This is fundamentally indistinguishable from the situation involving only the gluing of $\up V 0$ and $\up V 1$, as the junction condition under consideration exclusively depends on the infinitesimal vicinity of $\Sigma$. Therefore, when gluing together three or more bulks, it is inadequate to possess solely a $d+1$-dimensional differential structure at $\Sigma$. 

To collect pertinent information for a better understanding of the structure at $\Sigma$, we examine the consequences when Eq.~\eqref{dim2continu} fails to hold for gluing 2 bulks. First and foremost, it is evident that in this scenario, $\up g0\big|_\Sigma\not=\up g 1\big|_\Sigma$, signifying a jump in the metric at $\Sigma$.  {Nevertheless}, uncertainty remains regarding the existence of a well-defined curvature tensor $\up R\Sigma$ because the curvature $\up R\Sigma$ is not directly contingent on the metric but rather on the connection. However, even with a well-defined curvature, a question persists: how do we define the Ricci tensor, Ricci scalar, and Einstein tensor when there is a jump in the metric? One possible solution is to fix a standard metric, such as $\up g 0$, then calculate the Einstein tensor and derive the junction condition. Subsequently, one would need to demonstrate that the junction condition is independent of the choice of the standard metric.

Now we assume that $\up n0+\up n1$ is not necessarily equal to $0$ everywhere, then two possibilities arise. The first scenario is that $\up n 0$ and $\up n 1$ continue to be linearly dependent [$\mathrm{mod}\,\mathscr T(\Sigma)$], implying the existence of a function $\beta$ and a vector field $\bar w$ on $\Sigma$ such that 
\begin{equation}
	\label{n0betan1barw}
	\up n 0 + \beta\up n1 = \bar w.
\end{equation}
In this instance, $\mathscr T_\Sigma(\up V 0) = \mathscr T_\Sigma(\up V 1)$ remains valid. The second scenario occurs when $\up n 0$ and $\up n 1$ are linearly independent at some points of $\Sigma$. In such cases, $\mathscr T_\Sigma(\up V 0)$ and $\mathscr T_\Sigma(\up V 1)$ are not identical, resulting in the absence of a differential structure for $\mathscr V$ at $\Sigma$.

Let us commence by examining the first scenario. It can be established that either $\beta > 0$ holds at every point on $\Sigma$, or $\beta < 0$ applies to all points. To substantiate this, we assume that $\beta = 0$ for certain points on $\Sigma$. This directly leads to $\up n 0 = \bar w\in \mathscr T(\Sigma)$ for these points, which contradicts the initial assumption that $\up V 0$ is a $(d+1)$-dimensional manifold with boundaries. Consequently, $\beta\not= 0$ everywhere. Furthermore, due to the continuity of the function $\beta$, it is impossible for there to be a portion where $\beta > 0$ and another portion where $\beta < 0$ simultaneously. This establishes the assertion that $\beta$ maintains the same sign for all points on $\Sigma$. The continuity of the function $\beta$ implies the continuity of the distribution of normal vectors. It is challenging to envision a sudden change in the relative direction between $\up n 0$ and $\up n 1$. Furthermore, we can directly stipulate $\beta>0$ since the cases of $\beta>0$ and $\beta<0$ are equivalent. If $\beta<0$ everywhere, it indicates an ``acute angle'' between $\up n 0$ and $\up n 1$ (regardless of which metric is used for measurement). This implies that $\up V 0$ and $\up V 1$ overlap in the infinitesimal neighborhood of $\Sigma$, as the direction of the normal vectors is from the boundary to the interior of the bulk. Therefore, we can smoothly extend $\up V 1$ along the opposite direction of $\up n 1$, redefine the extended part as $\up V 1$, and thus redefine $-\up n1$ as $\up n 1$. Obviously, the geometry of $\up V 1$ is completely equivalent to the original. At this point, $-\beta$ is redefined as $\beta$, resulting in $\beta > 0$. Now we can proceed to define the curvature tensor at $\Sigma$ once again imitating Eq.~\eqref{addpathSigma},  
\begin{widetext}
\begin{eqnarray}
	\label{n0betan1RXn0U}
	\notag
	\up R\Sigma(\bar X,\up n 0)\bar U&=&\lim_{s,\iota,t\to 0^+}\frac{\up\tau0_{-t\vec  n^{[0]}}\up\tau0_{s\bar X}\up\tau0_{t\up n 0}\bar U-\up\tau1_{\iota\vec  n^{[0]}}\up\tau1_{s\bar X}\up\tau1_{-\iota\up n 0}\bar U}{(t+\iota)s} \\
	\notag
	&=&\up\delta 0(0)\left(\up \II 1(\bar X,\bar U)-\up \II 0(\bar X,\bar U)\right) +\up\Theta 0(0)\up R0(\bar X,\up n0)\bar U+\big(1-\up\Theta 0(0)\big)\up R1(\bar X,\up n0)\bar U\\
	\notag
	&=&\up\delta 0(0)\cdot\mathsf s\left(\up K 0(\bar X,\bar U)+\frac{1}{\beta}\up K 1(\bar X,\bar U)\right)\up n 0 -\up\delta 0(0)\cdot\frac{ \mathsf s}{\beta}\up K 1(\bar X,\bar U)\bar w \\
	&&+\bigg\{\up\Theta 0(0)\up R0+\big(1-\up\Theta0(0)\big)\up R1\bigg\}(\bar X,\up n0)\bar U.
\end{eqnarray}
\end{widetext}
Because of the non-coincidence of $-\up n 1$ and $\up n 0$, we have introduced the index $[i]$ for the distributions. $\up\Theta 0(0)$ is an alternative of $\lim_{t,\iota\to 0^+}\frac{t}{t+\iota}$, and $\up\delta 0(0)$ is an alternative of $\lim_{t,\iota\to 0^+}\frac{1}{t+\iota}$. Both $t$ and $-\iota$ are parameters of the integral curve of $\up n 0$. Then $\up\Theta 1(0)$ and $\up\delta 1(0)$ can be defined similarly. Defining the dual $1$-forms $n^{*[i]}$ of normal vectors, we have $n^{*[i]}(\bar X)=0$ and $n^{*[i]}(\up n i)=1$. Therefore, $\up n 0+\beta\up n1=\bar w$ implies
\begin{equation}
	n^{*[1]}=-\beta n^{*[0]}.
\end{equation}
Then, we introduce the functions $\up\ell 0$ on $\up V0$ and $\up\ell 1$ on $\up V1$, ensuring that both $\up\ell 0=0$ and $\up\ell 1=0$ on $\Sigma$, while satisfying
\begin{equation}
	n^{*[1]}=\mathrm d\up\ell 1\big|_\Sigma, \quad n^{*[0]}=\mathrm d\up\ell 0\big|_\Sigma.
\end{equation}
Consequently, $\up\ell 0$ and $-\up\ell 1/\beta$ serve as parameters for the integral curve of $\up n 0$, while $\up\ell 1$ and $-\beta\up\ell 0$ act as parameters for the integral curve of $\up n 1$. Clearly,
\begin{eqnarray}
	\notag
	\up\Theta 0(0)&=&\lim_{\up\ell 0,\up\ell 1\to 0^+}\frac{\up\ell 0}{\up\ell 0+\up\ell 1/\beta},\\
	 \up\Theta 1(0)&=&\lim_{\up\ell 0,\up\ell 1\to 0^+}\frac{\up\ell 1}{\up\ell 1+\beta\up\ell 0},
\end{eqnarray}
which immediately leads to
\begin{equation}
	\label{Theta0+Theta1}
	\up\Theta 0(0)+ \up\Theta 1(0)=1.
\end{equation}
More importantly,
\begin{eqnarray}
	\notag
	\up\delta 0(0)&=&\lim_{\up\ell 0,\up\ell 1\to 0^+}\frac{1}{\up\ell 0+\up\ell 1/\beta}\\
	&=&\lim_{\up\ell 0,\up\ell 1\to 0^+}\frac{\beta}{\up\ell 1+\beta\up\ell 0}=\beta\up\delta 1(0).
\end{eqnarray}
Thus we can observe that
\begin{equation}
	\label{delta0=b.delta1}
	\up\delta 0(0)=\beta\up\delta 1(0).
\end{equation}
The above equation indicates that we do not need to concern ourselves with the jump in the metric causing the Einstein tensor $\up G \Sigma$ to be ill-defined. The discrepancy between $\up\delta 0(0)$ and $\up\delta 1(0)$ can offset the metric jump. In fact, for the matter $\bar {\mathcal S}$ within $\Sigma$, to an observer inside $\up V i$, it corresponds to the energy-momentum tensor $\up\delta i(\ell)\bar{\mathcal S}$. Thus, we can arbitrarily choose a metric, such as $\up g 0$, as a standard. Then, using this standard metric, we can calculate the Ricci curvature and Einstein tensor to obtain the Einstein field equations. Eq.\eqref{delta0=b.delta1} ensures that the junction condition is independent of the choice of the standard metric.

On the other hand, by substituting $\up n 0$ with $\up n1$ in the definition \eqref{n0betan1RXn0U}, we can directly obtain
\begin{eqnarray}
	\label{betan1n0RXn0U}
	\notag
	&&\up R\Sigma(\bar X,\up n 1)\bar U\\
	\notag
	&&=\up\delta 1(0)\cdot\mathsf s\bigg(\up K 1(\bar X,\bar U)+\beta\up K 0(\bar X,\bar U)\bigg)\up n 1\\
	\notag
	&& \quad -\up\delta 1(0)\cdot\mathsf s\,\up K 0(\bar X,\bar U)\bar w\\
	 &&\quad+\bigg\{\up\Theta 1(0)\up R1+\big(1-\up\Theta 1(0)\big)\up R 0\bigg\}(\bar X,\up n 1)\bar U.\qquad \;
\end{eqnarray}
Furthermore, it is clear that the component $\up R\Sigma(\bar X,\bar Y)\bar U$ will remain unaffected by metric jumps; therefore, it continues to be defined according to Eq.~\eqref{RSigmaXYU}. The linearity of $\up R\Sigma$ requires
\begin{eqnarray}
	\label{RSigmalinear}
	\notag
	&&\up R\Sigma(\bar X,\up n 0+\beta\up n 1)\bar U\\
	\notag
	&&=\up R\Sigma(\bar X,\up n 0)\bar U+\beta\up R\Sigma(\bar X,\up n 1)\bar U\\
	&& =\up R\Sigma(\bar X,\bar w)\bar U.
\end{eqnarray}
By substituting Eqs.~\eqref{n0betan1RXn0U} and \eqref{betan1n0RXn0U}, along with Eqs.~\eqref{Theta0+Theta1} and \eqref{delta0=b.delta1}, the above equation holds. Proceed with the definition of the component $\up R\Sigma(\bar X,\up n 0)\up n0$. Replace $\bar U$ with $\up n 0$ in Eq.~\eqref{n0betan1RXn0U}, 
\begin{widetext}
\begin{eqnarray}
	\label{n0betan1RXn0n0}
	\notag
	\up R\Sigma(\bar X,\up n 0)\up n0&=&-\up\delta0(0)\left(\up\nabla0_{\bar X}\up n0-\up\nabla1_{\bar X}\up n0\right)+\bigg\{\up\Theta0(0)\up R0+\up\Theta1(0)\up R1\bigg\}(\bar X,\up n0)\up n0\\
	\notag
	&=&-\up\delta 0(0)\left(\up\nabla0_{\bar X}\up n0+\beta\up\nabla1_{\bar X}\up n1+\big(\bar X\beta\big)\up n 1-\up\nabla 1_{\bar X}\bar w\right)\\
	&&+\bigg\{\up\Theta0(0)\up R0+\up\Theta1(0)\up R1\bigg\}(\bar X,\up n0)\up n0.
\end{eqnarray}
\end{widetext}
Here, $\bar X\beta$ represents the directional derivative of the function $\beta$ along the vector $\bar X$. For the final component $\up R\Sigma(\bar X,\bar Y)\up n0$, it remains impervious to metric jumps and is consequently defined in accordance with Eq.~\eqref{RSigmaXYn}. Easy to verify, for $i=0,1$,
\begin{eqnarray}
	\notag
	&&\up R\Sigma(\bar X,\up n i)\up n0+\beta\up R\Sigma(\bar X,\up n i)\up n 1=\up R\Sigma(\bar X,\up n i)\bar w,\\
	\notag
	&&\up R\Sigma(\bar X,\up n 0)\up n i+\beta\up R\Sigma(\bar X,\up n 1)\up n i=\up R\Sigma(\bar X,\bar w)\up n i.\\
\end{eqnarray}

Next, we investigate whether $\up R\Sigma$ satisfies all the algebraic properties expected of the curvature tensor. Firstly, the antisymmetry $\up R\Sigma(X,Y)=-\up R\Sigma(Y,X)$ is automatically satisfied according to our definition of $\up R\Sigma$. Then, we consider the cyclic symmetry, $\up R\Sigma(X,Y)Z+\up R\Sigma(Y,Z)X+\up R\Sigma(Z,X)Y=0$, a condition easily confirmed to hold. Therefore, only the following property remains,
\begin{equation}
	\up g0\big(\up R\Sigma(X,Y)U,Z\big)=-\up g0\big(\up R\Sigma(X,Y)Z,U\big),
\end{equation}
where we have chosen $\up g 0$ as the standard metric. We will analyze all components, investigating the constraints on $\bar w$ and $\beta$ imposed by the above equation. Taking $X, Y$ as $\bar X, \bar Y$ which tangent to $\Sigma$, and utilizing Eqs.~\eqref{RSigmaXYU} and \eqref{RSigmaXYn}, the mentioned equation is automatically satisfied. Moving forward,
\begin{eqnarray}
	\notag
	&&\up g0\big(\up R\Sigma(\bar X,\up n0)\up n0,\up n 0\big)\\
	&&=\up\delta 0(0)\cdot\up g 0\bigl(\up \nabla 1_{\bar X}\bar w-(\bar X\beta)\up n 1,\up n 0\bigr)=0\qquad 
\end{eqnarray}
will yield
\begin{equation}
	\label{antisymXnnn}
	\bar X\beta+\mathsf s\,\up K 1(\bar X,\bar w)=0.
\end{equation}
Then, 
\begin{equation}
	\up g0\big(\up R\Sigma(\bar X,\up n0)\bar U,\bar Y\big)+\up g0\big(\up R\Sigma(\bar X,\up n0)\bar Y,\bar U\big)=0
\end{equation}
will yield
\begin{equation}
	\label{antisymXnUY}
	\up K 1(\bar X,\bar U)h(\bar w,\bar Y)+\up K 1(\bar X,\bar Y)h(\bar w,\bar U)=0.
\end{equation}
Finally, according to the following equation,
\begin{eqnarray}
	\notag
	&&\up g0\big(\up R\Sigma(\bar X,\up n0)\bar U,\up n 0\big)\\
	&&=\up\delta 0(0)\bigg(\up K 0+\frac{1}{\beta}\up K 1\bigg)(\bar X,\bar U)+\text{others},
\end{eqnarray}
\begin{widetext}
\begin{eqnarray}
	\notag
	-\up g0\big(\up R\Sigma(\bar X,\up n0)\up n 0,\bar U\big)&=&\up\delta 0(0)\bigg\{h\bigl(\up\nabla 0_{\bar X}\up n 0,\bar U\bigr) +\beta h\bigl(\up\nabla 1_{\bar X}\up n 1,\bar U\bigr) +\frac{\bar X\beta}{\beta}\up g 0\bigl(\bar w-\up n 0,\bar U\bigr)\\
	\notag
	&&-\up g 0\bigl(\bar\nabla_{\bar X}\bar w-\mathsf s\,\up K 1(\bar X,\bar w)\up n 1,\bar U\bigr)\bigg\}+\text{others}\\
	\notag
	&=&\up\delta 0(0)\bigg\{\big(\up K 0+\beta\,\up K 1\big)(\bar X,\bar U)+\frac{\bar X\beta}{\beta}h\bigl(\bar w,\bar U\bigr)-h\bigl(\bar\nabla_{\bar X}\bar w,\bar U\bigr)\\
	\notag
	&&+\frac{\mathsf s}{\beta}\up K 1(\bar X,\bar w) h(\bar w,\bar U)\bigg\}+\text{others}\\
	&=&	\up\delta 0(0)\bigg\{\big(\up K 0+\beta\,\up K 1\big)(\bar X,\bar U)-h\bigl(\bar\nabla_{\bar X}\bar w,\bar U\bigr)\bigg\} +\text{others},
\end{eqnarray}
\end{widetext}
where Eq.~\eqref{antisymXnnn} has been utilized, and ``others'' denotes terms related to $\up\Theta 0(0)$.  The constraint
\begin{equation}
	\up g0\big(\up R\Sigma(\bar X,\up n0)\bar U,\up n 0\big)+\up g0\big(\up R\Sigma(\bar X,\up n0)\up n 0,\bar U\big)=0
\end{equation}
will yield
\begin{equation}
	\label{antisymXnnU}
	(\beta^2-1)\up K 1(\bar X,\bar U)=\beta h\bigl(\bar\nabla_{\bar X}\bar w,\bar U\bigr).
\end{equation}
If $\beta^2\not=1$, the equation above entirely defines the values of $\up K 1$. Recognizing the symmetry of $\up K 1(\bar X,\bar U)$ with respect to $\bar X,\bar U$, we can further derive a constraint equation for $\bar w$,
\begin{equation}
	h\bigl(\bar\nabla_{\bar X}\bar w,\bar U\bigr)=h\bigl(\bar\nabla_{\bar U}\bar w,\bar X\bigr).
\end{equation}
Under this condition, it is easy to prove that
\begin{eqnarray}
	\notag
	(\beta^2-1)\up K 1(\bar X,\bar U)&=&\frac{1}{2}\beta\bigg\{h\bigl(\bar\nabla_{\bar X}\bar w,\bar U\bigr)+h\bigl(\bar\nabla_{\bar U}\bar w,\bar X\bigr)\bigg\}\\
	&=&\frac{1}{2}\beta\,\mathfrak L_{\bar w} h(\bar X,\bar U).
\end{eqnarray}
Notice that Eq.~\eqref{antisymXnnn}\eqref{antisymXnUY} and \eqref{antisymXnnU} involve only $\up K 1$, which is a consequence of the choice of the standard metric. By substituting $[1]$ with $[0]$, $\beta$ with $1/\beta$, and $\bar w$ with $\bar w/\beta$, Eqs.~\eqref{antisymXnnn}\eqref{antisymXnUY} and \eqref{antisymXnnU} undergo transformation into
\begin{eqnarray}
	\notag
	&&\beta\up K 0(\bar w,\cdot)=\mathrm d\beta,\\
	\notag
	&& \up K 0(\bar X,\bar U)h(\bar w,\bar Y)+\up K 0(\bar X,\bar Y)h(\bar w,\bar U)=0,\\
	\notag
	&&2\beta(1-\beta^2)\up K 0=\beta \mathfrak L_{\bar w}h-\mathrm d\beta\otimes h(\bar w,\cdot)-h(\bar w,\cdot)\otimes \mathrm d\beta.\\
\end{eqnarray}
We examine whether there exist nontrivial vector field $\bar w$ and function $\beta$ that satisfy all these constraint conditions. Taking $\bar U=\bar Y$ in Eq.~\eqref{antisymXnUY}, we can observe that
\begin{equation}
	\up K 1(\bar X,\bar Y) h(\bar w,\bar Y)=0
\end{equation}
holds for any $\bar X,\bar Y$. We aim to avoid restrictions on the values of $\up K 1(\bar X,\bar Y)$, as it would limit our utilization of the Einstein field equations. Consequently, we must ensure that $h(\bar w,\bar Y)=0$ holds for any $\bar Y$. The non-degeneracy of the metric $h$ implies that
\begin{equation}
	\bar w=0.
\end{equation}
Utilizing Eq.~\eqref{antisymXnnU}, we obtain
\begin{equation}
	\beta^2=1,
\end{equation}
namely $\beta=1$ everywhere since we have established the convention that $\beta>0$. Therefore Eq.~\eqref{n0betan1barw} returns to the scenario where the continuity condition \eqref{dim2continu} holds. In turn, when $\beta=1$ and $\bar w=0$, we have demonstrated the well-defined nature of the curvature $\up R\Sigma$ and the junction condition. It is evident that for the entire spacetime formed by gluing together two bulks, if there exists a consistent differential structure at $\Sigma$, then, concerning the junction condition, the continuity condition is both necessary and sufficient.

Consider another possibility where $\up n 0$ and $\up n 1$ are linearly independent. In this context, $\up n 0$ and $\up n 1$ belong to distinct vector spaces, precluding linear operations between them. Consequently, the definition method of Eq.~\eqref{n0betan1RXn0U} has proven ineffective, thus resulting in the absence of a corresponding junction condition. From a physical perspective, if only two bulks exist and the normal vectors remain linearly independent, then there is necessarily no interaction occurring between $\up V 0$ and $\up V 1$ (as well as the fields they carry). Otherwise, the interaction between two bulks that do not satisfy the junction condition would inevitably result in the generation of an additional bulk. This is due to the junction condition implying ``momentum conservation.'' Following this line of thought, we propose a possible solution. The normals vectors $\up n 0$ and $\up n 1$ can form a $2$-dimensional vector space denoted as $\Span\big\{\up n0,\up n 1\big\}$.   Then $\up V 0$ and $\up V 1$ can be embedded into a $d+2$-dimensional differential manifold $\mathscr M$. At $\Sigma$, the tangent space $\mathscr T_\Sigma(\mathscr M)$ should be isomorphic to $\mathscr T(\Sigma)\oplus \Span\big\{\up n0,\up n 1\big\}$. Consequently, we can arbitrarily select a vector $\up n 2\in \Span\big\{\up n0,\up n 1\big\}$, which can be expressed as a linear combination of $\up n 0$ and $\up n 1$, i.e., $\up n 2=\beta_0\up n 0+\beta_1\up n 1$. Clearly, neither $\beta_0$ nor $\beta_1$ is equal to $0$. Starting from $\Sigma$ and extending along $\up n 2$, we can construct a bluk $\up V 2$ and define a metric $\up g 2$, in which $\up n 2$ becomes the normal vector of $\Sigma$ within $\up V 2$. At the moment, an extrinsic curvature $\up K 2$ exists. Therefore, the problem now revolves around finding the junction conditions satisfied by $\up K i$ when gluing $\up V i$ together along $\Sigma$ where $i=0,1,2$. This is also one of the motivations behind our efforts to generalize the classical Israel junction condition.

Further clarification is required, as $\up n 0$ and $\up n 1$ may not be linearly independent at all points of $\Sigma$. A plausible scenario could unfold: at specific points within $\Sigma$, $\up V 0$ and $\up V 1$ are smoothly joined, indicating that their normal vectors satisfy the relation $\up n 0+\beta\up n 1 =\bar w$. This assumption holds true for points forming the set $\mathcal V$. Conversely, at points within the complement $\mathcal W=\Sigma-\mathcal V$, $\up n 0$ and $\up n 1$ demonstrate linear independence. To be compatible with this particular situation, we need to amend the condition $\up n 2=\beta_0\up n 0+\beta_1\up n1$ to
\begin{equation}
	\label{dim3contigen}
	\beta_0\up n 0+\beta_1\up n1+\beta_2\up n2=\bar w.
\end{equation}
Within the set $\mathcal V$, $\up n2$ is linearly independent of $\up n0,\up n1$, implying that $\beta_2=0$. Inside $\mathcal W$, we can set $\beta_0,\beta_1,\beta_2$ all not equal to $0$ to achieve a well-defined joining structure. However, a crucial prerequisite is necessary --- $\mathcal W$ must be an open set. This is because $\beta_2$ is a continuous function, and $\mathcal V=\beta_2^{-1}\big(\{0\}\big)$ is a closed subset. Additionally, the junction condition is a local property and can be determined within an open set, and the continuity at the boundary between $\mathcal W$ and $\mathcal V$ needs to be ensured. Fortunately, Sec.~\ref{subsec5A} rigorously proves that $\mathcal W$ is an open set. Then some natural questions arise: what constraints should the functions $\beta_i$ and the vector field $\bar w$ satisfy? Emulating \eqref{dim2continu}, is the extension of continuity conditions, as in 
\begin{equation}
	\up n 0+\up n 1+\up n 2=0
\end{equation}
necessary? We provide rigorous answers to these questions in Sec.~\ref{subsec5A}. Currently, to explore methods in simpler scenarios, we temporarily assume that the gluing of the three bulks satisfies the above continuity condition, initiating the discussion in Sec.~\ref{sec4}.

\section{The relation among the energy-momentum tensors\label{subsec4B}}
In the Sec.~\ref{sec4}, we observe that the component $\up\tau{-0}_{\ell \up n {-0}}{\up n {-0}}$ of the connection is contingent on the specific extension of $\up V{-0}$. This implies that different extensions of $\up V{-0}$ might give rise to distinct Riemann curvatures at $\Sigma$, which in turn results in different Einstein tensors $\up G{-0}$. Therefore, it becomes imperative to scrutinize the relationship between $\up G{-0}$ and the energy-momentum tensors $\up T1$ and $\up T2$.

We will commence by calculating such a component of $\up R{-0}$ whose all directions are tangent to $\Sigma$. This will be done by employing the Gauss equation, while abbreviating $\up g i\big(\up R i(\bar X,\bar Y)\bar U,\bar W\big)$ and $h\big(\bar{\mathcal R}(\bar X,\bar Y)\bar U,\bar W\big)$ as $\up R i(\bar X,\bar Y,\bar U,\bar W)$ and $\bar{\mathcal R}(\bar X,\bar Y,\bar U,\bar W)$,
\begin{eqnarray}
	\label{GaussEq}
	\notag
	&&\big(\up R i-\bar{\mathcal R}\big)(\bar X,\bar Y,\bar U,\bar W)
	\\
	&&=\up K{i}(\bar X,\bar U)\up K{i}(\bar Y,\bar W)-\big(\bar U \longleftrightarrow \bar W\big)\,,
\end{eqnarray}
where $\mathcal{\bar R}$ represent the intrinsic Riemann curvature tensor of $\Sigma$, and $[i]\in\big\{[-0],[1],[2]\big\}$. While using Eq.~\eqref{K-0}, we obtain
\begin{eqnarray}
	\label{R0=R1+R2-R+K}
	\notag
	&&\up R{-0}(\bar X,\bar Y,\bar U,\bar W)\\
	\notag
	&&=\big(\up R{1}+\up R2-\bar {\mathcal R}\big)(\bar X,\bar Y,\bar U,\bar W)\\
	\notag
	&&\quad +\bigg\{\up K{1}(\bar X,\bar U)\up K{2}(\bar Y,\bar W) +\up K{2}(\bar X,\bar U)\up K{1}(\bar Y,\bar W)\\
	&&\quad - \big(\bar U \longleftrightarrow \bar W\big)\bigg\}.
\end{eqnarray}
Next, we compute such a curvature component in which, among the 4 indices only one aligns with the direction of the normal vector. This calculation involves Gauss equation
\begin{widetext}
\begin{eqnarray}
	\label{GaussEqforM12}
	\big(\up R {1,2}-\up R i\big)(\bar X,\up n i,\bar U,\bar W)=\up {\mathscr K}i(\bar X,\bar U)\up {\mathscr K}i(\up n i,\bar W)-\big(\bar U \longleftrightarrow \bar W\big)
\end{eqnarray}
for $\up R{1,2}$, where $[i]\in\big\{[-0],[1],[2]\big\}$. While considering Eq.~\eqref{scrK0=K1=K2} and the first equality in Eq.~\eqref{scrKcalKXU}, we have
\begin{eqnarray}
	\label{R0-R1-R2}
	\notag
		\big(\up R{1,2}-\up R{-0}\big)(\bar X,\up n{-0},\bar U,\bar W)&=&\up {\mathscr K}{-0}(\bar X,\bar U)\up {\mathscr K}{-0}(\up n{-0},\bar W)-\big(\bar U\longleftrightarrow\bar W\big)\\
	\notag
	&=&\up \K1(\bar X,\bar U)\up \K1(\up n1,\bar W)+\up \K2(\bar X,\bar U)\up \K2(\up n2,\bar W)-\big(\bar U\longleftrightarrow \bar W\big)\\
	&=&\up R{1,2}(\bar X,\up n1+\up n2,\bar U,\bar W)-\up R{1}(\bar X,\up n1,\bar U,\bar W) -\up R{2}(\bar X,\up n2,\bar U,\bar W),\quad
\end{eqnarray}
\end{widetext}
which leads to 
\begin{eqnarray}
	\label{R0n0=R1n1+R2n2}
	\notag
	&&\up R{-0}(\bar X,\up n{-0},\bar U,\bar W)\\
	&&=\up R1(\bar X,\up n1,\bar U,\bar W)+\up R2(\bar X,\up n2,\bar U,\bar W).\qquad 
\end{eqnarray}

We will momentarily suspend our calculations to provide some insights. The components $\up R{-0}(\bar X,\up n{-0},\bar U,\bar W)$ exhibit distinct behavior when compared to the components $\up R{-0}(\bar X,\bar Y,\bar U,\bar W)$. According to Eq.~\eqref{R0=R1+R2-R+K}, we observe that, in addition to the curvature tensors $\up R1$ and $\up R2$, the coupling of extrinsic curvatures $\up K1$ and $\up K2$, together with the intrinsic curvature $\bar{\mathcal R}$ of $\Sigma$, all contribute to $\up R{-0}(\bar X,\bar Y,\bar U,\bar W)$. Note that all of these contributions are local. Specifically, the contribution of the coupling term of extrinsic curvatures stems from the interaction between $\up V1$ and $\up V2$ at $\Sigma$. 

However, as per Eq.~\eqref{R0n0=R1n1+R2n2}, when dealing with $\up R{-0}(\bar X,\up n{-0},\bar U,\bar W)$,  it simply amounts to the sum of  $\up R1$ and $\up R2$. This is attributed to the fact that the area formed by $\bar X\otimes\bar U$ is independent of the area formed by $\up n i\otimes\bar W$. One is tangent to $\Sigma$, and the other is normal to $\Sigma$. Each tensor $\mathscr K^{[i]}$ acting on these two regions will exhibit different properties. As previously mentioned in Sec.~\ref{sec4}, the normal space exhibits rotational symmetry. Therefore, Eq.~\eqref{scrK0=K1=K2} allows us to rotate $\up n i\otimes\bar W$ to align with $\up n{-0}\otimes\bar W$, where $i=1,2$. Subsequently, the contributions $\mathscr K^{[i]}(\bar X,\bar U)$ of $\bar X\otimes\bar U$ within different Pages $\up V i$ can be directly summed according to the first equality of Eq.~\eqref{scrKcalKXU}, where $i=1,2$. 

The remaining curvature components include two $\up n{-0}$ directions. This particular component is fundamentally different from $\up R{-0}(\bar X,\bar Y,\bar U,\bar W)$ and $\up R{-0}(\bar X,\up n{-0},\bar U,\bar W)$ that can be entirely determined by the local geometric of $\up V1$ and $\up V2$ at $\Sigma$. That is to say, $\up R{-0}(\bar X,\up n{-0},\up n{-0},\bar W)$ is non-local and depends on how $\up M{1,2}$ are constructed or how $\up V{-0}$ is extended. We can only calculate this component using the definition \eqref{defineRi}. To simplify the calculation, for any vector field $\bar X$ tangent to $\Sigma$ and the normal vector $\up n{-0}$, we extend them to other regions of $\up V{-0}$ in a way that $\up\nabla{-0}_{\up n0}\up n 0=0$ and $\big[\up n{-0},\bar X\big]=0$. This extension method is different from the one used in Sec.~\ref{subsec5B}. It can be easily proved that such an extension exists, but outside of $\Sigma$, $\bar X$ and $\up n{-0}$ may not remain orthogonal. Let $\Sigma$ perform a Lie translation along the extended vector field $\up n{-0}$ whose parameter is referred to as $\ell$, generating a family of hypersurfaces $\Sigma_\ell$.  Note that at $\ell\not=0$, the $\up K{-0}(\bar X,\bar Y)$ cannot be interpreted as the extrinsic curvature of $\Sigma_\ell$. Then,
\begin{eqnarray}
	\notag
	&&\up R{-0}\big(\bar X,\up n{-0},\up n{-0},\bar W\big)\\
	\notag
	&&=-\up g{-0}\big(\up\nabla{-0}_{\up n{-0}}\up\nabla{-0}_{\bar X}\up n{-0},\bar W\big)\\
	&&=\big(\up K{-0}\cdot \up K{-0}-\mathfrak L_{\up n{-0}}\up K{-0}\big)(\bar X,\bar W),
\end{eqnarray}
where $\up K{-0}\cdot \up K{-0}$ denotes the contraction of two tensors with adjacent indices, using the metric tensor $h$. The same result holds for $\up R{i}\big(\bar X,\up n{i},\up n{i},\bar W\big)$. Combining these outcomes, we obtain
\begin{eqnarray}
	\label{R0XnnW}
	\notag
	&&\up R{-0}\big(\bar X,\up n{-0},\up n{-0},\bar W\big)\\
	\notag
	&&=\big(\mathfrak L_{\up n1} \up K1+\mathfrak L_{\up n2}\up K2-\mathfrak L_{\up n{-0}} \up K{-0}\big)(\bar X,\bar W)\\
	\notag
	&&\quad +\up R1(\bar X,\up n1,\up n1,\bar W)+\up R2(\bar X,\up n2,\up n2,\bar W)\\
	&&\quad +\big(\up K1\cdot \up K2 +\up K2\cdot\up K1\big)(\bar X,\bar W),
\end{eqnarray}
here, the term $\mathfrak L_{\up n{-0}} \up K{-0}$ of Eq.~\eqref{R0XnnW} is entirely non-local. Its behavior is contingent upon the variations in extrinsic curvature along the direction of $\up n{-0}$. Consequently, it is intimately connected to the overall ``shape'' of $\up V{-0}$. To support this claim, note that $\mathfrak L_{\up n{-0}} \up K{-0}$ operates independently of $\mathfrak L_{\up n i} \up Ki,\;i=1,2$. The part in the third and fourth rows of Eq.~\eqref{R0XnnW} originates from the contributions of the curvatures of $\up V1$ and $\up V2$, as well as the interactions between them, both of which are local.

Using all components of the curvature tensor, the Einstein tensor can be computed,
\begin{widetext}
\begin{subequations}
	\begin{eqnarray}
		\notag
		&&\up G{-0}(\bar X,\bar Y)=\bigg\{\up G 1+\up G 2-\bar{\mathcal G}+2\bigl(\up K1\cdot\up K2+\up K 2\cdot \up K1\bigr)-\big(\mathrm{tr}_h\up K2\big)\up K1-\big(\mathrm{tr}_h\up K1\big)\up K2
		+\big\{\mathrm{tr}_h\up K1\,\mathrm{tr}_h \up K2
		\\
		\notag
		&&\hspace{2.3cm} -3\,\mathrm{tr}_h\big(\up K 1\cdot \up K2\big)\,\big\} h-\big\{\mathfrak L_{\up n {-0}}\up K {-0}-\mathrm{tr}_h\bigl(\mathfrak L_{\up n {-0}}\up K {-0}\bigr)h\big\}+\sum_{i=1}^2\mathfrak L_{\up n i}\up K i-\mathrm{tr}_h\bigl(\mathfrak L_{\up n i}\up K i\bigr)h\bigg\}(\bar X,\bar Y),\\
		\\
		&&\up G{-0}(\bar X,\up n{-0})=\up G{1}(\bar X,\up n1)+\up G{2}(\bar X,\up n2),\\
		&&\up G{-0}(\up n{-0},\up n{-0})=\up G{1}(\up n{1},\up n{1})+\up G{2}(\up n{2},\up n{2})+\frac{1}{2}\mathrm{tr}_h\overline{\Ric}+\mathrm{tr}_h\up K1\,\mathrm{tr}_h \up K2-\mathrm{tr}_h\big(\up K 1\cdot \up K2\big).
	\end{eqnarray}
\end{subequations}
\end{widetext}
Because $\up V{-0}$ serves as a purely auxiliary manifold introduced by us, its geometric or gravitational structure results from an extension, rather than being induced by actual matter. Conversely, by relying on the gravitational structure we have established for $\up V{-0}$, we can introduce an equivalent energy-momentum tensor denoted as $\up T{-0}=\frac{1}{8\pi G}\up G{-0}$.

The equivalent energy-momentum tensor $\up T{-0}$ naturally cannot provide additional dynamics, but it enables us to analyze the composition of gravitational interactions among the individual pages. We observe that, apart from along the direction $(\bar X,\up n{-0})$, $\up T{-0}$ does not exhibit linear dependence on $\up T1$ and $\up T2$. This is because, $\up T i(\up n i,\bar X)$ represents the energy flux of the matter within $\up V i$ passing through the hypersurface $\Sigma$, which can be directly summed. The interaction between $\up V1$ and $\up V2$ at $\Sigma$, manifested as the coupling of extrinsic curvatures, also contributes to $\up T{-0}$. Additionally, the contributions of non-local terms are present, expressed in the form of $\mathfrak L_{\up n {-0}}\up K {-0}-\mathrm{tr}_h\big(\mathfrak L_{\up n {-0}}\up K {-0}\big)h$. The non-local contribution to the equivalent matter at $\Sigma$ originates from the influence of adjacent regions of $\up V {-0}$ outside of $\Sigma$. Consequently, the inclusion of these terms is not only reasonable but also essential. In bulk/brane structures, non-local gravity is prevalent, and the reference \cite{PhysRevD.62.024012} provides a more simplified example. 

In summary, $\up V 1$ and $\up V 2$ can induce gravity on the higher-dimensional manifold $\mathscr M$, and then $\mathscr M$ will reduce to an effective gravity in the auxiliary bulk $\up V{-0}$. In this appendix, we analyzed the composition of this effective gravity.

\section{Proof that booklet cannot be a manifold}\label{AppenManifold}
In this appendix, we will prove the following proposition:
\begin{theorem}
	\label{bookletmanifold}
	When $m>1$, namely the number of pages is greater than $2$, the booklet $\mathscr V$ cannot be a manifold.
\end{theorem}
\begin{proof}
	Using a proof by contradiction, let us assume $\mathscr V$ is a manifold. We first determine the dimension of this manifold. Consider any open set $\up W0$ of $\up V0$ such that $\up W0\cap\Sigma=\varnothing$, making $\up W0$ an open set of $\mathscr V$. Since $\up W0$ is $d+1$ dimensional, according to the definition of a manifold, open sets of a manifold should have the same dimension as the manifold itself. Hence, $\mathscr V$ should also be $d+1$ dimensional (alternatively, one could directly show that the Lebesgue covering dimension of $\mathscr V$ is $d+1$, which should be equal to its manifold dimension~\cite{van1988infinite}). Therefore, a homeomorphism $\psi$ can be established between the local open sets $\mathscr W$ of $\mathscr V$ and $\mathbb W$ of $\mathbb R^{d+1}$, where $\widetilde\Sigma=\mathscr W\cap\Sigma$ is defined. We can make $\mathscr W$ small enough so that $\widetilde\Sigma$ is connected. The coordinate homeomorphism $\psi$ maps $\widetilde\Sigma$ into $\mathbb R^{d+1}$, resulting in a $d$-dimensional hypersurface $\psi(\widetilde\Sigma)$. Furthermore, for any point $o\in\psi(\widetilde\Sigma)$, we take a ball neighborhood $\mathbb B_o(\epsilon)$ of radius $\epsilon$. When $\epsilon$ is sufficiently small, $\mathbb B_o(\epsilon)\cap\psi(\widetilde\Sigma)$ is orientable, so the small open ball $\mathbb B_o(\epsilon)$ will be divided by the hypersurface $\widetilde\Sigma$ into two disjoint parts, $\mathbb B_o(\epsilon)-\psi(\widetilde\Sigma)=\mathbb B_+\cup \mathbb B_-$, and $\mathbb B_\pm$ are both connected. This is a corollary of the Jordan-Brouwer Separation Theorem\cite{lima1988jordan,Vizoso2020}.

	Now we will prove that $\psi^{-1}(\mathbb B_+)$ must be entirely contained within one of the $\up Vi-\Sigma$. If this is not true, consider two points $p,q\in\mathbb B_+$ such that their inverse-images $\psi^{-1}(p)\in\up Vi-\Sigma$ and $\psi^{-1}(q)\in\up Vj-\Sigma$. We can then connect $p$ and $q$ with a continuous curve $\mathscr X$ in $\mathbb B_+$, ensuring that $\mathscr X$ does not intersect $\psi(\widetilde\Sigma)$. The curve $\psi^{-1}(\mathscr X)$ then connects $\psi^{-1}(p)$ and $\psi^{-1}(q)$, avoiding intersection with $\Sigma$. However, this is impossible, as a curve connecting points in different pages must cross $\Sigma$. To be more precise, since $\psi^{-1}(\mathscr X)\cap\Sigma=\varnothing$, it must be $\psi^{-1}(\mathscr X)=\bigcup_{k=0}^m \psi^{-1}(\mathscr X)\cap(\up Vk-\Sigma)$. However, this implies $\psi^{-1}(\mathscr X)$ is a disconnected curve, as all $\up Vk-\Sigma$ are disjoint (noting that $\mathscr V\coloneqq \bigsqcup_{k=0}^m(\up Vk-\Sigma)\bigsqcup\Sigma$). But this is impossible, as $\psi$ is a continuous map, which preserves connectivity. Hence, $\psi^{-1}(\mathbb B_+)$ must be entirely contained in a certain $\up Vi-\Sigma$, and similarly, $\psi^{-1}(\mathbb B_-)$ must be entirely contained in another $\up Vj-\Sigma$. This implies $\psi^{-1}(\mathbb B_o(\epsilon))\subset\up Vi\cup\up Vj$. However, $\psi^{-1}(\mathbb B_o(\epsilon))$ should be an open set in $\mathscr V$ and $\psi^{-1}(\mathbb B_o(\epsilon))\cap\Sigma\neq\varnothing$, indicating $\psi^{-1}(\mathbb B_o(\epsilon))$ must intersect a third page $\up Vk-\Sigma$ at an open set, leading to a contradiction. Thus, $\mathscr V$ cannot be a manifold.
\end{proof}

The gluing given by the definition $\mathscr V=\bigsqcup_{i=0}^m\up V i/\sim$ only describes $\mathscr V$ as a set, and the pages remain independent in practice. To further define consistent gravity on $\mathscr V$, there must be certain constraints between the normal vectors. For the case of gluing two pages, this constraint is $\up n 0+\up n 1=0$; however, in the general case, each normal vector $\up n i$ belongs to a different tangent space $\mathscr T_\Sigma(\up V i)$. This indicates that the booklet has a branching structure at the junction, making it impossible to define algebraic operations or linear combinations between different normal vectors in the union of these tangent spaces. This further demonstrates that $\mathscr V$ is not a differential manifold. Now, starting from this idea, we further provide an alternative proof from the perspective of differential structure.

\begin{proof}
The booklet $\mathscr V$ has a branching structure at $\Sigma$, meaning all tangent spaces $\mathscr T_\Sigma(\up V i)$ are not uniform. The union $\bigcup_{i=0}^m\mathscr T_\Sigma(\up V i)$ cannot form a vector space, hence linear combinations of normal vectors cannot be tangent to $\mathscr V$. To demonstrate this, consider an arbitrary linear combination of two normal vectors, for instance, $\beta_0\up n 0+\beta_1\up n 1$, which cannot belong to $\mathscr T_\Sigma(\up V 0)$; otherwise, we would have $\beta_0\up n 0+\beta_1\up n 1=\alpha\up n 0+\bar w$, implying $\up n 1\in \mathscr T_\Sigma(\up V 0)$ and consequently, $\up V 0$ and $\up V1$ are tangent at $\Sigma$, i.e., they coincide in an infinitesimal neighborhood, leading to the degeneration of the gluing structure. Similarly, $\beta_0\up n 0+\beta_1\up n 1$ cannot belong to $\mathscr T_\Sigma(\up V 1)$; thus, it must belong to some other $\mathscr T_\Sigma(\up V i),\,i\not=0,1$. However, since the number of pages is finite, fixing $\beta_1$ and varying $\beta_0$ continuously will inevitably yield two distinct $\beta_0$ values making $\beta_0\up n 0+\beta_1\up n 1$ belong to the same $\mathscr T_\Sigma(\up V i)$, as $\beta_1$ is fixed, $\up n 1$ can be eliminated, indicating $\up n 0\in\mathscr T_\Sigma(\up V i)$, and $i\not=0$. This again leads to tangency between $\mathscr T_\Sigma(\up V i)$ and $\mathscr T_\Sigma(\up V 0)$. Ultimately, through induction, the gluing structure degenerates into the case of two pages, where all other pages coincide infinitesimally with these two pages. At this point, the booklet naturally becomes a differential manifold, which is evidently not what we desire.
\end{proof}

We have clearly recognized the fact that there is a lack of differential structure at the interface. To address this issue, we need a unified vector space that contains all the normal vectors as well as $\mathscr T(\Sigma)$. Naturally, we can embed the booklet $\mathscr V$ into an external manifold $\mathscr M$ of sufficient dimension, so that its tangent space $\mathscr T_\Sigma(\mathscr M)$ provides the stage for algebraic operations of the normal vectors. This naturally brings us back to the method used in Sec.~\ref{subsec5A}.

\section{Attempt to construct a closed loop\label{appx1a}}
In this appendix, we will present some preliminary ideas regarding the extension of the junction condition to the case of gluing together three bulks.
\begin{figure}[htbp!]
	\centering
	\includegraphics[width=0.55\columnwidth]{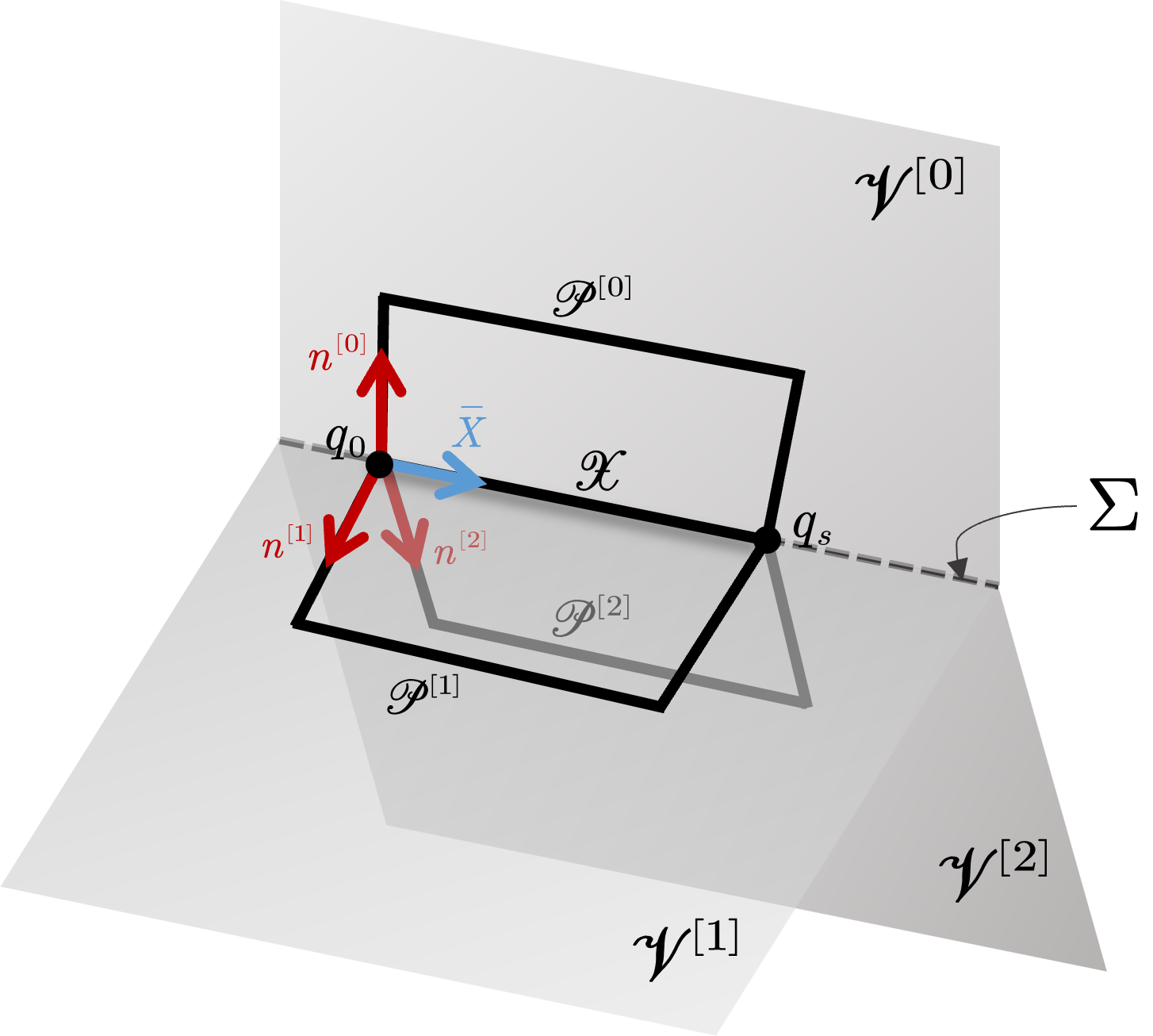}
	\caption{Constructing a closed loop by utilizing paths from 3 bulks.\label{3bulks}}
\end{figure}

Not difficult to discern, the booklet $\mathscr V=\bigcup_{i=0}^2\up V i$ lacks a differential structure at $\Sigma$, making it quite a formidable task to define $\up R\Sigma$. Our initial idea was to directly extrapolate the geometric construction depicted in Fig.~\ref{2bulks}, just like in Fig.~\ref{3bulks}. Let us assume that $\mathscr X$, which connects points $q_0$ and $q_s$, represents a curve tangent to $\Sigma$ and is parametrized by $s$, with its tangent vector at point $q_0$ denoted as $\bar X$. We establish three paths, denoted as $\up P i$, where $i=0,1,2$,  each of them connecting points $q_0$ and $q_s$, and located within their respective $\up V i$. Each path $\up P i$ is segmented into three sections. At point $q_0$, the tangent vector of $\up P i$ is $\up n i$, and at point $q_s$, the tangent vector is given by $-\mathfrak l_{s\bar X}\up n i$. However, the algebraic sum of these three paths, $\up{\mathscr P}0-\up{\mathscr P}1+\up{\mathscr P}2$, does not forms a closed loop. Hence, to complete the loop, we must include the integral curve $\mathscr X$ of the vector $\bar X$. This completes the loop, denoted as $\mathscr O = \up{\mathscr P} 0-\up{\mathscr P} 1+\up{\mathscr P}2-\mathscr X$. Subsequently, we parallel-translate the vector $\bar U$ tangent to $\Sigma$ or the vector $\up n i$ normal to $\Sigma$ along the loop $\mathscr O$, thereby defining $\up R\Sigma$. However, such construction poses numerous challenges in its finer details, and even if we disregard these issues, it still falls short of providing junction conditions that encompass the contribution of  all $\up Ki$. In addition, alternative methods for constructing a loop can be considered, such as  $\mathscr O^\prime = \up{\mathscr P} 0-\up{\mathscr P}2+\up{\mathscr P} 1-\mathscr X$. However, the $\mathscr O^\prime$ is not topologically equivalent to $\mathscr O$, disrupting the uniqueness of the definition of $\up R\Sigma$. After various attempts, we conclude that the aforementioned approach is ineffective. Hence, we opt for an alternative approach presented in the main text of Sec.~\ref{sec4}.

\bibliographystyle{apsrev4-2.bst}
\bibliography{Refs.bib}

\end{document}